\documentclass[aps,preperint,notitlepage,superscriptaddress]{revtex4-1}
\usepackage{graphicx}   
\usepackage{subfigure}  
\usepackage{algorithm}
\usepackage{algorithmic}
\usepackage{color}
\usepackage{soul}

\graphicspath{{./Figures/}}

\begin{document}
\title{Dark matter phenomenology of high speed galaxy cluster collisions}
\author{Yuriy Mishchenko}
\affiliation{Faculty of Engineering, Izmir University of Economics, Izmir 35330, TURKEY}

\author{Chueng-Ryong Ji}
\affiliation{Department of Physics, North Carolina State University, Raleigh NC 27695-8202, USA}

\begin{abstract}
We perform a general computational analysis of possible post-collision mass distributions in high-speed galaxy cluster collisions in the presence of self-interacting dark matter. Using this analysis, we show that astrophysically weakly self-interacting dark matter can impart subtle yet measurable features in the mass distributions of colliding galaxy clusters even without significant disruptions to the dark matter halos of the colliding galaxy clusters themselves. Most profound such evidences are found to reside in the tails of dark matter halos' distributions, in the space between the colliding galaxy clusters.
Such features appear in our simulations as shells of scattered dark matter expanding in alignment with the outgoing original galaxy clusters, contributing significant densities to projected mass distributions at large distances from collision centers and large scattering angles of up to $90^\circ$. 
Our simulations indicate that as much as 20\% of the total collision's mass may be deposited into such structures without noticeable disruptions to the main galaxy clusters.
Such structures at large scattering angles are forbidden in purely gravitational high-speed galaxy cluster collisions.
Convincing identification of such structures in real colliding galaxy clusters would be a clear indication of the self-interacting nature of dark matter.
Our findings may offer an explanation for the ring-like dark matter feature recently identified in the long-range reconstructions of the mass distribution of the colliding galaxy cluster CL0024+017. 
\end{abstract}

\maketitle

\section{Introduction}
\label{sec:introduction}
Dark matter and dark energy, comprising together 95\% of the energy budget in the Universe, remain among the biggest unsolved mysteries of modern physics. Dark matter (DM) has been described conventionally using the Cold Dark Matter (CDM) model, where the primary candidate for DM is an extremely massive ($m_{DM}\approx 10-1000$ GeV) particle interacting exclusively via the weak interaction - the so-called Weakly Interacting Massive Particle (WIMP) \cite{Bertone2004,Baudis2013}. In recent years, observations began to suggest that DM can be interacting with the cross-sections large enough to influence the formation of small-scale cosmological structures, the so called Self-Interacting Dark Matter (SIDM)  \cite{Buckley2010,Cyburt2002,Kochanek2000b,Pontzen2014,Wandelt2000a,Yoshida|2000|,Massey2015,Mishchenko2003,Williams2011,Weinberg2015}. Recent theoretical works put forth various models of such self-interacting DM including mirror DM \cite{Foot2014}, flavor-oscillating DM \cite{Medvedev2014}, SIDM \cite{Dawson2013,Hochberg2014,Jacobs2014,Robertson2017}
, etc.

Recent observations of colliding galaxy clusters provided a unique opportunity for gaining additional insights about the properties of DM empirically 
\cite{Bradac2009,Bradac2008,Clowe2006,Clowe2007}. 
The observed galaxy clusters may be regarded as natural astrophysical accelerators for high-energy DM particles' collisions. 
The observations of two or more galaxy clusters undergoing a high-speed central or near-central passage through each other after a gravitational in-fall
can offer new clues about the microscopic properties of DM
\cite{Clowe2003,Markevitch2004,Clowe2006,Bradac2006,Angus2006,
Springel2007,Mahdavi2007,Jee2007,Clowe2007,Randall2008,
Okabe2008,Nusser2008,Mastropietro2008,Deb2008,Bradac2008,Bradac2009,Coe2010,
Umetsu2011,Merten2011a,Ragozzine2012,Kneib2012,Jee2012,Clowe2012,
Lage2013,Paredes2015,Guzman2016}.
The bullet-type galaxy cluster collisions are such collisions that involve a smaller galaxy cluster, sometimes called the ``bullet'', falling onto a much larger cluster. In several cases of such collisions, a bullet-type galaxy cluster collision has been observed shortly after the passage of the bullet through the main cluster \cite{Clowe2003,Bradac2006}. Those observations evinced that the galaxy groups in the bullet-type galaxy cluster collisions exhibit a collisionless behavior, namely, passing through each other essentially freely  without interactions. On the other hand, the gas component of such colliding galaxy clusters---the intracluster medium (ICM)---exhibits a drastically different behavior with significant ram friction, super-sonic bow-shocks, and strong heating accompanied by X-ray emission as witnessed \cite{Angus2006,Okabe2008,Bradac2008,Bradac2009}.
One may ask which of these components the DM halos are co-localized with.
Subsequent reconstructions of the mass distribution in some of such bullet-type collisions by means of strong and weak gravitational lensing showed that the DM halos in these collisions are co-localized with the collisionless galaxy groups but not with the collisional ICM gas
\cite{Clowe2003,Markevitch2004,Clowe2006,Bradac2006,Angus2006,
Springel2007,Mahdavi2007,Jee2007,Clowe2007,Randall2008,
Okabe2008,Nusser2008,Mastropietro2008,Deb2008,Bradac2008,Bradac2009,Coe2010,
Umetsu2011,Merten2011a,Ragozzine2012,Kneib2012,Jee2012,Clowe2012,
Lage2013}. This co-localization led to conclusion that the material in the DM halos is collisionless much like the galaxy groups, rather than collisional like the ICM. Arguments such as the preservation of mass-to-light ratios and the coincidence of the centroids of DM halos with that of galaxy groups led to the constraints on the cross-section of possible SIDM particles at approximately $\sigma_{DM}/m_{DM} < 1\ cm^2g^{-1}$ 
\cite{Clowe2003,Markevitch2004,Mastropietro2008,Randall2008,Bradac2009,Merten2011a,Harvey2015}.

In this work, we perform a computational study of possible post-collision mass distributions that may be realized in high-speed collisions of galaxy clusters in the presence of weakly self-scattering DM. 
Several past computational studies focused on simulating the known colliding galaxy clusters and estimating the properties of the dark matter particles from that analysis
\cite{Sijacki2006,Springel2007,Angus2008,Mastropietro2008,
Nusser2008,Randall2008,Lage2013}. 
Here, we survey rather different possibilities for post-collision mass distributions in high-speed collisions of galaxy clusters under a variety of scenarios
in the presence of self-interacting DM. In that respect, our study does not focus on any specific galaxy cluster collision but aim at an explorative analysis of possible configurations that can be realized in galaxy cluster collisions in the presence of self-interacting DM.
For instance, we do not specifically simulate the classical Bullet cluster 1E 0657-56 although such Bullet cluster observations motivated 
our study.
A similar work in spirit has been recently published by Robertson, Massey and Eke \cite{Robertson2016,Robertson2017}, where the focus has been on the changes in the shape of DM halos in the presence of DM self-interactions. 
In particular, it has been found that shape changes can significantly affect and make unreliable simple analyses of SIDM effects such as the calculation of DM halo centroids' lag. 
Our work can be viewed in a complementary light as such investigating the effects of DM self-interactions in the space around the colliding galaxy clusters, in the tails of DM mass distributions, arising due to 
astrophysically weak 
DM self-interactions and manifesting in the projected mass density maps of galaxy cluster collisions in the regions between and around the colliding galaxy clusters. 
We find that while the DM particle interactions with $\sigma_{DM}/m_{DM}\approx 1\ cm^2g^{-1}$ and above cause severe disruption of colliding galaxy clusters, possibly leading to complete destruction and merger of their DM halos over cosmologically short time scales,
a range of weaker DM self-interactions $\sigma_{DM}/m_{DM} \approx 0.1-0.5\ cm^2g^{-1}$ can create weak yet detectable features in DM mass distributions around the colliding galaxy clusters, while not causing major distortions in the main galaxy clusters themselves. One such feature is the shell of scattered DM material that forms due to the scattering of DM particles off each other during the passage of the DM halos of colliding galaxy clusters through each another. Such shells can produce noticeable differences in the projected mass density maps of SIDM galaxy cluster collisions, in the form of extended concentrations of DM at large distances either from the collision center or the outgoing galaxy groups and large scattering angles. 
It is interesting that strikingly resembling structures can be spotted in many mass reconstructions of colliding galaxy clusters in the literature 
\cite{Kneib2012,Jee2012,Merten2011a,Okabe2008}.
A scattered DM shell may also explain the ring-like DM feature recently observed in a long-range reconstruction of the mass distribution in the galaxy cluster CL0024+17 \cite{Jee2007}.
Convincing observations of such features in high-speed galaxy cluster collisions can provide a clear evidence of the self-interacting nature of DM.

The remainder of the paper is organized as follows. In Section \ref{sec:methods}, we discuss the methodology of our study. In Section \ref{sec:results},  we survey different types of post-collision mass distributions with respect to the parameters such as  collision's speed, mass, DM self-interaction strength, etc. 
In this Section (subsection \ref{sec:observability}), we discuss the conditions necessary for the astrophysical observation of the effects associated with the self-interacting nature of DM in galaxy cluster collisions.
The summary and conclusions follow in Section \ref{sec:conclusions}. 
In Appendix \ref{AppA}, we provide the summary of the algorithms used in this work.
In Appendix \ref{AppB}, we present the numerical checks related to the convergence and accuracy of our numerical simulations.


\section{Methodology}
\label{sec:methods}

The bulk of our study focused on carrying out a set of simulations of galaxy cluster collisions using Particle Mesh method and collisional DM particles. In this section, we discuss the details of these simulations' initialization, evaluating DM particle collisions, evaluating gravitational evolution, and selecting the simulation parameters.

\subsection{Simulation's initial conditions}\label{sec:alg-initial-profile}

\begin{figure}[h!]
\begin{center}
\includegraphics[width=0.4\textwidth]{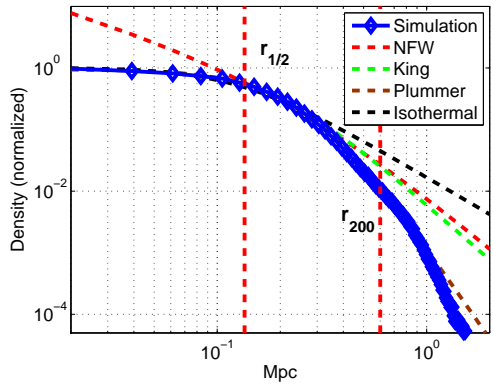}
\end{center}
\caption{\label{fig:iniprofile} The initial mass profile of SIDM halos used in this work, blue diamonds, is the the Plummer density shown with brown dashed line.
Also shown is the approximate King model's profile used in \cite{Clowe2003, Randall2008}, green dash-dotted line.
NFW profile \cite{nfw1996} (red dashed line) and the isothermal profile \cite{Bertin2000} (black dashed line, $\rho(r)=\rho_0/(1+r^2/r_c^2)$) are also shown for reference. The profile's 1/2-level half-width and the virial radius are indicated with labels $r_{1/2}$ and $r_{200}$, respectively.}
\end{figure}

We use Plummer profile \cite{Dejonghe1987,Bertin2000} obtained as the result of equilibrating a cloud of collisional DM particles in self-consistent gravitational potential to set up the initial particle distribution of colliding halos. Such Plummer profile reproduces closely the soft-core King model's mass profile, used in some literature in the past as an empirical model of SIDM halos \cite{Clowe2003, Randall2008}, up to the halo's virial radius $r_{200}$ and soft-truncates at $r_{200}$ as is seen Fig. \ref{fig:iniprofile}. 
Note that Navarro-Frenk-White (NFW) mass profile \cite{nfw1996}, popular in CDM literature, is unsuitable for modeling the collisional SIDM halos because NFW profile does not allow for the presence of soft halo cores necessarily present in SIDM halos \cite{Dave2001,Newman2013a,Newman2013b,Rocha2013,Peter2013,Weinberg2015,DelPopolo2016}.
Particle-particle scatterings of DM particles in the central regions of SIDM halos are known to reduce the central density of such halos resulting in an isothermal-like behavior of mass density at small radii, differently from the central cusp of NFW profile.
The so called approximate King model's profile, $\rho(r)=\rho_0/(1+r^2/r_c^2)^{3/2}$, had been used to {\em ad-hoc} model SIDM halos by introducing into NFW profile a soft core of radius $r_c$. 
The SIDM halo profile in this work has been obtained directly by equilibrating self-scattering halo of SIDM particles and can be indeed represented using such King profile very well.

The specific parameters of the initial particle distributions are as follows:
The profiles were created with $N=100,000$ particles with total mass of $2.5\cdot 10^{14}M_\odot$ at  per-particle mass of $1.125\cdot 10^{9}M_\odot$. The scale radius parameter of the Plummer density was $r_s=0.24$ Mpc, the core radius was $r_{c}=0.15$ Mpc, and the virial radius was $r_{200}= 0.6$ Mpc, corresponding to the effective concentration parameter of $c=r_{200}/r_c=4$.

\subsection{Simulation of SIDM particle-particle scattering}\label{sec:alg-dm-collision}

The non-gravitational interactions in SIDM halos, that is, the particle-particle scatterings of SIDM particles, were modeled by scattering simulated particles when they occupied the same cell of the simulation's spatial grid $\mathcal{G}$ (see the next subsection \ref{sec:alg-pmesh} for the definition of $\mathcal{G}$). 
That is, each pair of such particles scattered with the probability 
\begin{equation}\label{eq:collalpha}
P=\alpha V_{rel}\Delta t,
\end{equation}
where $V_{rel}=|\vec v_1-\vec v_2 |$ is the relative speed of the two particles, $\Delta t$ is the simulation time step, and $\alpha$ is an effective numerical parameter controlling the intensity of DM self-scattering. 
Robertson, Massey and Eke have addressed many of the science goals discussed in the present work by using essentially similar model with different anisotropic DM particle-particle interactions in their recent work \cite{Robertson2017}. In principle, it is conceivable that different particle physics models will predict very different scattering probabilities.
Fundamentally, in quantum field theoretic models, the coupling constant of particle interaction provided by a given model estimates the probability of 
particle scattering. The physical coupling constant in the renormalizable quantum field theory may be defined through multi-loop calculations 
with self-consistent regularization procedures. The scattering cross section predicted by such models would then be the key to test the validity of the models in comparison with experimental data. Lacking evidence for such sophisticated models at present, we assume in this work the simplest isotropic scattering model parametrized by a single cross-section parameter related to the numerical parameter $\alpha$ via
\begin{equation}\label{eq:alphasigmamass}
\frac{\sigma_{DM}}{m_{DM}}=\alpha \frac{N_{tot}d^3}{M_{tot}},
\end{equation}
where $M_{tot}$ is the total simulation mass, $N_{tot}$ is the total number of simulated particles, and $d$ is the resolution of the simulation grid, $\mathcal{G}$.
Then, the scattering of DM particles was evaluated as follows:
First, the Center-of-Mass velocity, $\vec V_{CM}=(\vec v_1+\vec v_2)/2$, and the relative speed, $V_{rel}=|\vec v_1-\vec v_2 |$, of the scattering particles were computed. 
Second, a new direction for the relative velocities, $\vec n$, was selected uniformly at random on a unit sphere, assuming elastic and isotropic scattering.
The velocities of the particles were then updated as 
\begin{equation}
\begin{array}{l}
\vec{v}'_{1}=\vec{V}_{CM}+ \frac12V_{rel} \vec n,\\
\vec{v}'_{2}=\vec{V}_{CM}- \frac12V_{rel} \vec n.\\
\end{array}
\end{equation}
Further details of this implementation of the DM particle-particle scattering in our simulations are presented in the algorithm in Appendix \ref{AppA}.

\subsection{Simulation of SIDM gravitational evolution}\label{sec:alg-pmesh}

Gravitational evolution of SIDM particles was implemented using Particle Mesh algorithm in Matlab. 
Namely, a continuous spatial distribution of mass $\rho(\vec x,t)$ was modeled by using a collection of $N$ particles $\vec r_i(t)$, $i=1\dots N$, distributed according to $\rho(\vec x,t)$. As the particles moved in common gravitational potential, $\Phi(\vec x,t)$, self-consistent $\Phi(\vec x,t)$ was calculated by approximating $\rho(\vec x,t)$ on a 3D grid $\mathcal{G}$ by counting the number of the particles in each cell of the grid, $n(\vec x_\mathcal{G},t)$, and numerically solving the Poisson equation
\begin{equation}\label{eqn:poisson}
\nabla^2 \Phi(\vec x,t) = 4\pi G n(\vec x,t),
\end{equation}
where $G$ is the gravitational constant.
The method of Fourier transform was used for the numerical solutions of Eq.~(\ref{eqn:poisson}), where for $\tilde \Phi(\vec k,t)=\int {d\vec x}{(2\pi)^{-3/2}}
e^{-i\vec k\cdot\vec x}\Phi(\vec x,t)$ we got
\begin{equation}\label{eqn:poissonmomentum}
\tilde \Phi(\vec k,t) = -4\pi G \frac{\tilde n(\vec k,t)}{\vec k^2}.
\end{equation}
Thus, $\Phi(\vec x,t)$ was computed by performing two discrete fast Fourier transforms, $n(\vec x,t)\rightarrow \tilde n(\vec k,t)$ and $\tilde \Phi(\vec k,t)\rightarrow \Phi(\vec x,t)$, and making use of Eq.~(\ref{eqn:poissonmomentum}) in between them. Once $\Phi(\vec x,t)$ had been computed, the speed and the positions of all the particles in the common gravitational potential were updated according to the regular Newtonian dynamics. 
A SIDM self-scattering step had been embedded into this algorithm as described in Section \ref{sec:alg-dm-collision}.
The simulation advanced by using an adaptive time step $\Delta t$ set from the restriction that the maximum change of the speed and the position of the simulated particles was below one grid-cell, and varied between 0.1 My and 10 My.
The detail of the Particle Mesh algorithm is also presented in Appendix \ref{AppA}.

The algorithm requires a set of parameters to be set, specifying the total mass of simulated collision, $M_{tot}$, the number of simulated particles in colliding SIDM halos, $N1$ and $N2$, the kinetic collision parameter defined as the square of the galaxy clusters' speed at infinity, $\Delta V2$, and controlling the collision velocity, the initial offsets of colliding halos, $\Delta R$, and the collision impact parameter $\Delta b$. The other algorithm parameters are the number of cells $D$ along each dimension in the cubic spatial grid $\mathcal{G}$ and the cells' dimension $d$, with the total of $D^3$ cells in the grid. These are related to the total linear size of the modeled region of space as $Dd$.

\subsection{Selection of simulation parameters}\label{sec:alg-parameters}

All simulations had been performed using the super-computing facility National Energy Research Scientific Computing Center (NERSC). The majority of simulations used $N=2\cdot 10^5$ total particles, the grid of $D^3=400^3$ cells with cell-resolution $d=15$ kpc and the region of space modeled being a cube of 6 Mpc on the side. Only the dynamics of SIDM halos was simulated and the ICM and the visible matter contributions were not included. All simulations spanned the duration of time $t_{max}$ of 1.5 Gy to 3 Gy chosen so as to cover a single passage of colliding SIDM halos through each other.

We simulated various collision scenarios with regard to the collision speed, collision centrality, symmetricity, and the SIDM self-interaction strength (expressed via the effective strength parameter $\alpha$ or equivalently, $\sigma_{DM}/m_{DM}$ given by Eq.~(\ref{eq:alphasigmamass})).

The simulations were initialized with all particles divided into two initial halos placed at separation from each other $\Delta R = 2 Mpc$, moving towards each other with initial relative inbound velocity between 500 kmps and 4400 kmps.
In the case of a symmetric galaxy cluster collision, the two halos were initialized with equal number of particles.
For asymmetric collisions, the particles were divided between the two halos in the ratio 5:1. 
This choice was motivated by the mass split in the Bullet cluster \cite{Clowe2006,Bradac2006,Deb2008}.
In the case of off-central collisions, the center of one of the halos was shifted with respect to the other halo in direction perpendicular to the collision axis by the amount specified by the impact parameter $\Delta b$ chosen equal to the core radius of the larger halo. This choice was motivated by maximizing the effect on non-centrality on the collision, whereas the larger values of the impact parameter resulted in halos missing each others' dense cores and smaller impact parameters resulting in post-collision mass distributions not much different from that of central collisions.

The total mass of simulation here was not varied. This was because the total mass can be reduced from gravitational dynamics by suitably re-scaling the distance and/or time variables in the simulation. 
Indeed, consider the Newtonian equations of motion of particles in self-consistent gravitational potential,
\begin{equation}
\begin{array}{l}
\frac{d\vec r_i}{dt}=\vec v_i,\\
\frac{d\vec v_i}{dt}=-G \nabla \int d^3z \frac{\rho(\vec z,t)}{|\vec r_i-\vec z|}.
\end{array}
\end{equation}
If one rescales distances, times, and velocities of all particles according to
\begin{equation}\label{eq:scalegravtransform}
\begin{array}{l}
\vec r= a\vec r\hspace{0.05cm}', \\
\vec v = ab^{-1}\vec v\hspace{0.05cm}', \\
t = bt', 
\end{array}
\end{equation}
then the equations of motion become respectively 
\begin{equation}
\begin{array}{l}
ab^{-1}\frac{d\vec r\hspace{0.05cm}'_i}{dt'}=ab^{-1}\vec v\hspace{0.05cm}'_i, \\
ab^{-2}\frac{d\vec v\hspace{0.05cm}'_i}{dt'}=-a^{-2}G \nabla' \int d^3z' \frac{\rho'(\vec z\hspace{0.05cm}',t)}{|\vec r\hspace{0.05cm}'_i-\vec z\hspace{0.05cm}'|},
\end{array}
\end{equation}
where we took into account that the element of mass, $dm=d^3r\rho(\vec r,t)$, remained invariant. Simplifying the above we obtain
\begin{equation}\label{eq:scaledgraveqn}
\begin{array}{l}
\frac{d\vec r\hspace{0.05cm}'_i}{dt}=\vec v\hspace{0.05cm}'_i, \\
\frac{d\vec v\hspace{0.05cm}'_i}{dt}=-a^{-3}b^{2}G \nabla \int d^3z' \frac{\rho'(\vec z\hspace{0.05cm}',t)}{|\vec r\hspace{0.05cm}'_i-\vec z\hspace{0.05cm}'|}.
\end{array}
\end{equation}
Therefore, we observe that the total simulation mass $\int d^3r \rho(\vec r,t)$ can be arbitrarily rescaled by scaling up or down either distances, times, or both. For example, $\int d^3r \rho(\vec r,t)$ can be brought to be equal to any (standard) mass $M_{std}$ simply by scaling the distances $\vec r=(M_{tot}/M_{std})^{1/3} \vec r\hspace{0.05cm}'$, where $M_{tot}$ is the total mass in $\rho(\vec r,t)$, or by similarly scaling the time variable, $t=(M_{std}/M_{tot})^{1/2} t'$. Note that the velocities also scale according to Eq.~(\ref{eq:scalegravtransform}).
Thus, the total mass can be reduced from the simulations and all our simulations were performed using a ``standard" total collision mass of $5\cdot 10^{14}M_\odot$.

The full list of the parameters used for each class of simulations discussed in this work is given in Table \ref{table:params}.

\begin{table}[ht]
\caption{The parameters of the simulations of different galaxy cluster collision scenarios performed in this work. 
}
\centering
\begin{tabular}{|p{2.8cm}|p{1.0cm}|p{0.5cm}|p{0.9cm}|p{0.9cm}|p{0.9cm}|p{1.15cm}|p{0.9cm}|p{0.9cm}|p{0.9cm}|p{1.2cm}|p{0.5cm}|p{0.5cm}|}
\hline
Type of scenario & $M_{tot}$ ($M_\odot$) & $D$ (\#) & $Dd$ (Mpc)& $N1$ ($10^5$)& $N2$ ($10^5$)& $\Delta V2$ (kmps$^2$)& $\Delta R1$ (Mpc) & $\Delta R2$ (Mpc) & $\Delta b$ (Mpc) & $\frac{\sigma_{DM}}{m_{DM}}$ ($cm^{-2}g$) & $k$ & $a$ (\%)\\
\hline\hline
fast CDM & $5\cdot 10^{14}$ & 400 & 6.0 & 1.0 & 1.0 & $1300^2$  & 1.0 & 1.0 & 0.0 & 0.0  & 1.6 & -\\
free-fall CDM & $5\cdot 10^{14}$ & 400 & 6.0 & 1.0 & 1.0 & 0.0  & 1.0 & 1.0 & 0.0 & 0.0  & 1.0 & - \\
slow CDM & $5\cdot 10^{14}$ & 400 & 6.0 & 1.0 & 1.0 & $-(700^2)$  & 1.0 & 1.0 & 0.0 & 0.0 & 0.8 & - \\
\hline
symmetric-central-weak & $5\cdot 10^{14}$ & 400 & 6.0 & 1.0 & 1.0 & $1300^2$ & 1.0 & 1.0 & 0.0 & 0.45  & 1.6 & 10 \\
symmetric-central-strong & $5\cdot 10^{14}$ & 400 & 6.0 & 1.0 & 1.0 & $1300^2$ & 1.0 & 1.0 & 0.0 & 1.80  & 1.6 & 40 \\
\hline
symmetric-noncentral-CDM & $5\cdot 10^{14}$ & 400 & 6.0 & 1.0 & 1.0 & $1300^2$ & 1.0 & 1.0 & 0.25 & 0.0 & 1.6 & - \\
symmetric-noncentral-weak & $5\cdot 10^{14}$ & 400 & 6.0 & 1.0 & 1.0 & $1300^2$ & 1.0  & 1.0 & 0.25 & 0.45  & 1.6 & 7 \\
symmetric-noncentral-strong & $5\cdot 10^{14}$ & 400 & 6.0 & 1.0 & 1.0 & $1300^2$ & 1.0 & 1.0 & 0.25 & 1.80  & 1.6 & 31 \\
\hline
asymmetric-central-CDM & $5\cdot 10^{14}$ & 400 & 6.0 & 1.5 & 0.3 & $800^2$ & 0.0 & 2.0 & 0.0 & 0.0 & 1.4 & - \\
asymmetric-central-weak & $5\cdot 10^{14}$ & 400 & 6.0 & 1.5 & 0.3 & $800^2$ & 0.0 & 2.0 & 0.0 & 0.50 & 1.4 & 7 \\
asymmetric-central-strong & $5\cdot 10^{14}$ & 400 & 6.0 & 1.5 & 0.3 & $800^2$ & 0.0 & 2.0 & 0.0 & 3.25 & 1.4 & 60 \\
\hline
asymmetric-noncentral-CDM & $5\cdot 10^{14}$ & 400 & 6.0 & 1.5 & 0.3 & $800^2$ & 0.0 & 2.0 & 0.25 & 0.0 & 1.4 & - \\
asymmetric-noncentral-weak & $5\cdot 10^{14}$ & 400 & 6.0 & 1.5 & 0.3 & $800^2$ & 0.0 & 2.0 & 0.25 & 0.80 & 1.4 & 9 \\
asymmetric-noncentral-strong & $5\cdot 10^{14}$ & 400 & 6.0 & 1.5 & 0.3 & $800^2$ & 0.0 & 2.0 & 0.25 & 4.05 & 1.4 & 58 \\
\hline
very fast collisions comparison & $10^{15}$ & 100 & 6.0 & 1.0 & 1.0 & $5400^2$ & 1.0 & 1.0 & 0.0 & 0.45 & 8.0 & 7 \\
very fast collisions comparison & $10^{15}$ & 100 & 6.0 & 1.0 & 1.0 & $3500^2$ & 1.0 & 1.0 & 0.0 & 0.45 & 4.0 & 8 \\
very fast collisions comparison & $10^{15}$ & 100 & 6.0 & 1.0 & 1.0 & $2000^2$ & 1.0 & 1.0 & 0.0 & 0.45 & 2.0 & 10 \\
very fast collisions comparison  & $10^{15}$ & 100 & 6.0 & 1.0 & 1.0 & $1400^2$ & 1.0 & 1.0 & 0.0 &  0.45 & 1.5 & 11 \\
\hline
\end{tabular}
\label{table:params}
\end{table}

\section{Results}
\label{sec:results}

\subsection{Characterization of possible galaxy cluster collision phenomenologies}
\label{sec:results:phevariables}

Our primary goal is to characterize possible effects of astrophysically-weak DM particle-particle self-interactions on the projected mass distributions appearing in collisions of galaxy clusters. 
By ``weak'', it is meant that DM self-interaction results in the fractions of DM particles scattered during the collisions that are significantly less than one. From the nature of that weak effect, such effects can be expected to appear in the tails of DM halos of the colliding galaxy clusters. To make those effects discernible, we choose a suitable way for visualizing the tails of DM halo distributions using the logarithmic scale for projected mass density maps, adopted for all illustrations in this section.

We point out that the general profile of post-collision mass distributions in colliding galaxy clusters is well characterized by three phenomenological parameters --- the collision's kinetic parameter as defined by the ratio of the relative kinetic energy of the colliding galaxy clusters and mutual gravitational energy, $k=|E_K/E_G|$; the fraction of  DM halos' non-gravitationally scattered mass, $a$; and the separation of post-collision galaxy clusters in terms of a certain typical radius here chosen as the core radius, $r/r_c$.

We first discuss the kinetic parameter $k=|E_K/E_G|$. This parameter characterizes the degree of the effect that the gravitational effects impart to the post-collision mass distributions. Intuitively, small values of $k$ imply slow collisions in which gravitational interactions have a long time to play dominant role. For large values of $k$, however, the galaxy clusters collide much more rapidly with little to no gravitational distortions. 
We specifically define the kinetic parameter $k$ as the ratio of the colliding galaxy clusters' kinetic to mutual gravitational energy at the point of closest approach. 


The fraction parameter of the DM halo mass scattered in DM particle-particle collisions, $a$, is another interesting parameter affecting the shape of post-collision galaxy clusters' mass distributions. 
The parameter $a$ quantifies the degree of effect that DM self-interactions have on the post-collision mass distribution by setting an upper limit on the relative weight that the non-gravitational scattering features can introduce into post-collision galaxy clusters' mass maps. For instance, for $a=0.1$ at most 10\% of the DM halo mass can contribute to self-interaction-related features in the post-collision mass distributions. 

The fraction $a$ can be directly related to the effective DM self-interaction parameter in simulations, $\alpha$, and the physical ratio $\sigma_{DM}/m_{DM}$. 
For concreteness we define
\begin{equation}\label{eq:colla}
 a=2N_{12}/(N_1+N_2)=(\delta M_1+\delta M_2)/M_{tot},
\end{equation} 
where $N_{12}$ is the number of DM particle-particle scattering events during the collision and $N_1+N_2$ is the total number of DM particles in the collision.
In terms of real masses of scattered DM, $\delta M_{1,2}$, the same parameter is given as their ratio with the total mass of the collision $M_{tot}$.

The post-collision separation of the galaxy clusters is another phenomenological descriptor of a galaxy cluster collision. We observe that the post-collision mass distributions share significant similarities for the same separations of outgoing galaxy clusters, if expressed in the terms of a typical distance scale associated with the galaxy clusters. 
Therefore, such a parameter is advantageous for characterizing the post-collision stage instead of regular time. Of course, post-collision galaxy clusters' separation is also a quantity that can be directly measured from astrophysical data.
We define that parameter as the ratio $r/r_c$ of the distance $r$ between the centers of the outgoing galaxy clusters and the core radius $r_c$ of the larger of the galaxy clusters (that is, the 1/2 half-width of the projected mass distribution of that cluster's DM halo).

\subsection{The phenomenology of CDM high-speed galaxy cluster collisions}
\label{sec:results:gravitational}

We first review the galaxy cluster collisions in standard CDM, that is, when the gravity is the only effect present. 
We inspect central collisions, in which case the only two free parameters controlling the post-collision mass distribution are $M_{tot}$ and 
$k=|E_K/E_G|$, of which the total mass can be excluded by rescaling the distances as discussed above and a single parameter $k$ is left to completely characterize the post-collision mass distribution's properties. 

The kinetic collision parameter $k$ has been defined in Section \ref{sec:results:phevariables} as the ratio of kinetic and gravitational energy in collision. 
By the conservation of energy, $k$ is thus related to the total energy of colliding system, $E$, as $k=1+E/|E_G|$, where $E_G$ is the mutual gravitational energy of colliding galaxy clusters. Three essentially different regimes therefore need to be distinguished:
fast collisions with $E>0$ and $k>1$, ``free-fall" collisions with $E=0$ and $k=1$, and slow collisions with $E<0$ and $k<1$.
In fast collisions, the clusters fall towards each other from a finite in-fall velocity at infinity. In ``free-fall" collisions, the clusters fall towards each other from zero speed at infinity. The case $k<1$ corresponds to the situation where the clusters fall towards each other from zero speed at a finite distance. 

\begin{figure}[h!]
\begin{center}
\subfigure{
\includegraphics[width=0.25\textwidth]{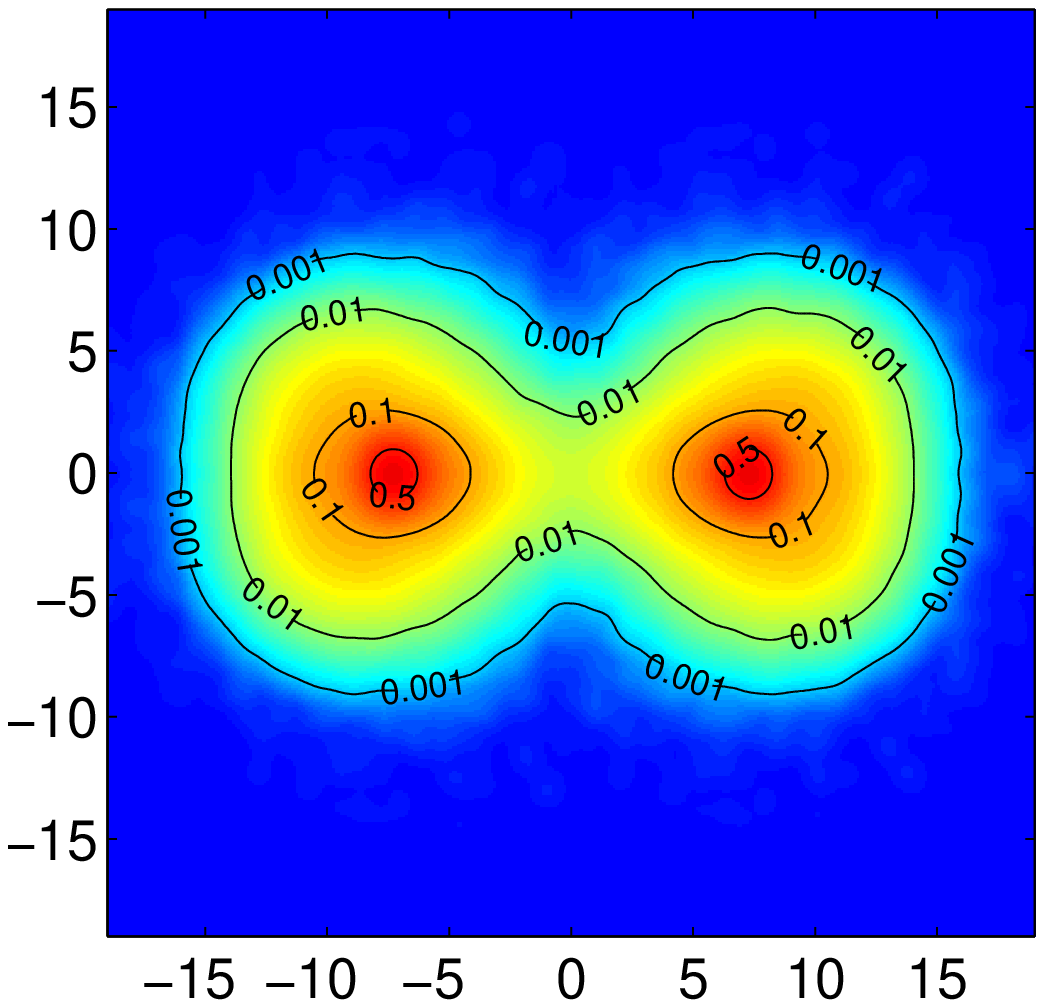}
}
\subfigure{
\includegraphics[width=0.25\textwidth]{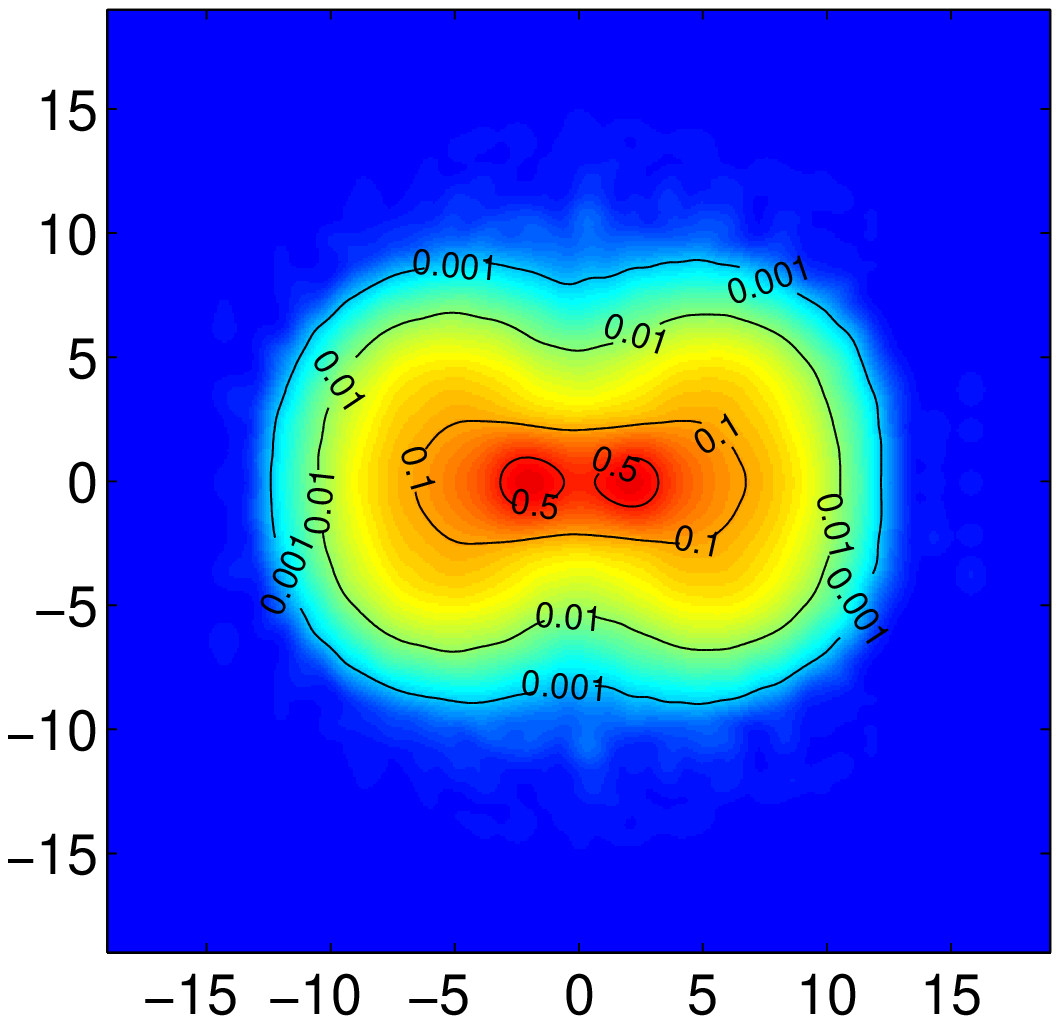}
}
\subfigure{
\includegraphics[width=0.355\textwidth]{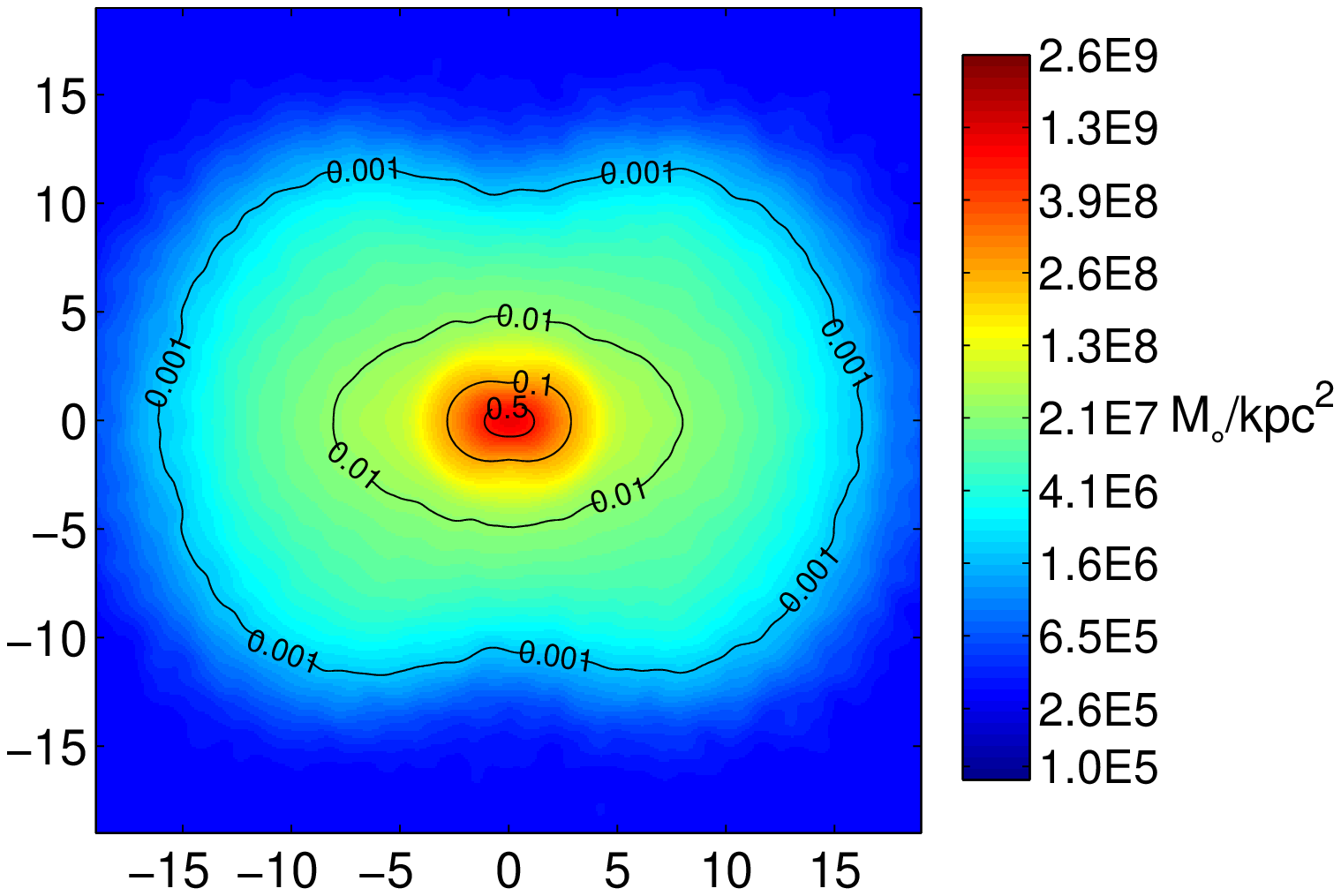}
}
\end{center}
\caption{\label{fig:std} Possible phenomenologies of the projected mass density profiles for galaxy cluster collisions in CDM model, for different values of the kinetic parameter $k$.
From left to right shown are the examples of a central symmetric fast collision ($k=1.6$, separation 15$r_c$), a ``free-fall" collision ($k=1.0$, maximum separation of approximately $6r_c$), and a slow collision ($k=0.8$, final merger configuration). 
The simulation parameters are as defined in Table \ref{table:params}.
The projected mass density maps are shown color-coded according to logarithmic scale, with the contour-lines defining the density levels of such maps normalized to the peak projected mass density of one. 
The colormap on the right shows the projected mass density in the physical units of  $M_\odot/kpc^2$, assuming the total collision mass of $5\cdot10^{14} M_\odot$.
The distance scales shown along x and y axes are in the units of $r_c$. 
The colorbar and the distance scale are for this and all similar figures except Fig. \ref{fig:ideal} (``ideal" case).}
\end{figure}

The typical shapes of post-collision mass distributions in CDM for fast collisions regime is explored in Fig. \ref{fig:std}. For very fast collision with $k\approx 2$ and above, we observe that the colliding galaxy clusters pass through each other without gravitational distortions, essentially maintaining their original shape and velocity after the collision.
For slower collisions, $2>k>1$, the galaxy clusters can substantially interact gravitationally, however, forming high-velocity ejecta in the form of forward conic jets around the clusters' initial velocity vectors. These ejecta give the projected mass density map a notable forward ``fan-out'' shape, as shown in the left panel of Fig. \ref{fig:std} in a log-scale. 
A narrow, weak ``central bridge" of slow trailing material is also observed in this regime in some simulations.

An important observation to make at this point is that such fast ejecta is always in forward and backward cone-directions and is restricted to small scattering angles. This is, in fact, a well-known and expected property of the differential cross-section of gravitational interaction, in which large cross-sections are observed for small scattering angles and very small cross-section are observed towards $90^\circ$ scattering angle. Gravitational interaction, therefore, does not produce orthogonal or ``equatorial'' ejecta during high-speed collisions.

For a slower case of ``free-fall" or $k=1$ collisions, we observe that the colliding galaxy clusters merge in a single passage producing large amount of ejecta in a characteristic ``butterfly'' shape, as shown in the central panel of Fig. \ref{fig:std}. 
In slow collisions, $k<1$, the clusters rapidly merge with a large amount of close to isotropic ejecta, as shown in the right panel of Fig. \ref{fig:std}. 


\subsection{The phenomenology of SIDM high-speed galaxy cluster collisions}

In the interacting DM case, the phenomenology of high-speed galaxy cluster collisions can be substantially different and is governed by two parameters: the kinetic parameter $k$ and the effective DM self-scattering intensity $a$.
%
%
%
With respect to the DM self-scattering strength, we inspect three regimes of strong self-scattering DM, $0.5\leq a$, weak self-scattering, $a\leq 0.2$, and intermediate scattering $0.2\leq a\leq 0.5$. These roughly correspond to $\sigma_{DM}/m_{DM}>2\ cm^{-2}g$ for strong, $\sigma_{DM}/m_{DM}<0.5\ cm^{-2}g$ for weak, and $0.5\ cm^{-2}g<\sigma_{DM}/m_{DM}<2\ cm^{-2}g$ for intermediate DM scattering (see Table \ref{table:params}).

The typical shapes of the post-collision mass distributions for all of these regimes are presented in Fig. \ref{fig:phe}. The respective shapes are shown using a table where the columns correspond to different DM self-interaction intensities $a$ and the rows corresponding to different collision scenarios such as symmetric central, asymmetric central, symmetric non-central and asymmetric non-central, as defined in the caption. Note here that the kinetic parameter in symmetric collision scenarios is $k=1.6$ and in asymmetric collision scenarios is $k=1.4$.

In all regimes, we observe that the DM self-interaction can result in additional mass components appearing as diffuse circular mass concentrations centered at the collision center and extending radially out to the distance defined by the outgoing galaxy clusters. The most significant difference is present at $90^\circ$ scattering angles, that is, the equatorial plane perpendicular to the axis of the collision. As have been discussed in the previous section, gravitational interactions cannot produce significant mass ejecta in equatorial plane in fast collisions. DM mass concentrations in that region is a unique consequence of  self-interactions of DM particles observed in the simulations.

The relative weight of this additional DM component is defined by the parameter $a$. In the case of strong DM particle-particle scattering, $a > 0.5$ and $\sigma_{DM}/m_{DM} > 2\ cm^{-2}g$, we observe that the mass distribution of colliding galaxy clusters is significantly disrupted in all collision scenarios (see the third column of Fig.~\ref{fig:phe}). A very wide approximately spherical hot cloud of DM material forms in that situation around the collision center. In the case  of an asymmetric bullet cluster-like collision, the halo of the ``bullet" cluster does not survive the passage through the main cluster and is completely dispersed after the first passage, only appearing in the projected mass density maps as a weak extrusion from the main cluster at 1\%-level (relative to the peak projected mass density).
We note that this dramatic effect is in contrast to relatively minor effects discussed in the literature for this collision regimes, whereas the effects of such relatively strongly interacting DM in galaxy cluster collisions had been greatly underappreciated in the literature.


In contrast to the previous regime, when $a\leq 0.2$ or $\sigma_{DM}/m_{DM}<0.5\ cm^{-2}g$, the second column of Fig.~\ref{fig:phe}, the original DM halos fail to get distorted significantly, consistent with existing observations. However, previously undisclosed features are observed in these simulations contrary to the pure CDM model. 
These differences include heavier and substantially wider central regions of projected mass density maps appearing as mass-bridges connecting the halos of outgoing galaxy clusters at 1\% to 10\% peak density-levels and DM densities appearing at $90^\circ$ scattering angles (see the second and fourth rows of Fig.~\ref{fig:phe}).
The latter feature, in particular, presents the greatest interest in that it is completely absent in CDM scenario. Such equatorial mass densities seen in the second column of Fig.~\ref{fig:phe} at 1\% peak density-levels appear in the projected mass density maps at distances from the center equal to that of the outgoing galaxy clusters. 

It is interesting to understand the nature of that latter feature. For that, we can inspect the process of DM halos passing through each other in a galaxy cluster collision. In the weak scattering regime, $a \ll 1$, the mean free path of DM particles is substantially greater than the diameter of DM halos. In that case, the DM particles that scatter leave their respective halos without secondary scattering with large probability. This forms a feature in the shell of scattered DM material expanding radially outwards from the collision center. The conservation of energy and momentum in elastic collisions of DM particles dictates that the speeds of such shell is the same with the original galaxy clusters. As time passes, that shell forms a DM distribution around the collision center at the observed distances.

\def \width1 {0.251}
\begin{figure*}[h!]
\begin{center}
\subfigure{
\includegraphics[width=\width1 \textwidth]{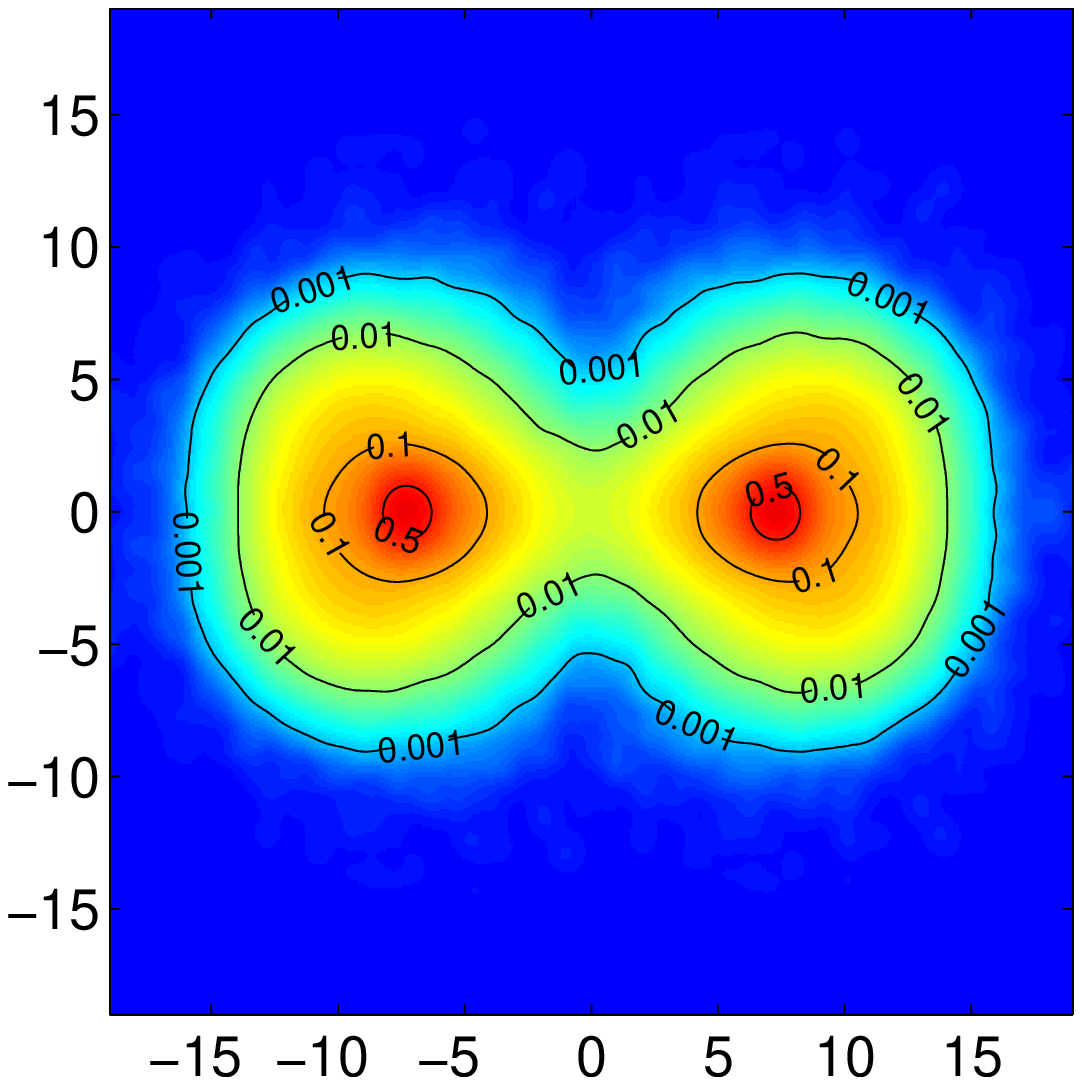}
}
\subfigure{
\includegraphics[width=\width1 \textwidth]{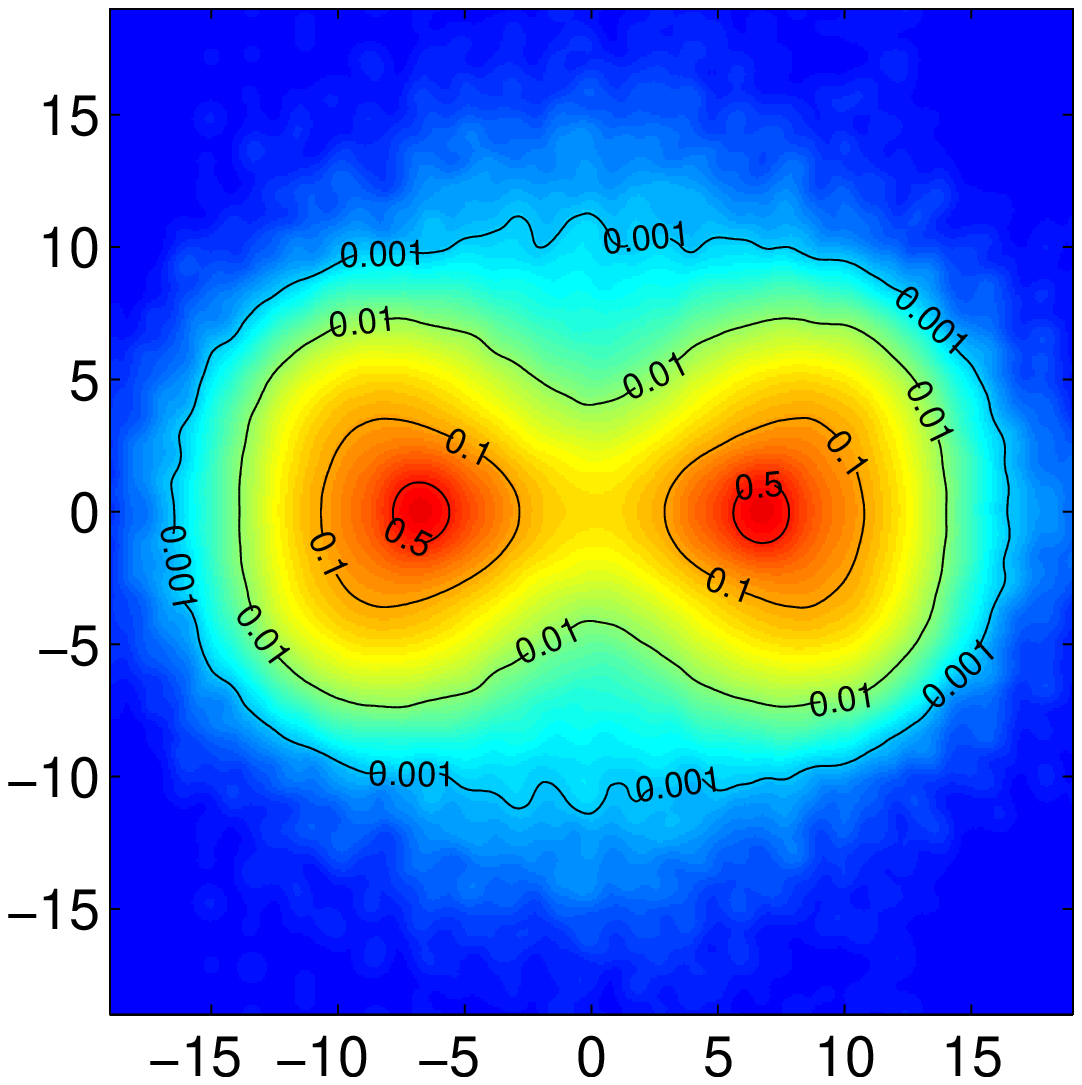}
}
\subfigure{
\includegraphics[width=\width1 \textwidth]{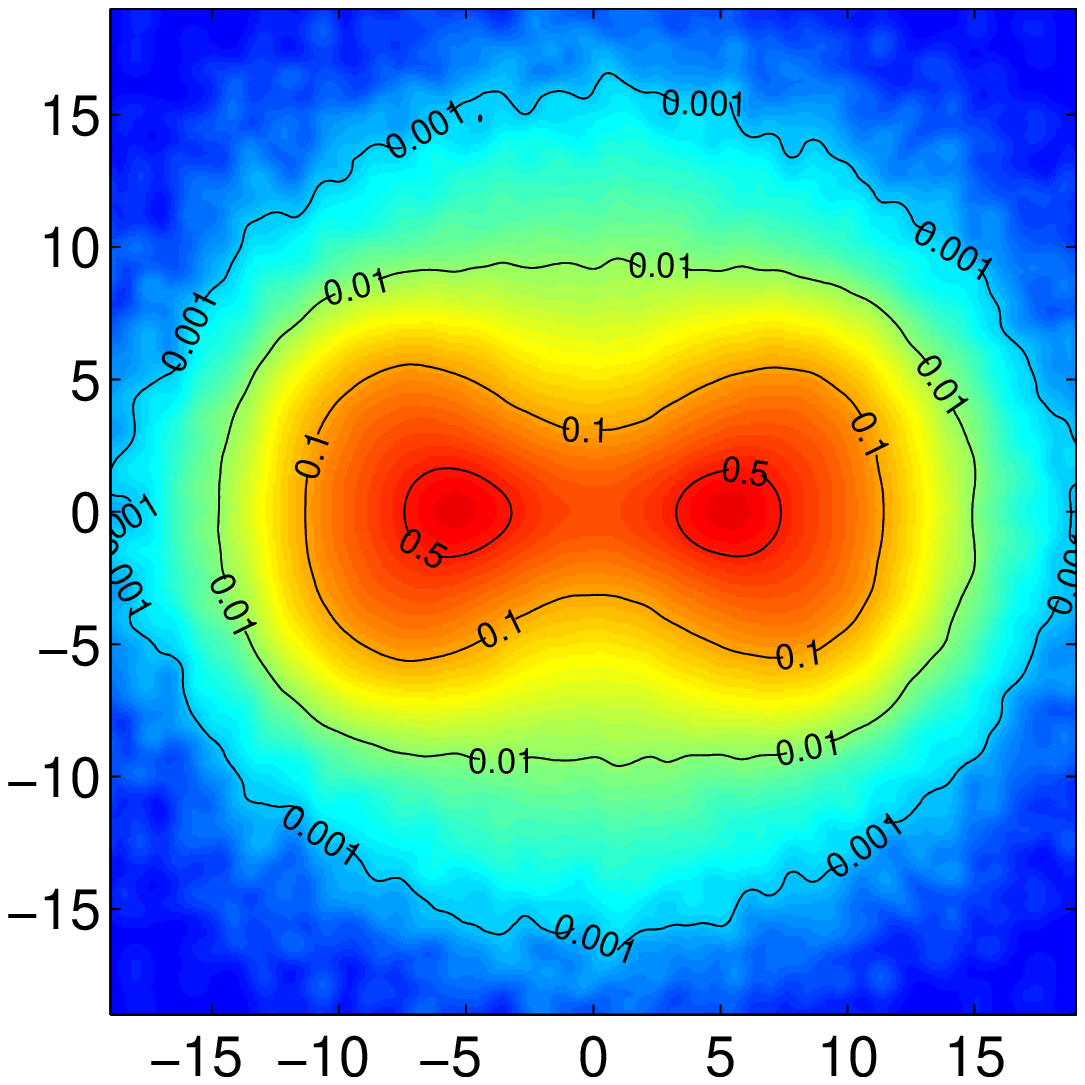}
}
\\
\subfigure{
\includegraphics[width=\width1 \textwidth]{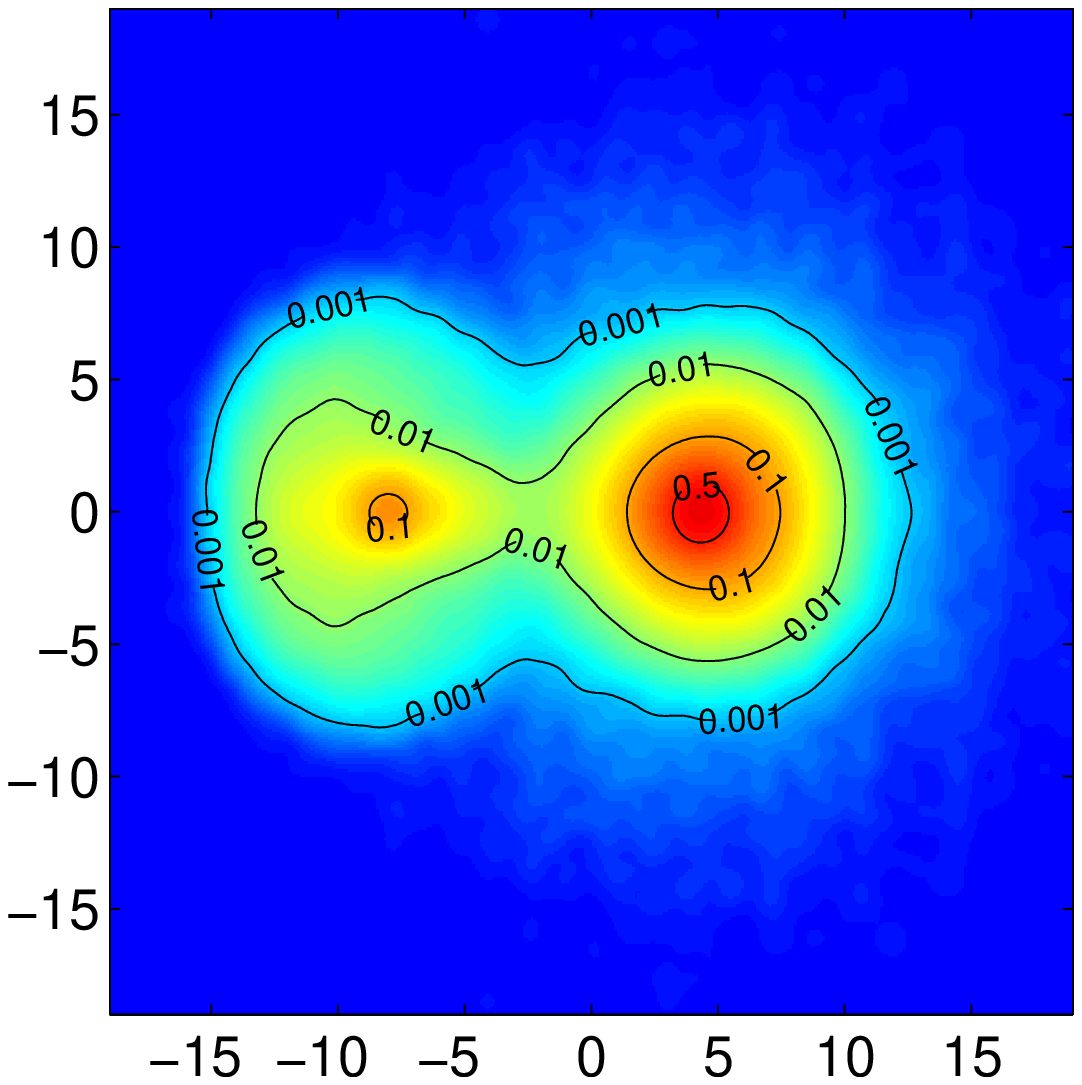}
}
\subfigure{
\includegraphics[width=\width1 \textwidth]{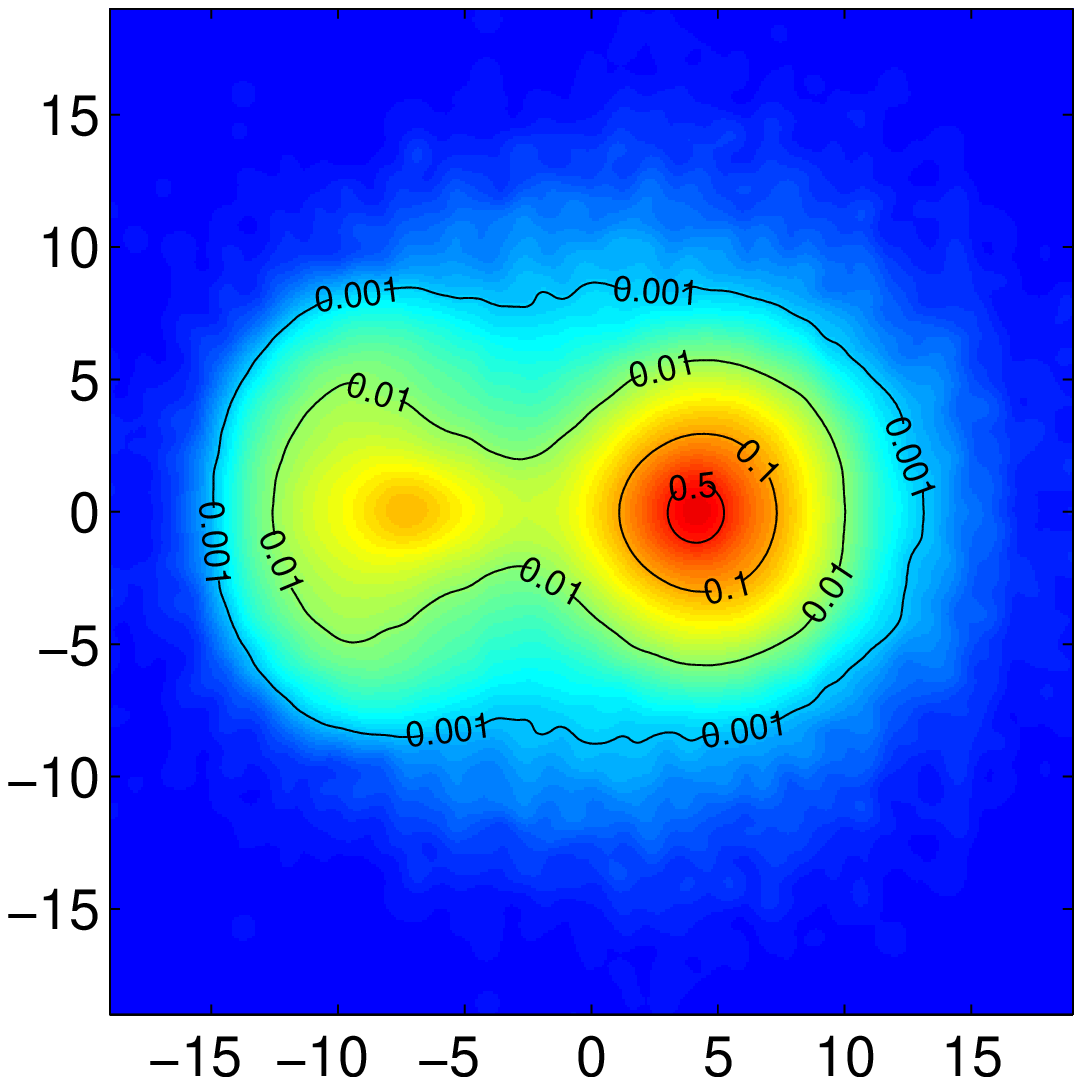}
}
\subfigure{
\includegraphics[width=\width1 \textwidth]{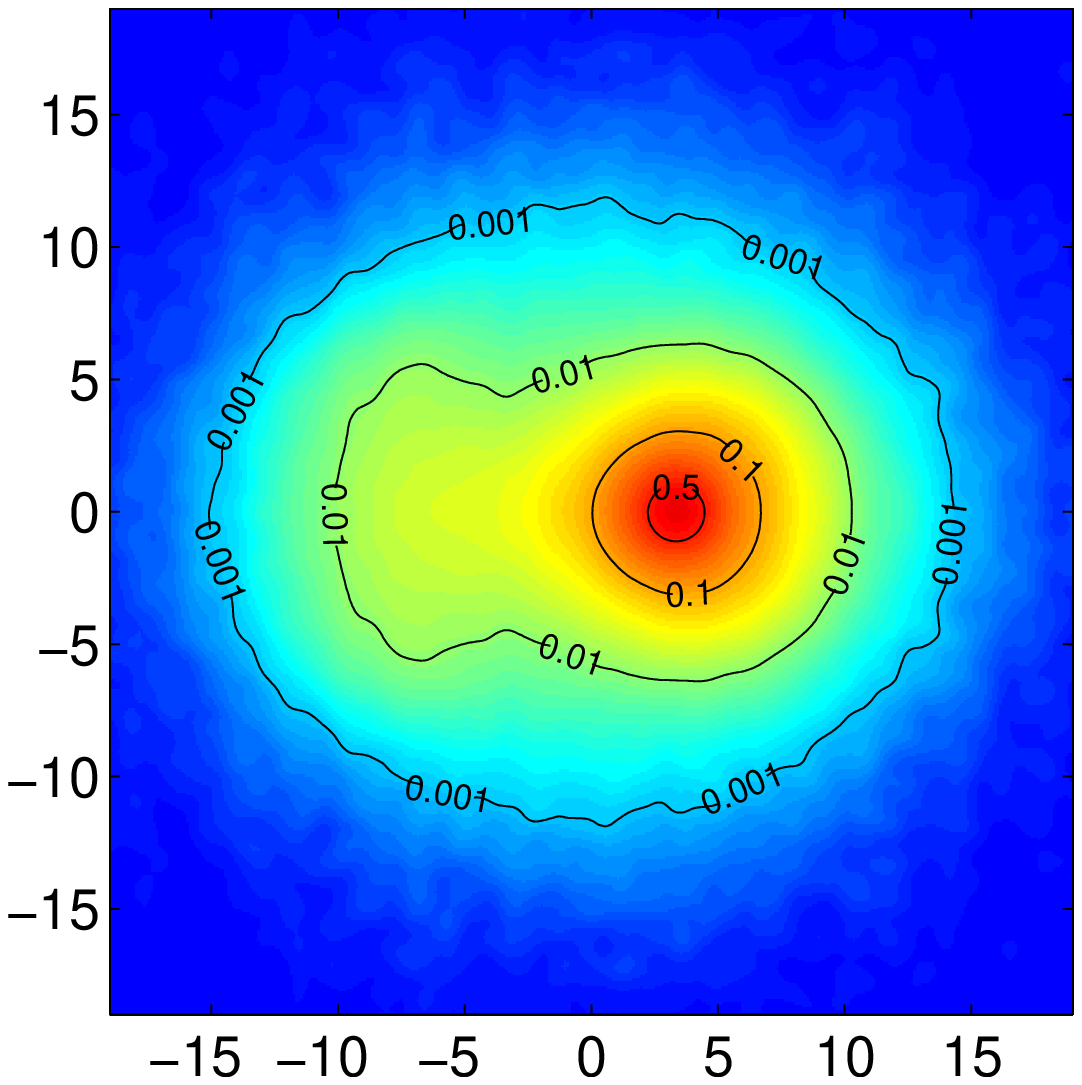}
}
\\
\subfigure{
\includegraphics[width=\width1 \textwidth]{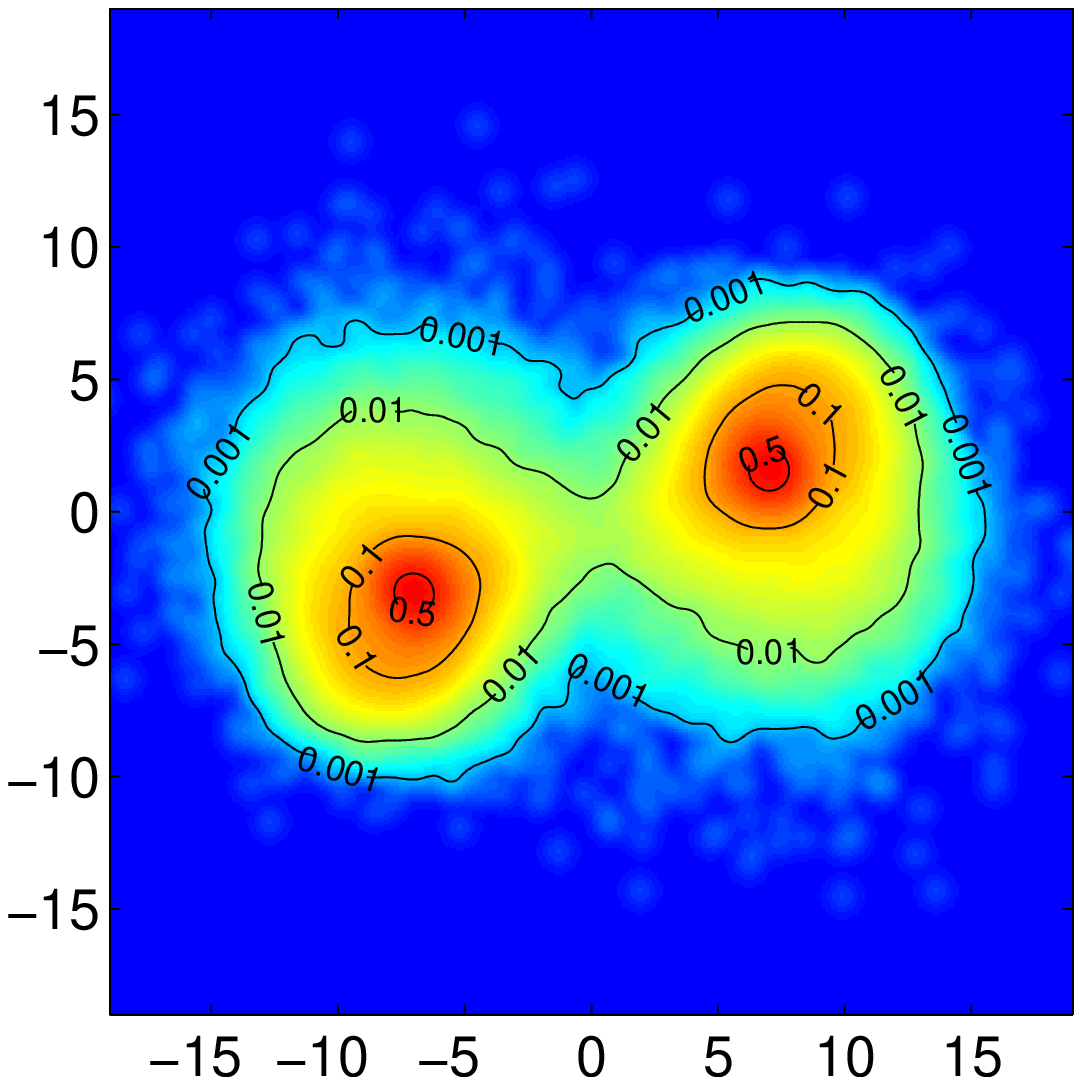}
}
\subfigure{
\includegraphics[width=\width1 \textwidth]{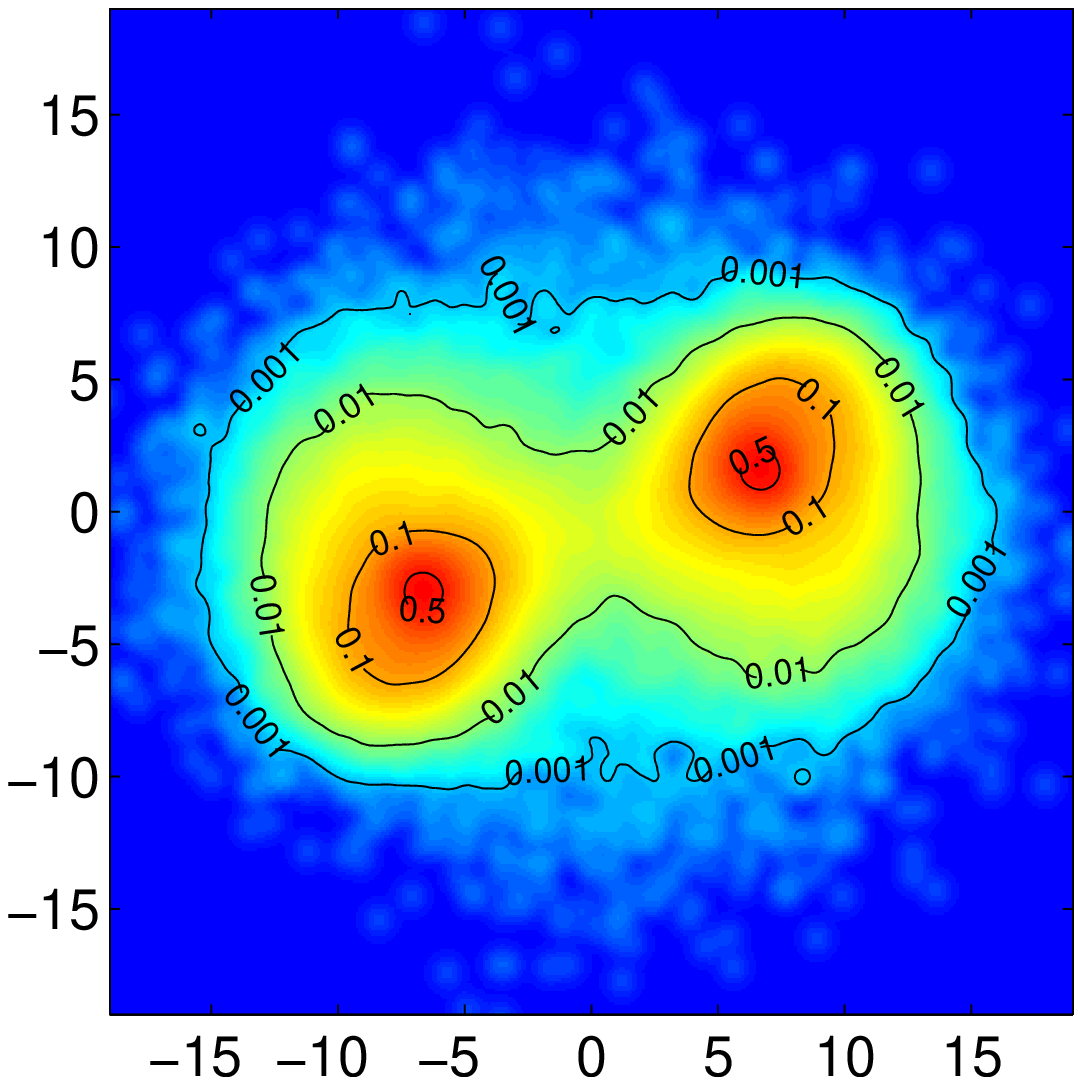}
}
\subfigure{
\includegraphics[width=\width1 \textwidth]{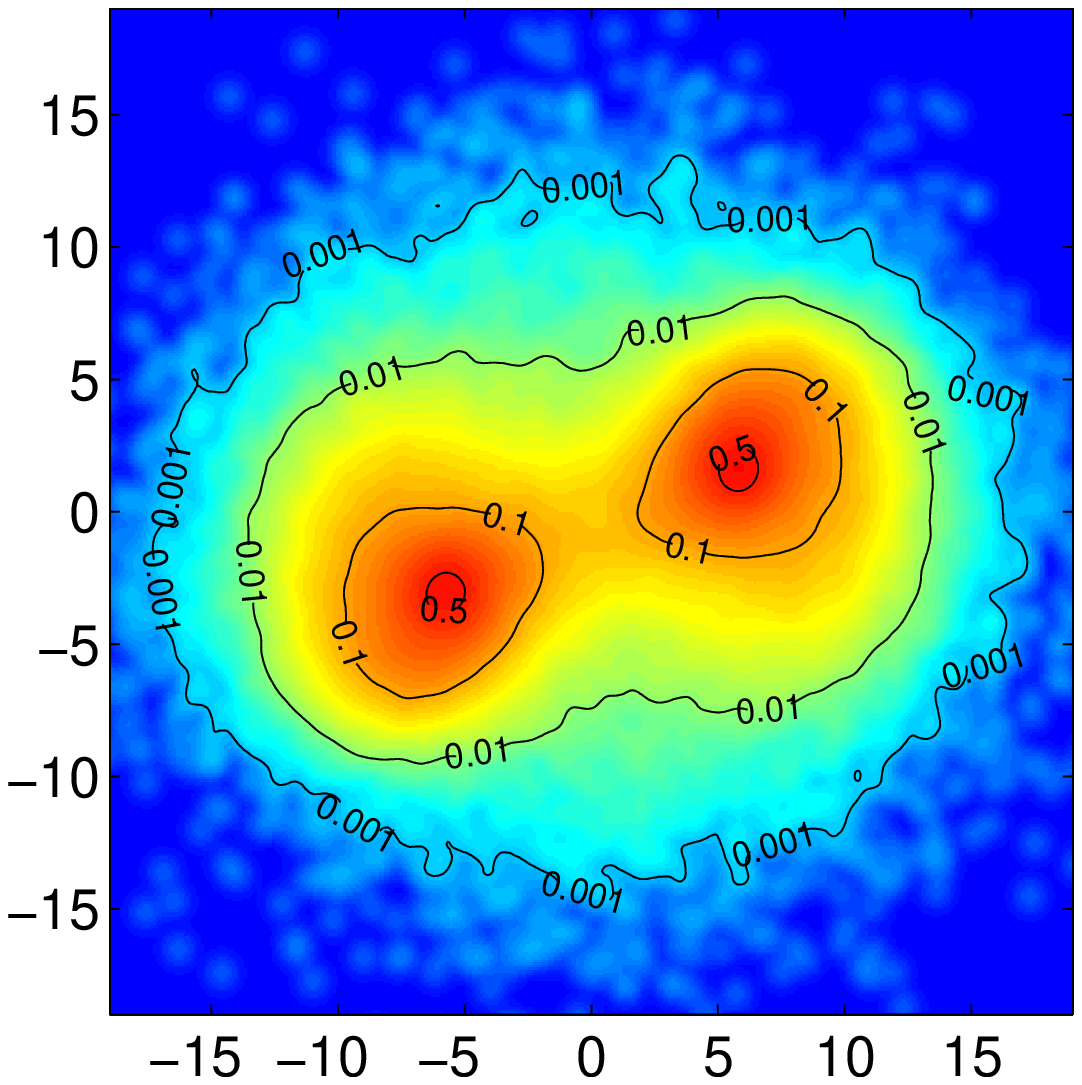}
}
\\
\subfigure{
\includegraphics[width=\width1 \textwidth]{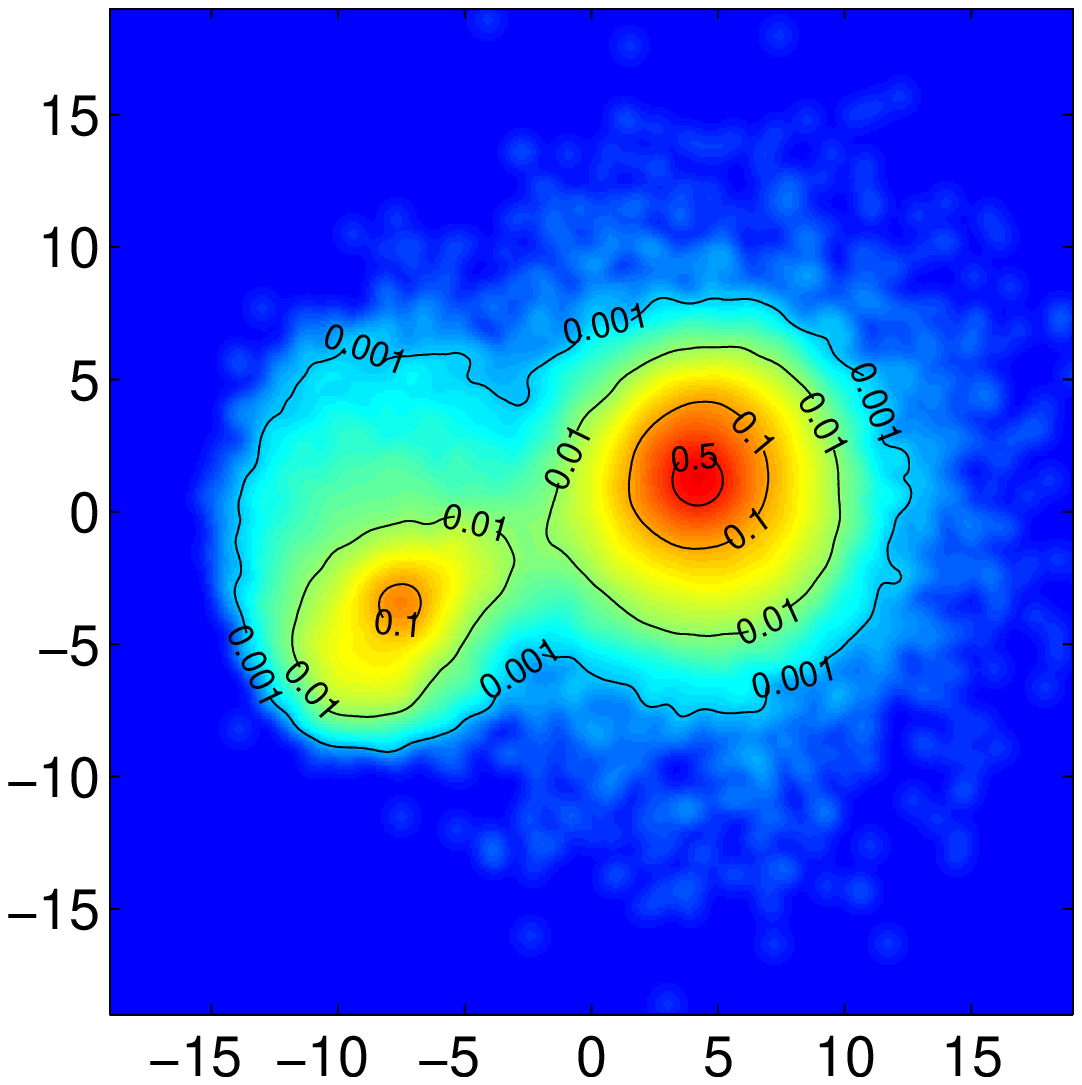}
}
\subfigure{
\includegraphics[width=\width1 \textwidth]{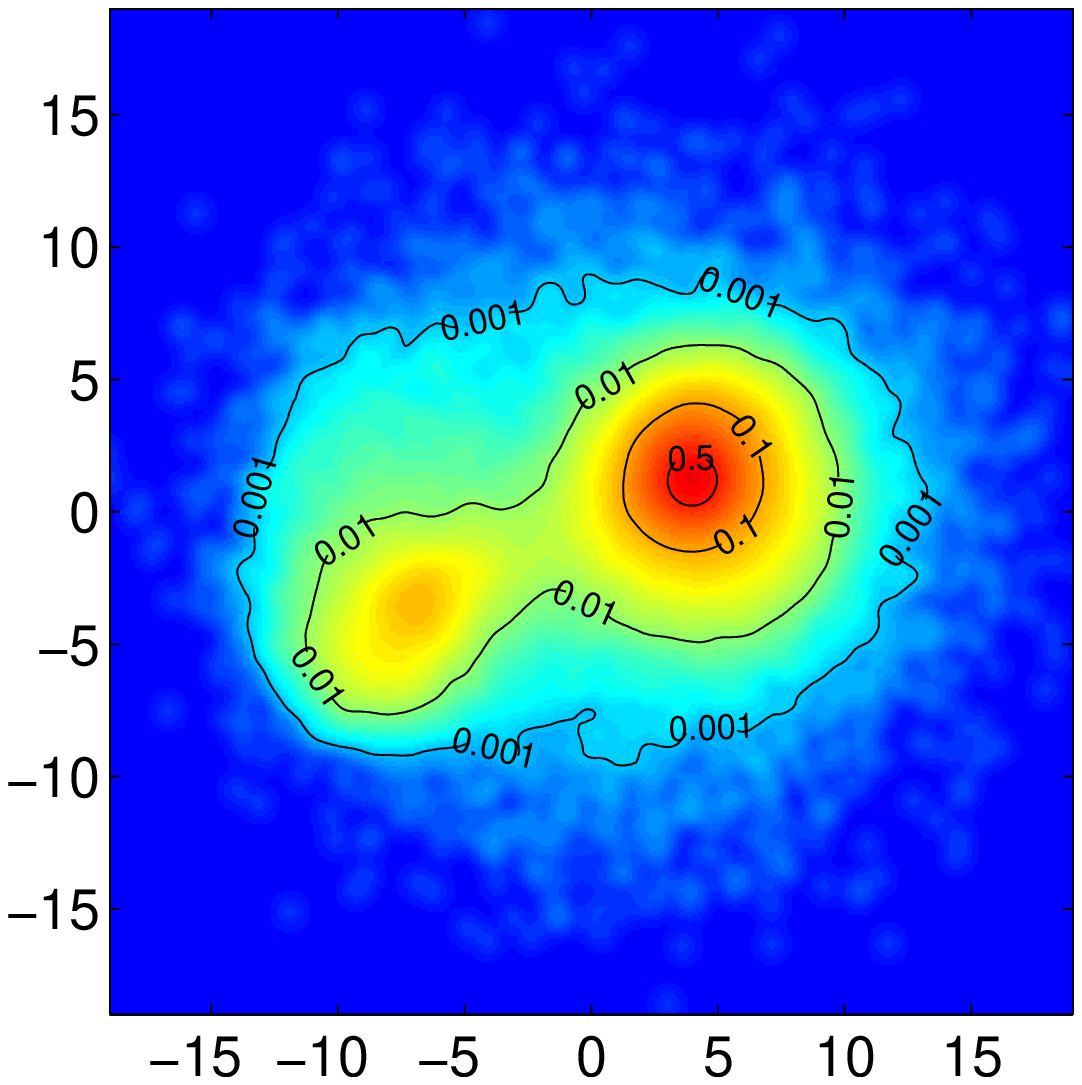}
}
\subfigure{
\includegraphics[width=\width1 \textwidth]{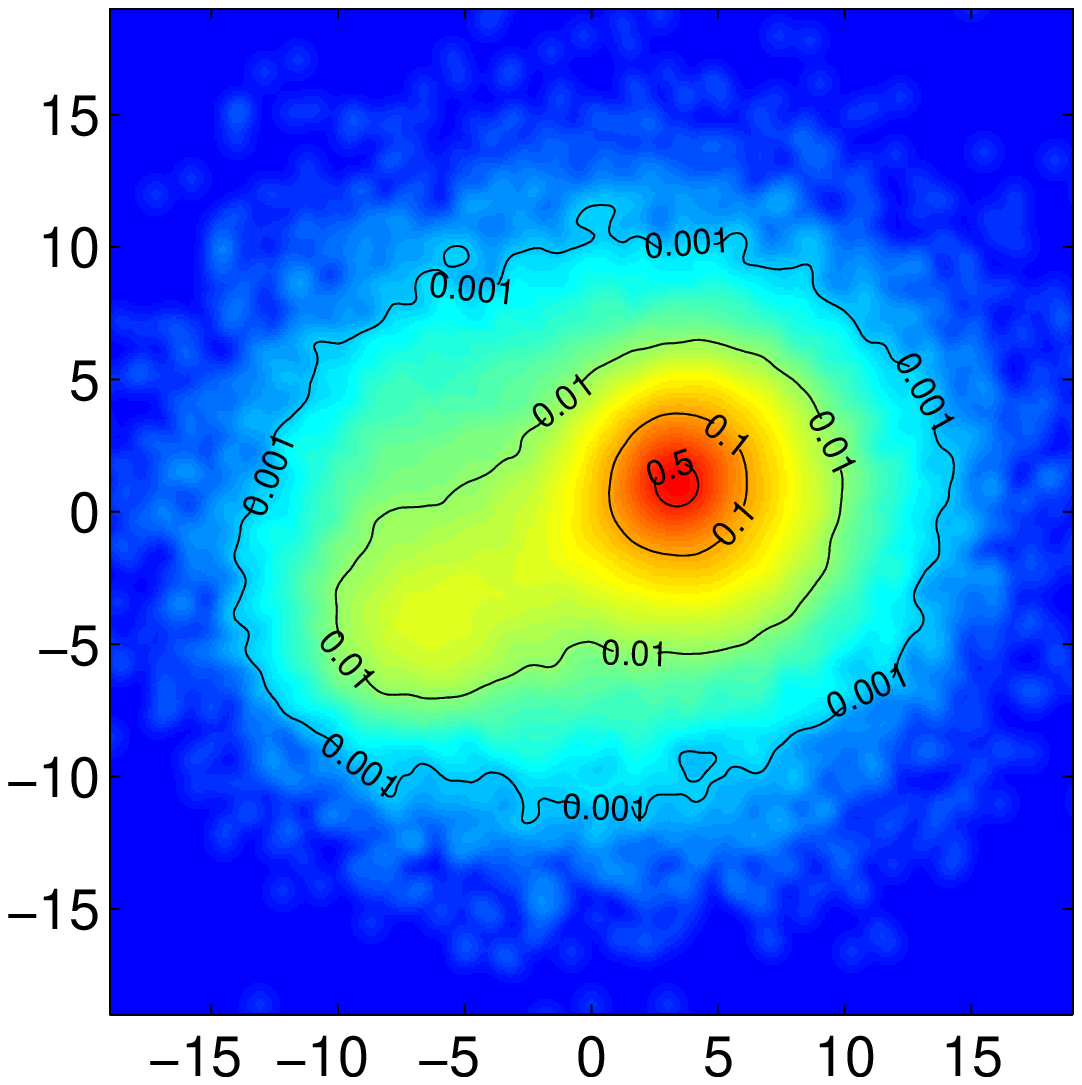}
}
\end{center}
\caption{\label{fig:phe}
The phenomenologies of possible post-collision projected mass distributions in collisions of galaxy clusters with self-interacting DM. Different galaxy cluster collision scenarios are shown in different rows: 
1st row is central symmetric collisions, 2nd row is central asymmetric collisions, 3rd row is non-central symmetric collisions, and 4th row is non-central asymmetric collisions. 
Different DM self-interaction strengths are shown in different columns:
left column - CDM regime, center column - weak DM scattering regime $a=0.1$, right column - strong DM scattering regime $a=0.5$; 
The kinetic parameter in symmetric collision scenarios is $k=1.6$ and in asymmetric collision scenarios is $k=1.4$.
The simulation parameters are as defined in Table \ref{table:params}.
}
\end{figure*}


Note that in that picture, the angular distribution of the material in the DM ejecta-shell should reflect the microscopic properties of the differential cross-section of DM particles.
It is reasonable to expect that such differential cross-section should be isotropic.
Indeed, the low energy spin-averaged cross-sections of all known short-range particle interactions are isotropic.
Under this assumption, the DM ejecta shell will form in a spherically symmetric way forming a narrow isotropic shell expanding radially in alignment with the outgoing galaxy clusters.

\begin{figure}[h!]
\begin{center}
\subfigure{
\includegraphics[width=0.3\textwidth]{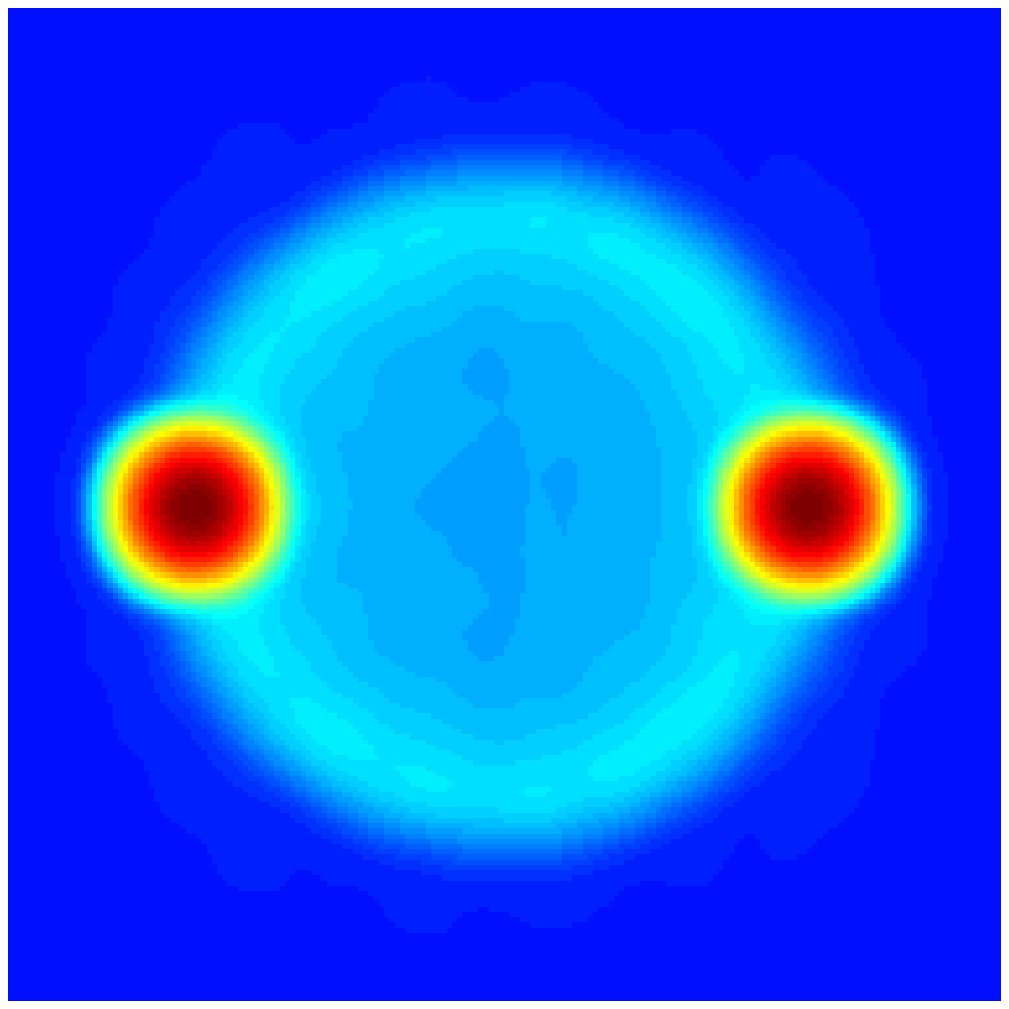}
}
\subfigure{
\includegraphics[width=0.3\textwidth]{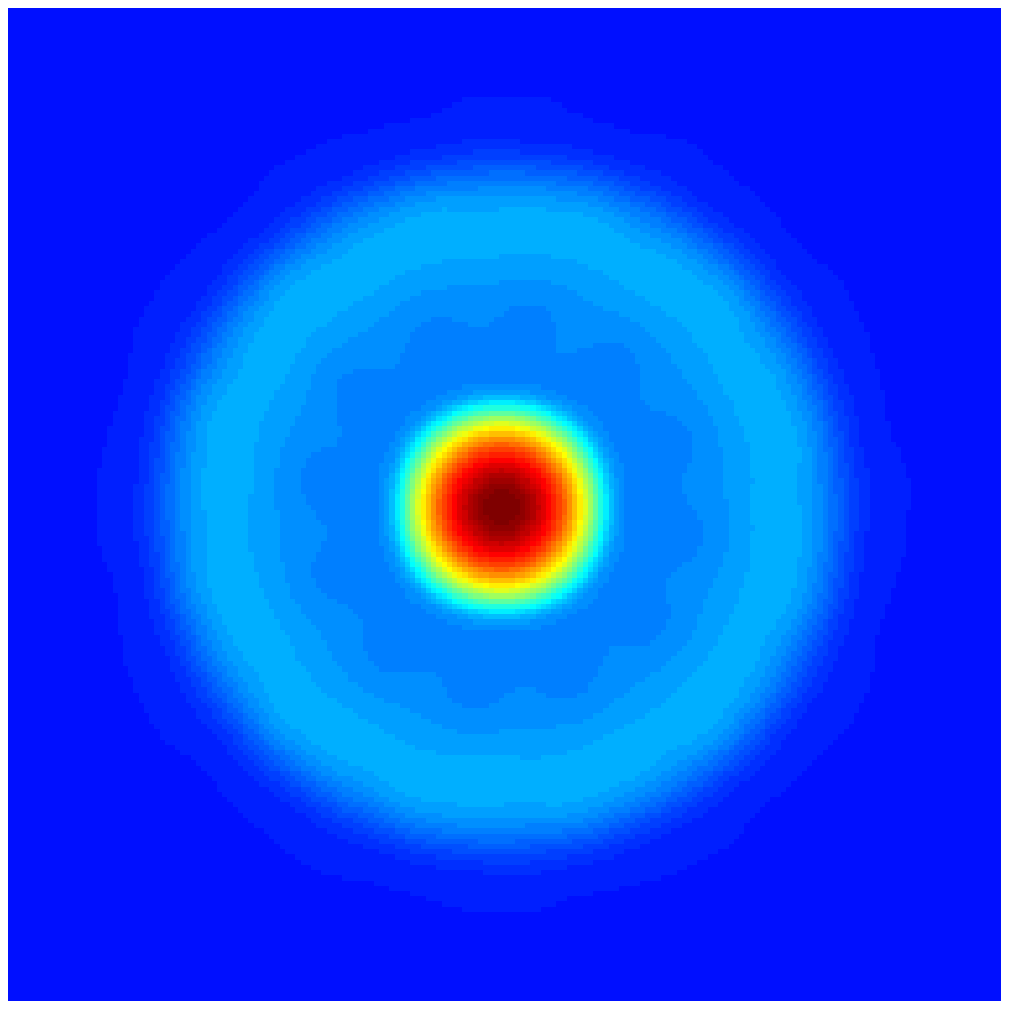}
}
\end{center}
\caption{\label{fig:ideal}The mass distribution for a self-interacting DM galaxy cluster collision in an ``ideal'' scenario, where the colliding clusters are very fast and compact. The left panel shows the projected mass density map illustrating the scattered DM shell observed along a direction perpendicular to the collision's axis. The right panel shows the same mass density observed along the collision's axis.
In the former, the DM shell appears as a weak disk-shaped mass distribution with a ring-like rim similar in width to the size of the colliding galaxy clusters, extending outwards from the center of the collision and linking the outgoing galaxy clusters into a ring-like structure. In the latter, the DM shell appears as a weak DM disk and a ring surrounding the centrally placed colliding galaxy clusters, now seen on top of each other.
}
\end{figure}

To summarize, the new features that we observe SIDM galaxy cluster collisions are the additional mass components introduced into the collision's mass distribution as a spherically symmetric shell radially expanding from the center of the collision in sync with the outgoing galaxy clusters, moving away in-lock with the galaxy clusters and linking them into a spherical-like structure. At larger separations and seen sideways, that feature appears in projected mass density maps as a disk-like diffuse DM concentration lying in between the outgoing galaxy clusters and featuring a thin, ring-like boundary of the width equal to the core-diameters of the collided galaxy clusters. 
If seen along the collision axis, the same feature appears as a disk and a ring surrounding the galaxy clusters now placed centrally.
These situations are illustrated ``ideally'' in Fig. \ref{fig:ideal}. 

\subsection{Observability conditions of SIDM effects in galaxy cluster collisions}
\label{sec:observability}


\subsubsection{Comparative study of projected mass distributions in CDM and weak SIDM scenarios}

\def \width1 {0.30}
\begin{figure}[h!]
\begin{center}
\subfigure{
\includegraphics[width=\width1 \textwidth]{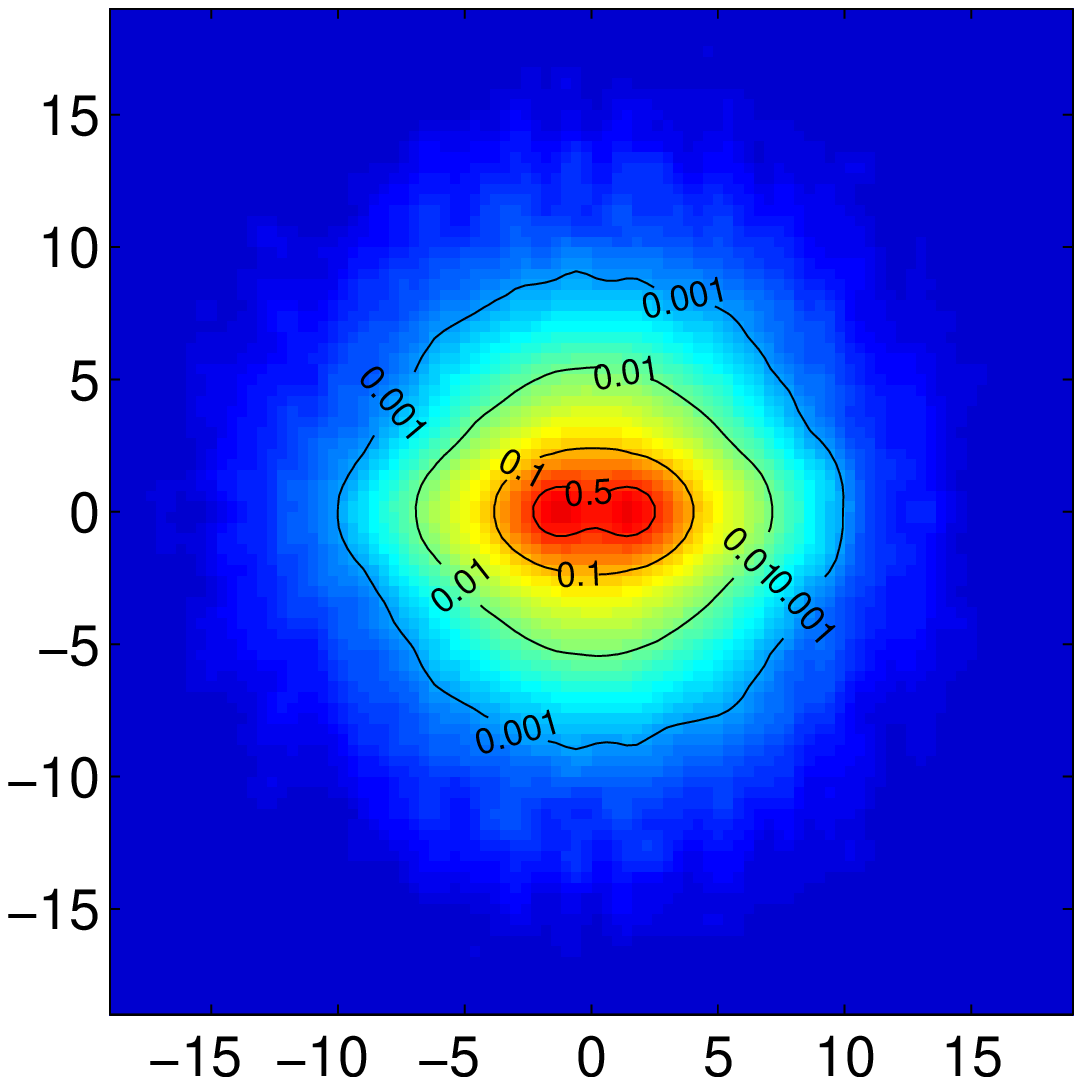}
}
\subfigure{
\includegraphics[width=\width1 \textwidth]{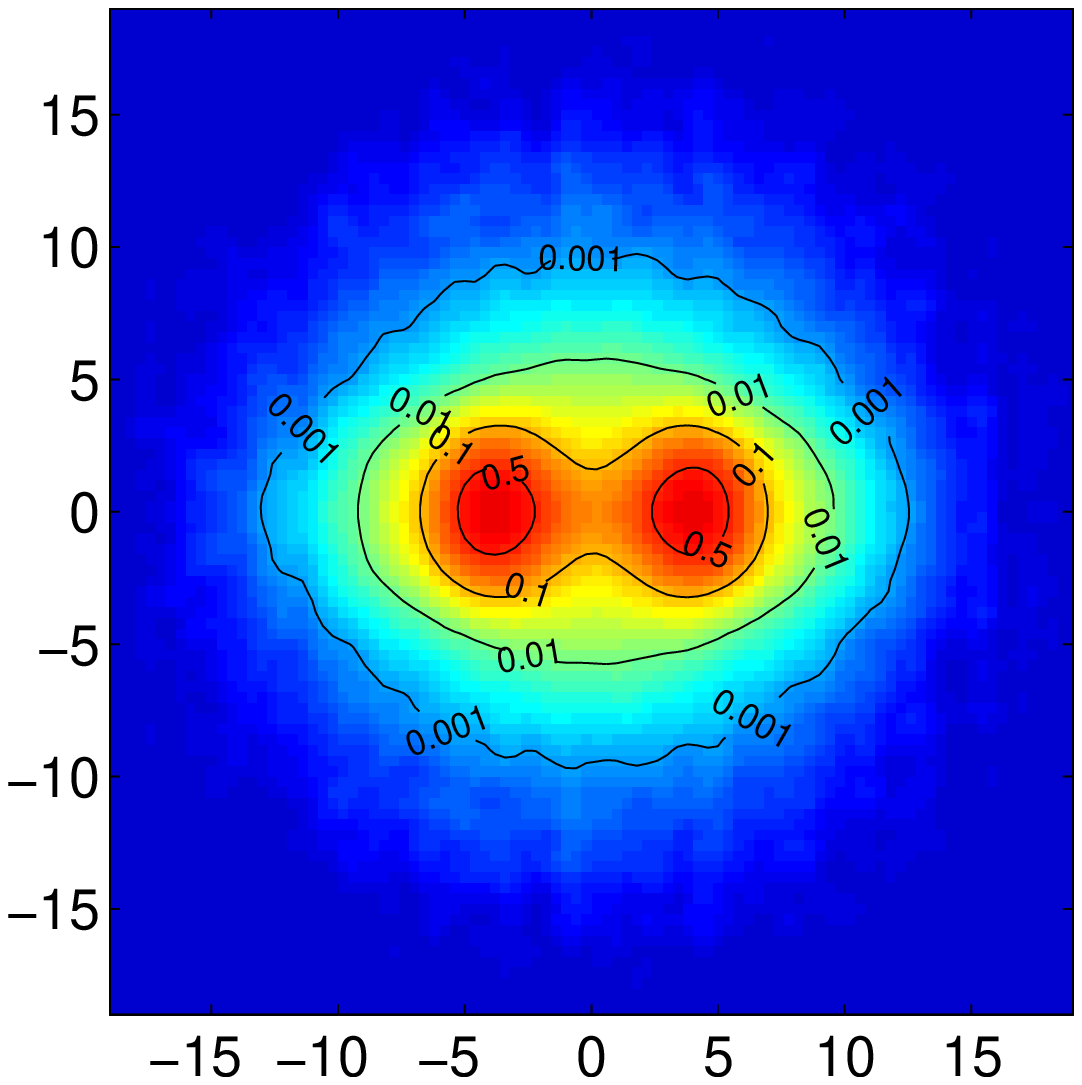}
}
\subfigure{
\includegraphics[width=\width1 \textwidth]{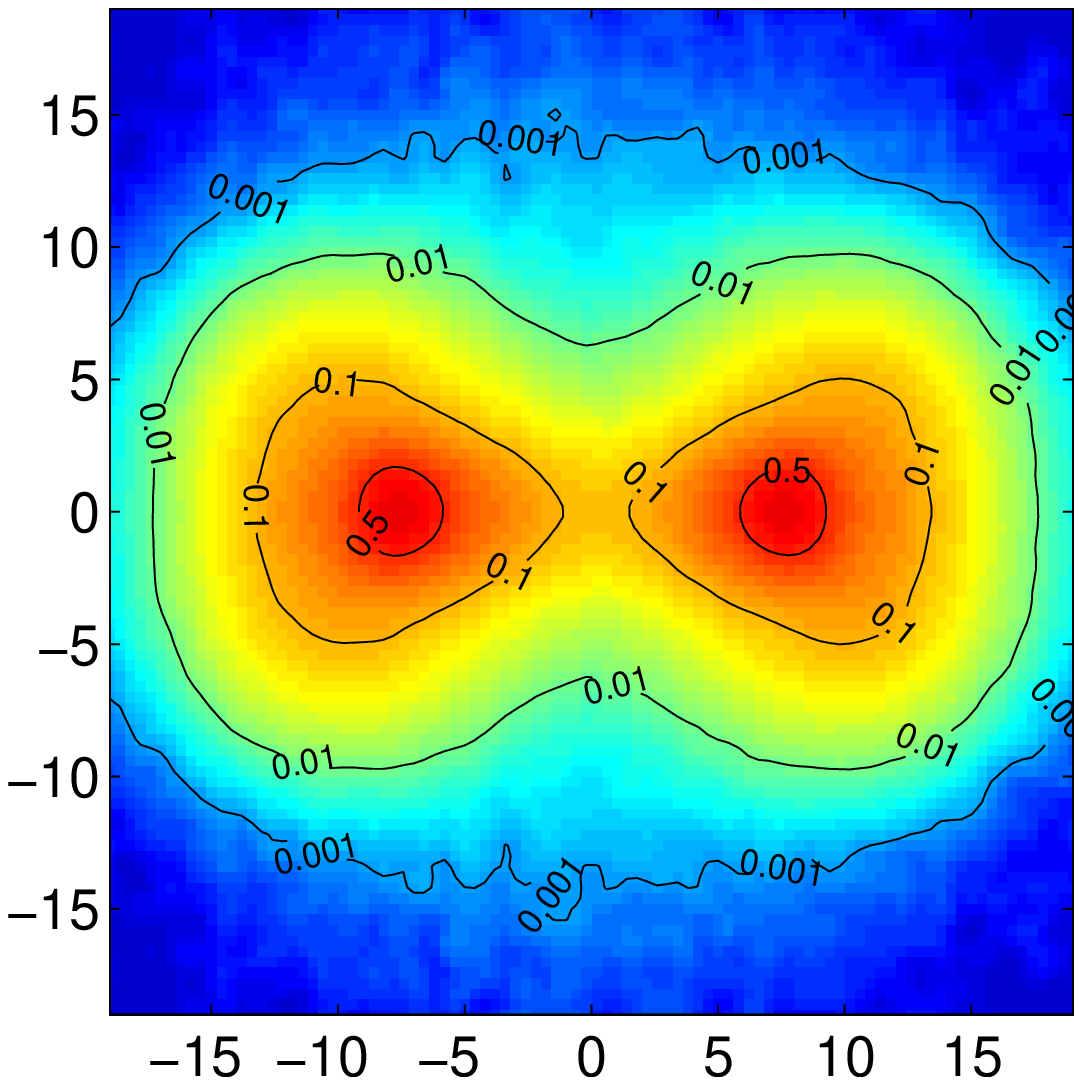}
}
\\
\subfigure{
\includegraphics[width=\width1 \textwidth]{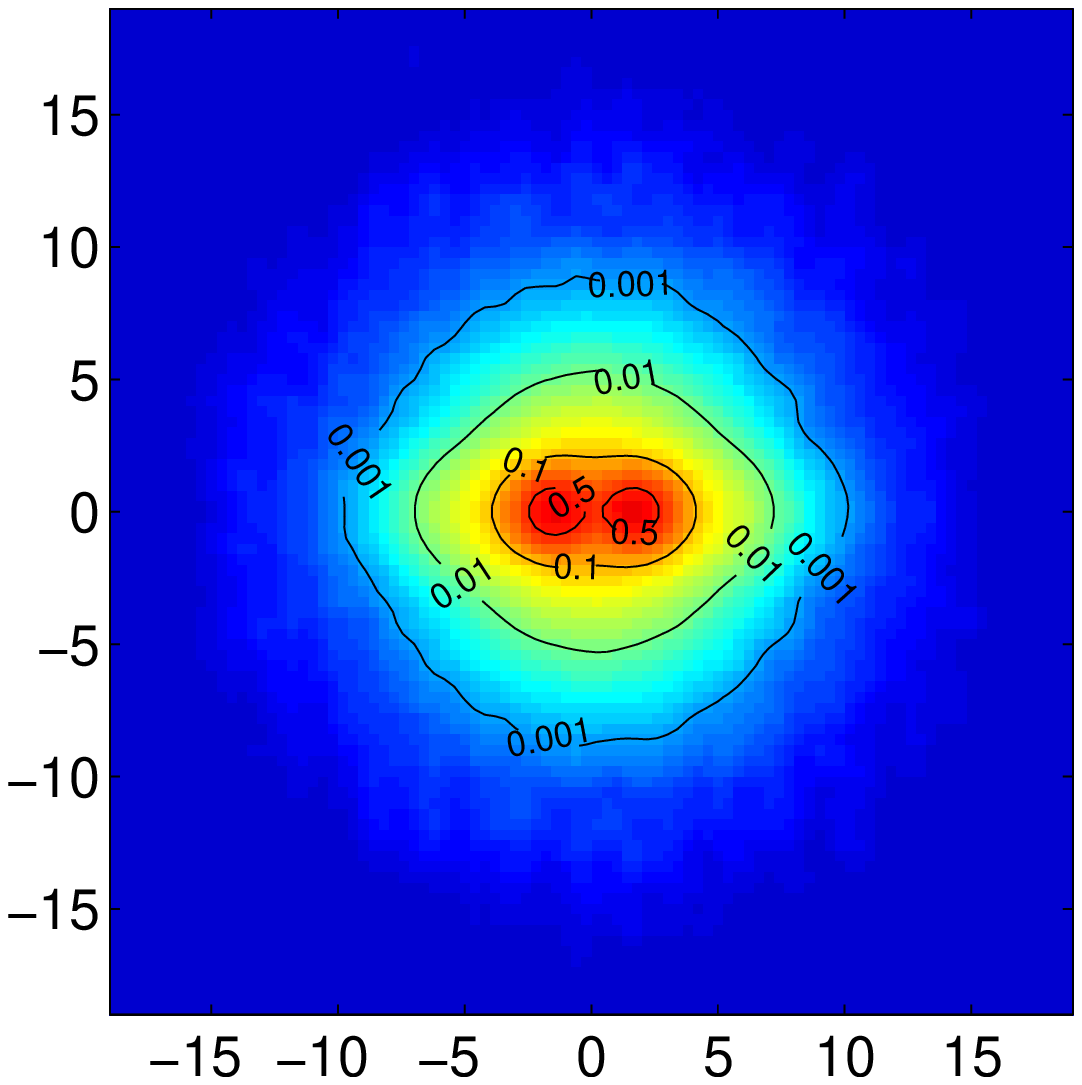}
}
\subfigure{
\includegraphics[width=\width1 \textwidth]{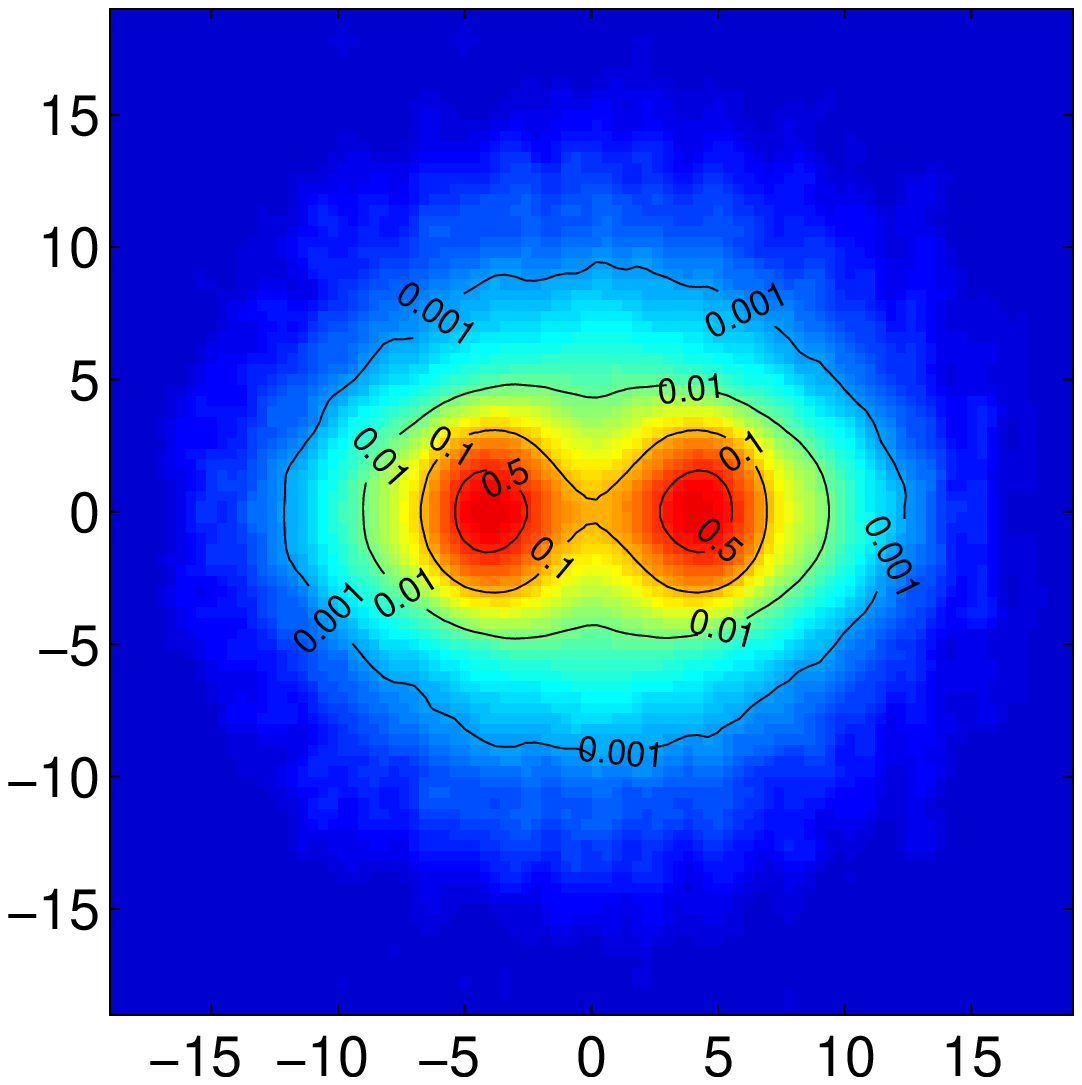}
}
\subfigure{
\includegraphics[width=\width1 \textwidth]{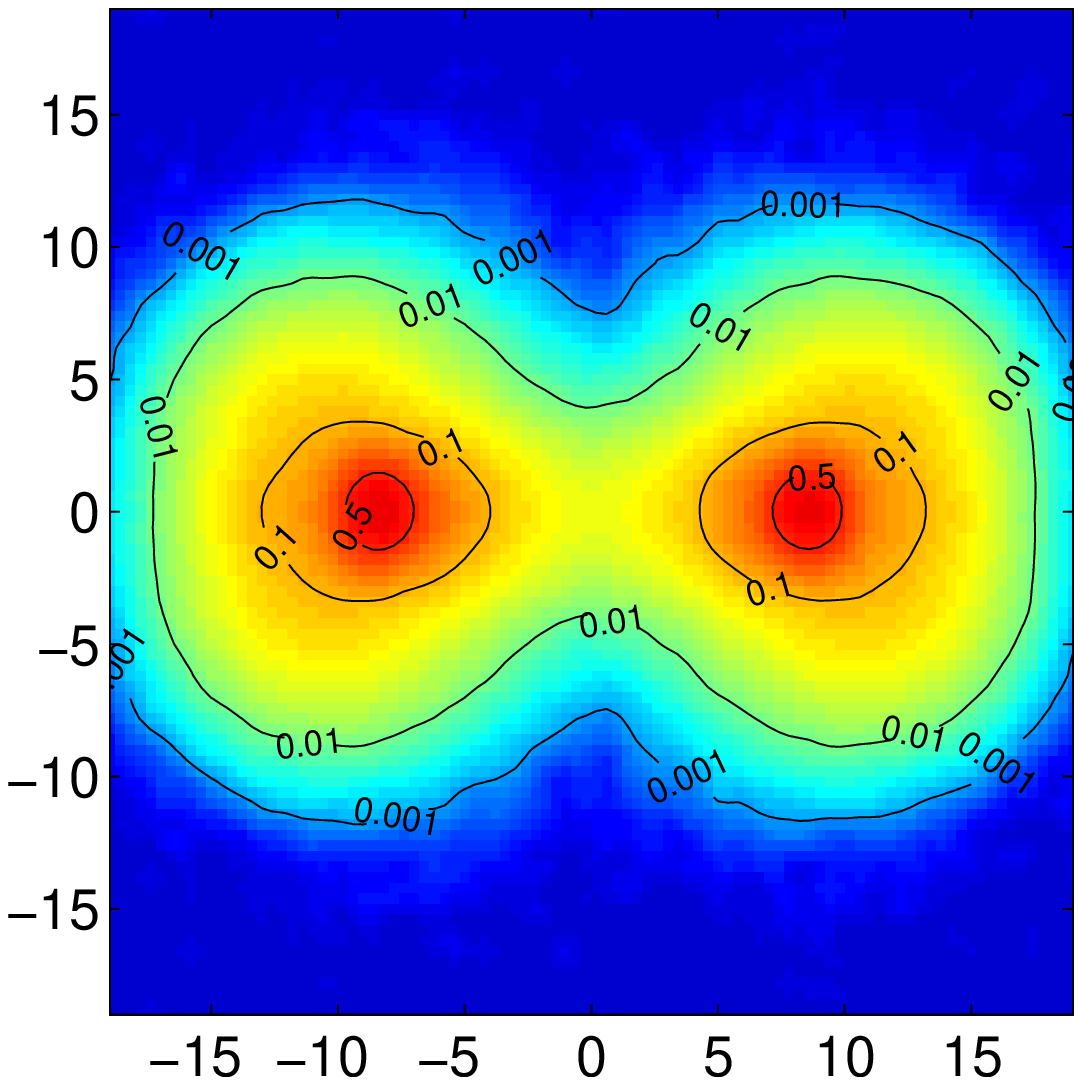}
}
\end{center}
\caption{\label{fig:obscmp15}
The difference in the post-collision mass distributions of fast symmetric galaxy cluster collisions with weakly scattering DM and $k=1.5$. Top row shows the post-collision mass distribution with weakly scattering DM and the bottom row shows the same for standard CDM, for comparison. Three post-collision stages are shown characterized by inter-cluster separation in units of $r_c$: early stage where the galaxy clusters are just barely separated ($\Delta r=2.5r_c$, left), intermediate stage where the galaxy clusters just recently became fully separated ($\Delta r=7.5r_c$, center), and late stage where the galaxy clusters fully separated and moved away to an appreciable distance ($\Delta r=15 r_c$, right).
The simulation parameters are as in Table \ref{table:params}.
}
\end{figure}

\def \width1 {0.30}
\begin{figure}[h!]
\begin{center}
\subfigure{
\includegraphics[width=\width1 \textwidth]{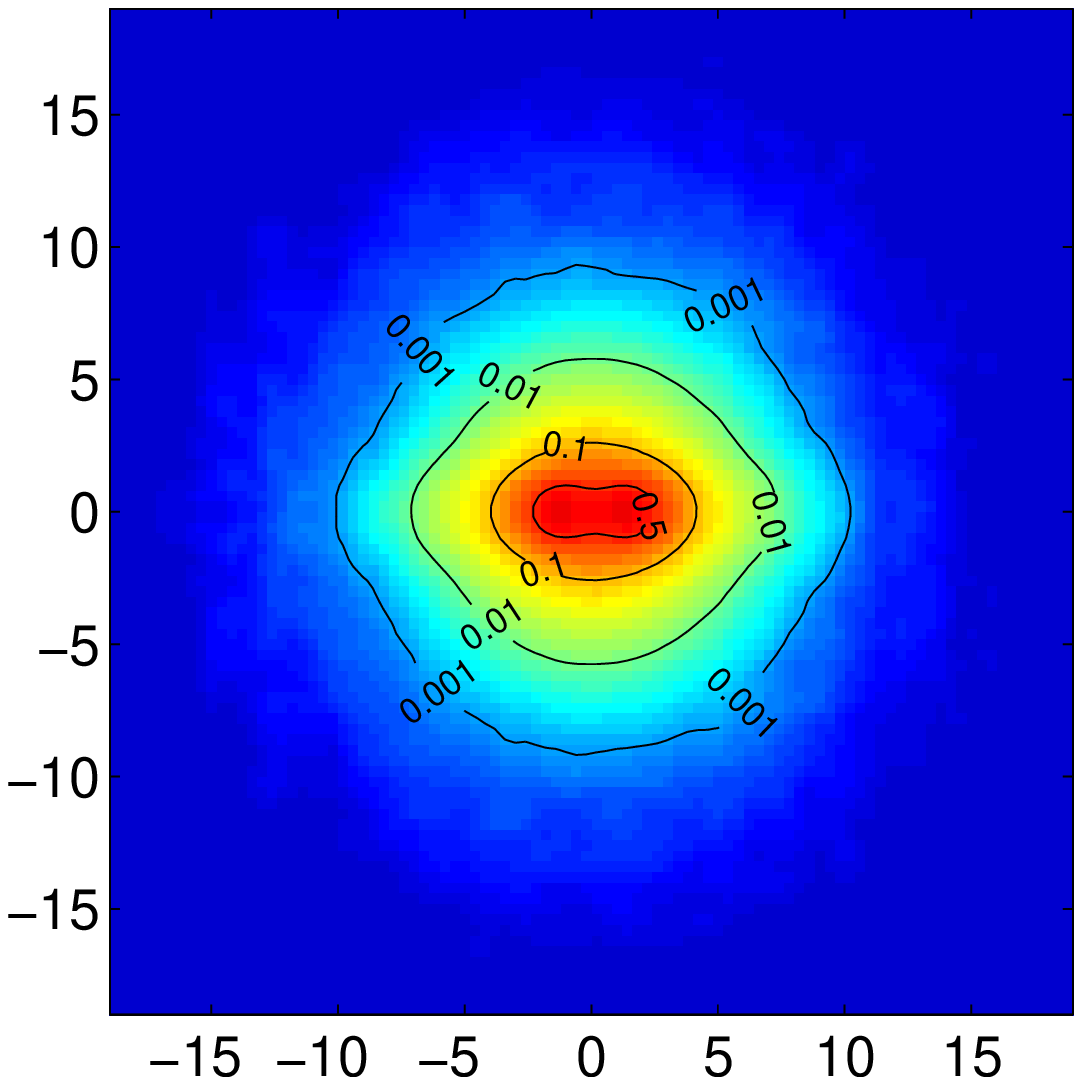}
}
\subfigure{
\includegraphics[width=\width1 \textwidth]{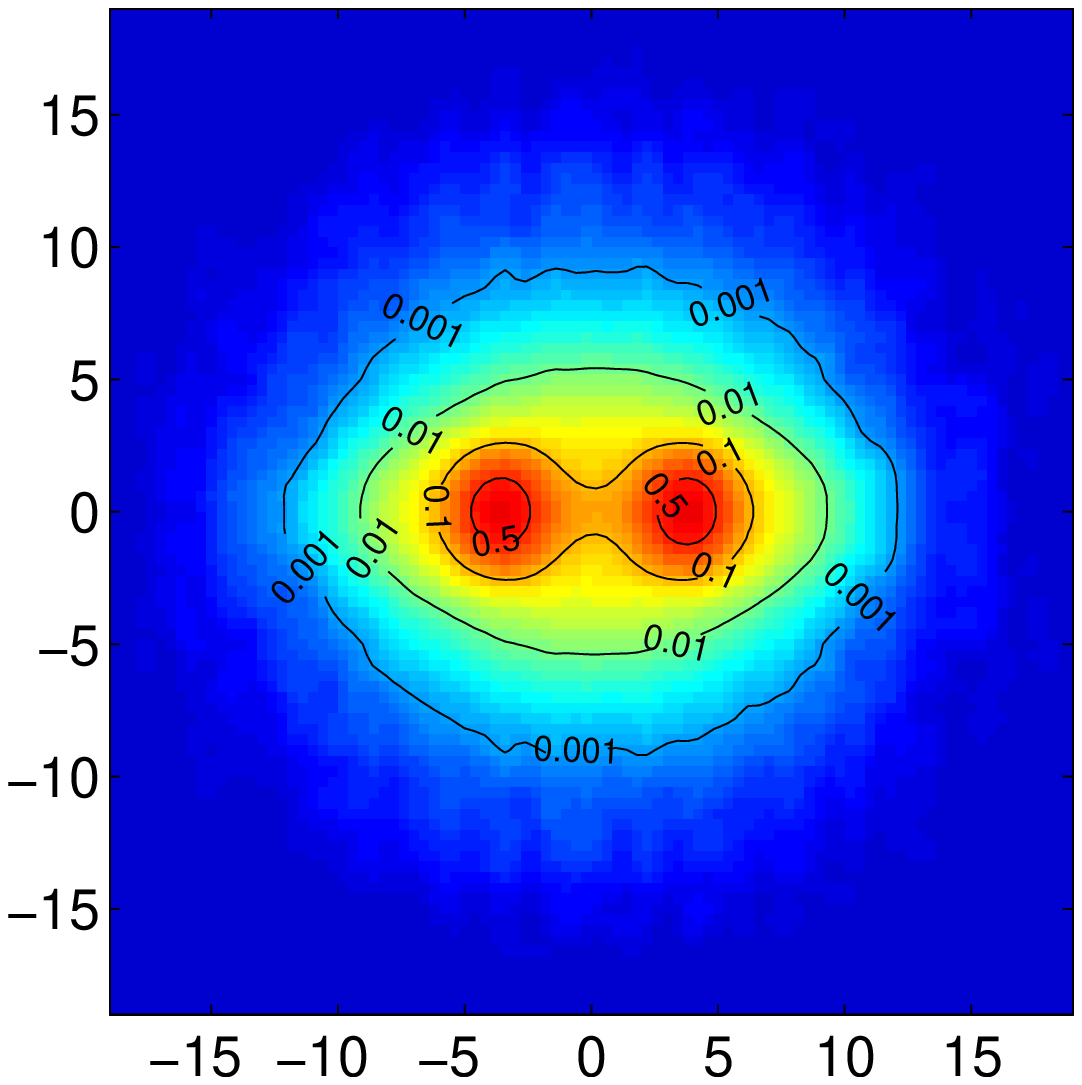}
}
\subfigure{
\includegraphics[width=\width1 \textwidth]{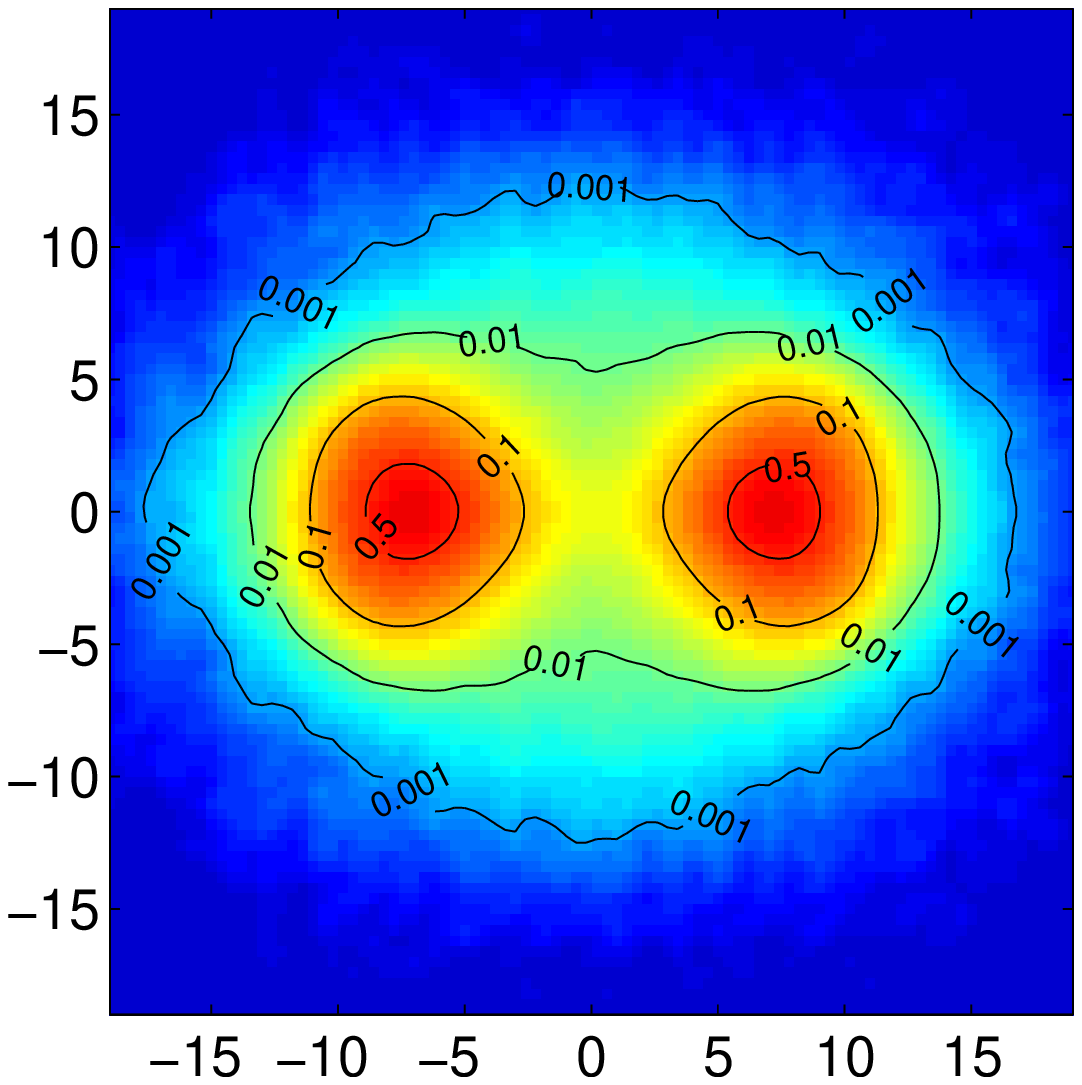}
}
\\
\subfigure{
\includegraphics[width=\width1 \textwidth]{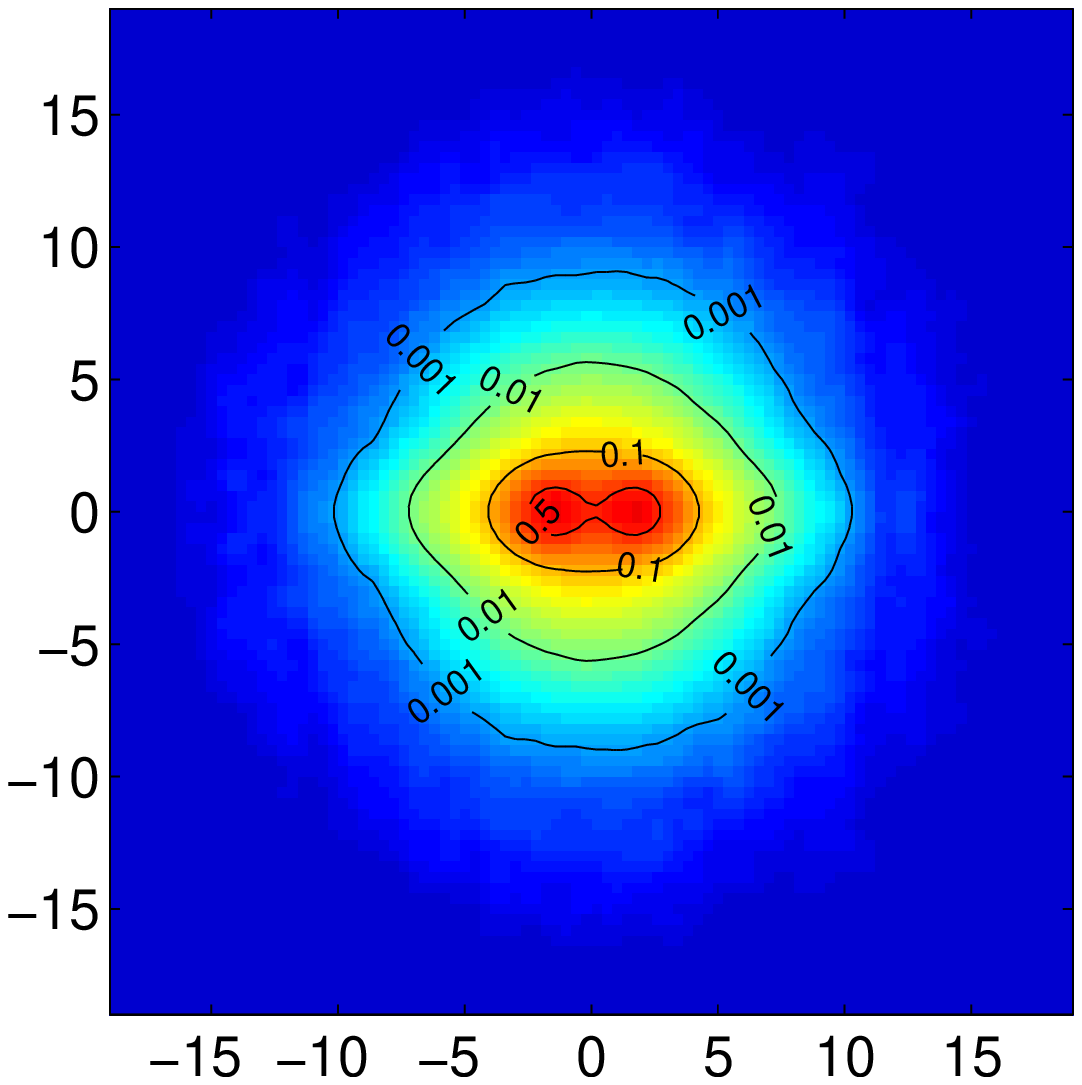}
}
\subfigure{
\includegraphics[width=\width1 \textwidth]{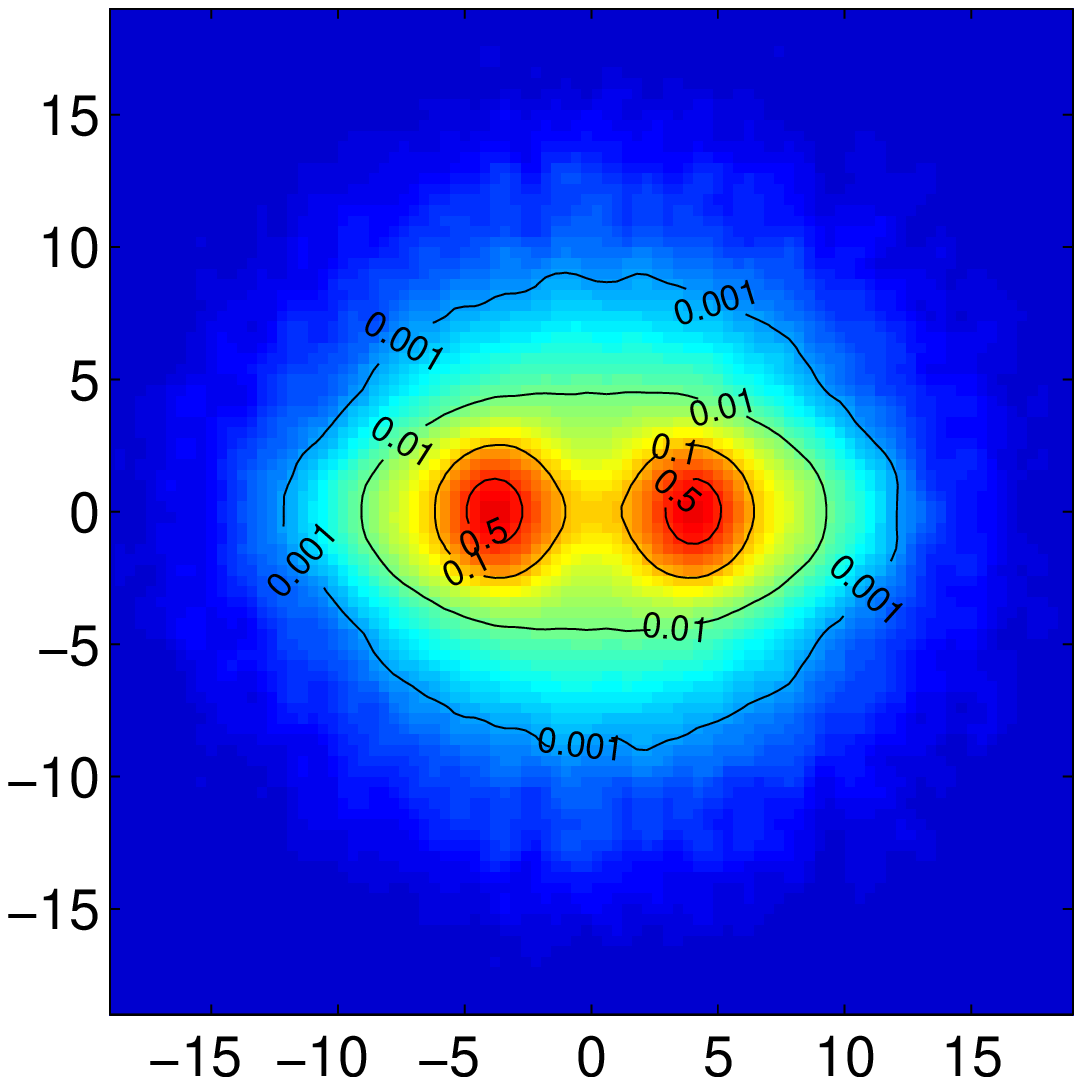}
}
\subfigure{
\includegraphics[width=\width1 \textwidth]{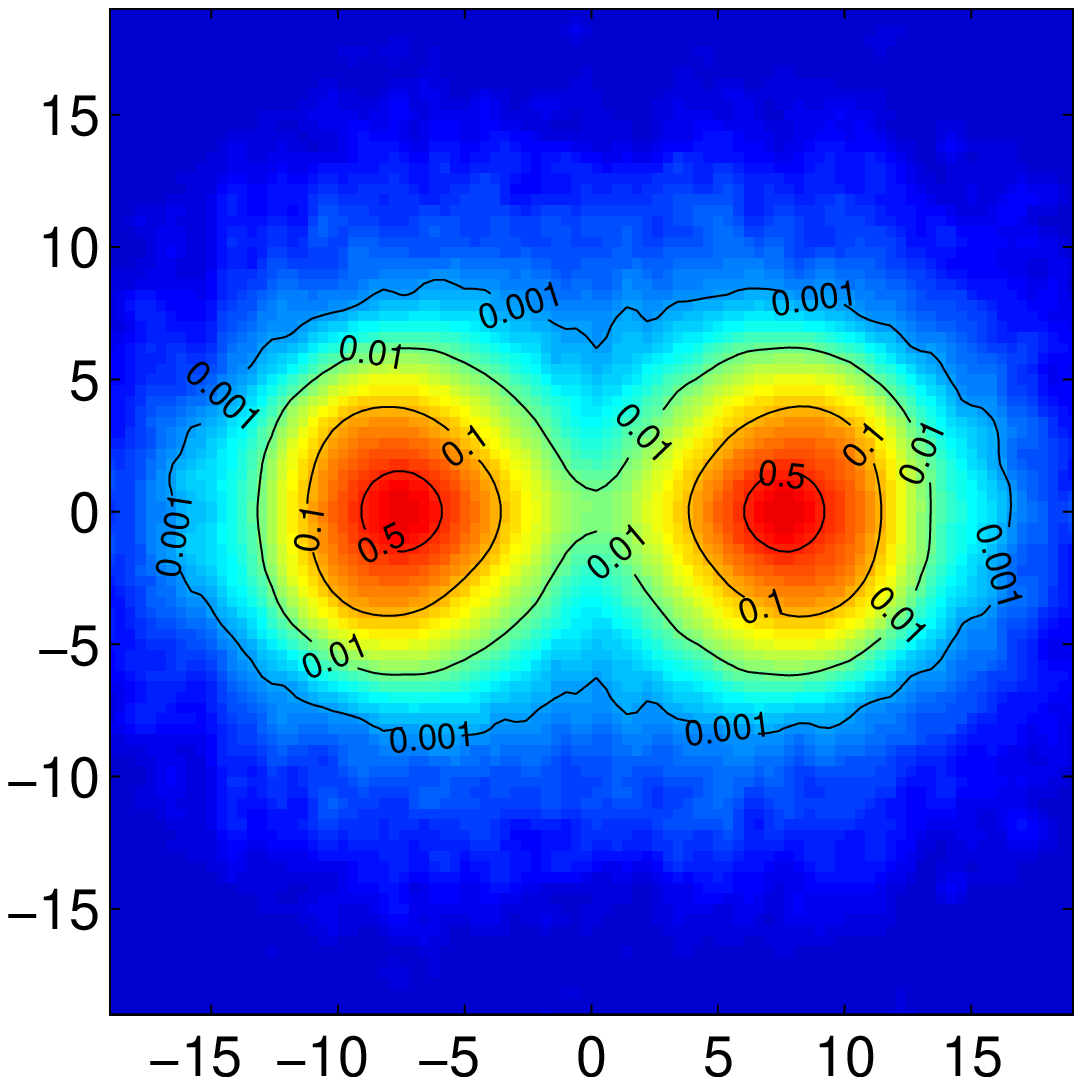}
}
\end{center}
\caption{\label{fig:obscmp20}
The difference in the post-collision mass distributions of fast symmetric galaxy cluster collisions with weakly scattering DM and $k=2.0$. Top row shows the post-collision mass distribution with weakly scattering DM and the bottom row shows the same for standard CDM. Left panels show the mass distributions at early separation stage ($\Delta r=2.5r_c$), center panels show the mass distributions at intermediate separation stage ($\Delta r=7.5r_c$), and right panels show the mass distributions at late separation stage  ($\Delta r=15 r_c$).
The distance scales are in the units of $r_c$.
The simulation parameters are as in Table \ref{table:params}.
}
\end{figure}

\def \width1 {0.30}
\begin{figure}[h!]
\begin{center}
\subfigure{
\includegraphics[width=\width1 \textwidth]{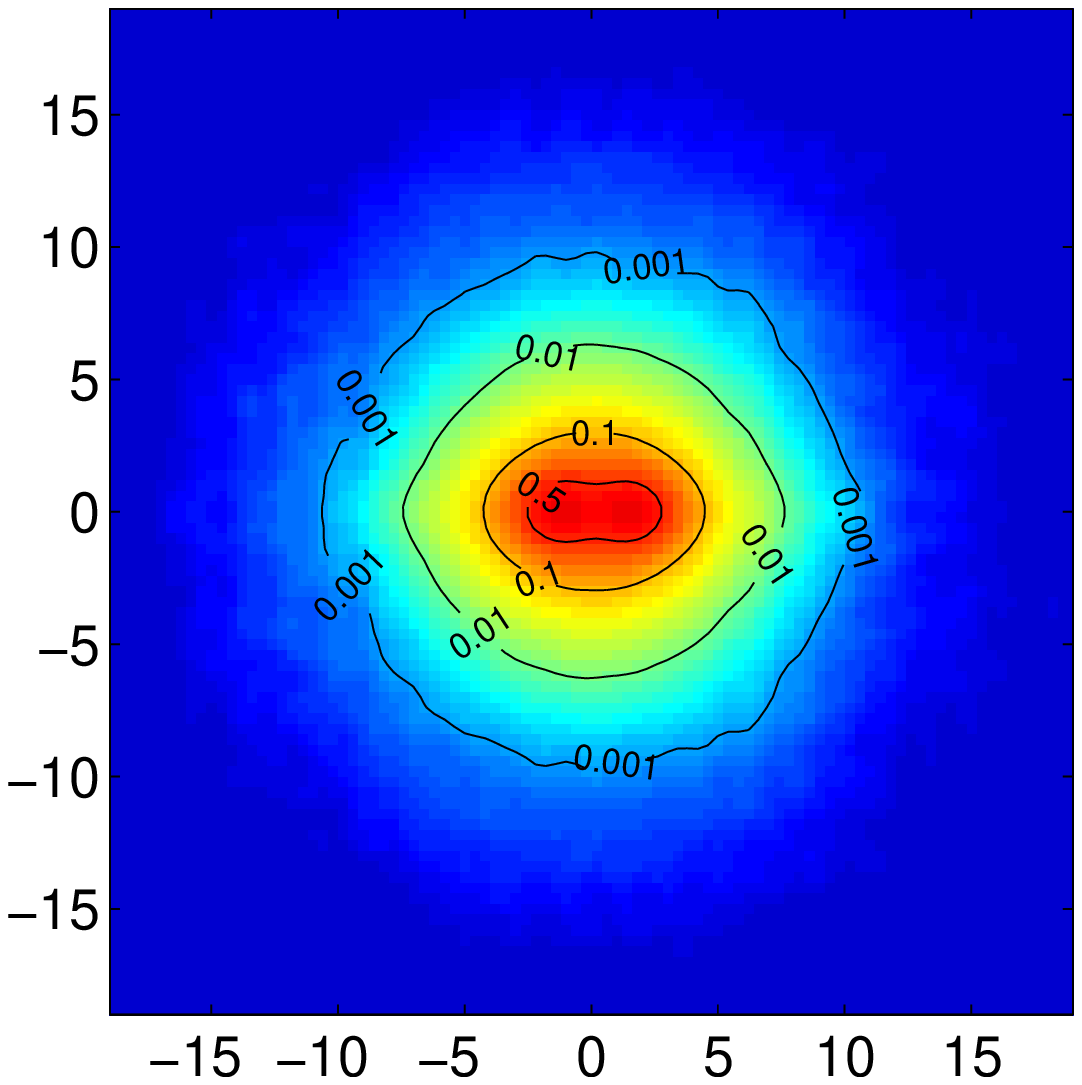}
}
\subfigure{
\includegraphics[width=\width1 \textwidth]{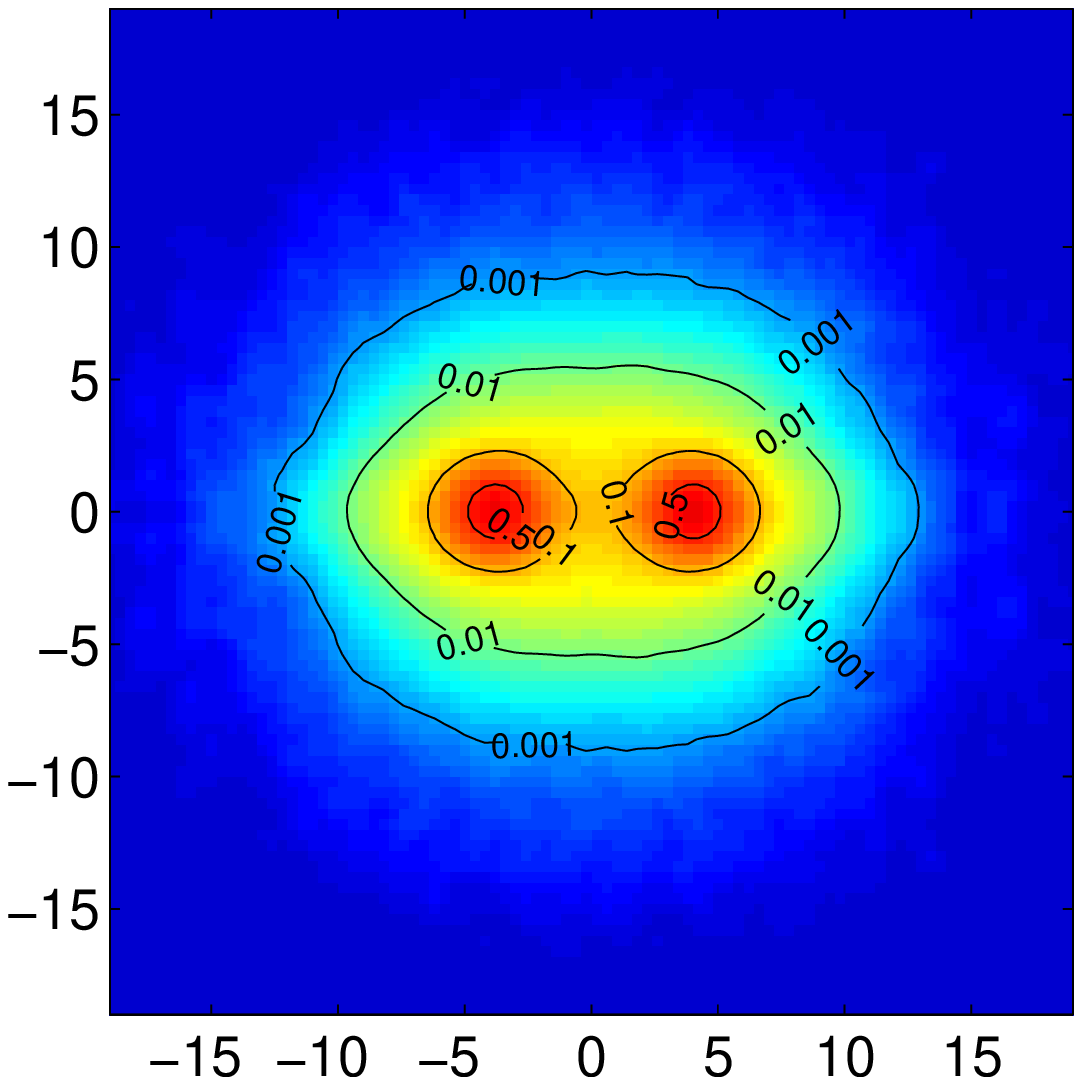}
}
\subfigure{
\includegraphics[width=\width1 \textwidth]{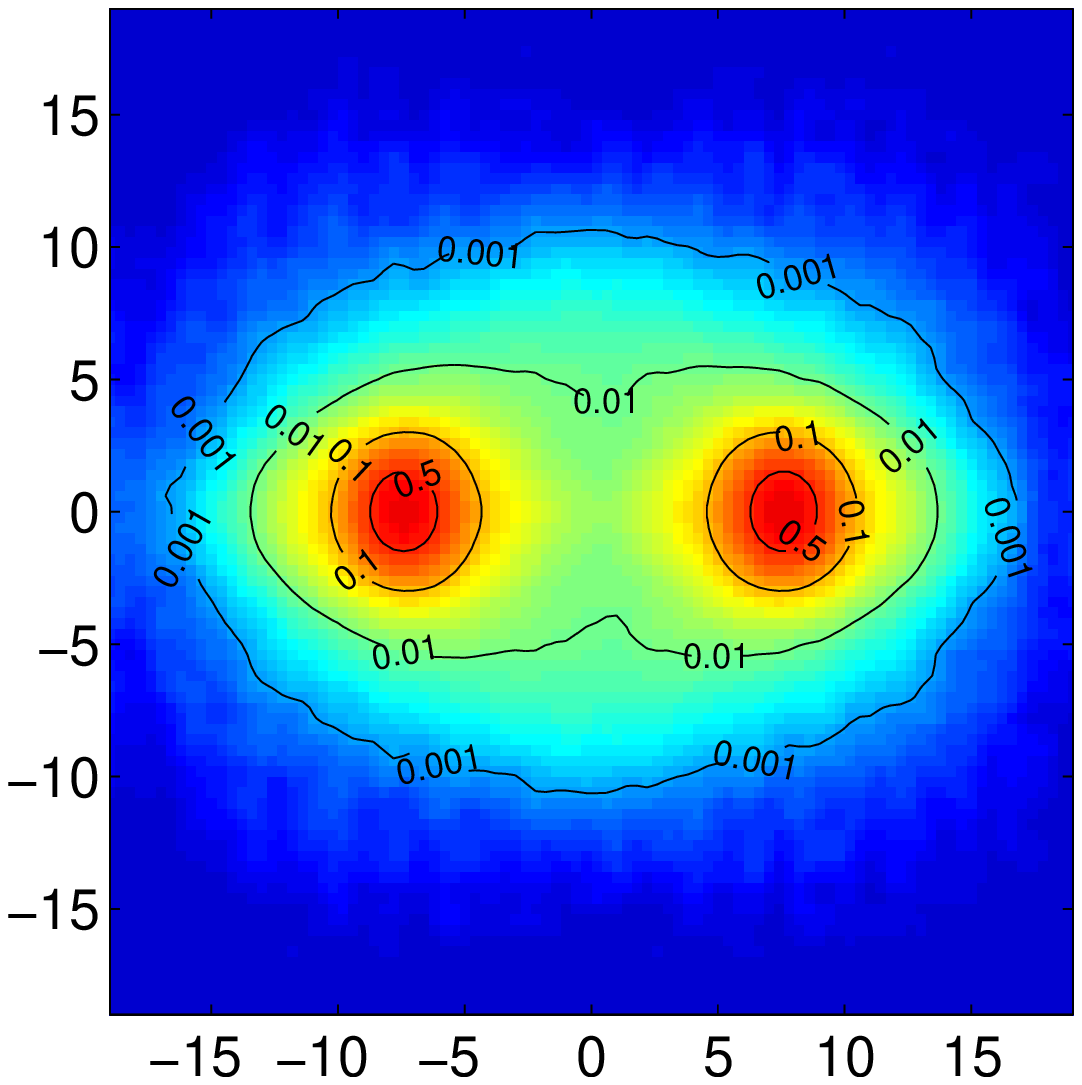}
}
\\
\subfigure{
\includegraphics[width=\width1 \textwidth]{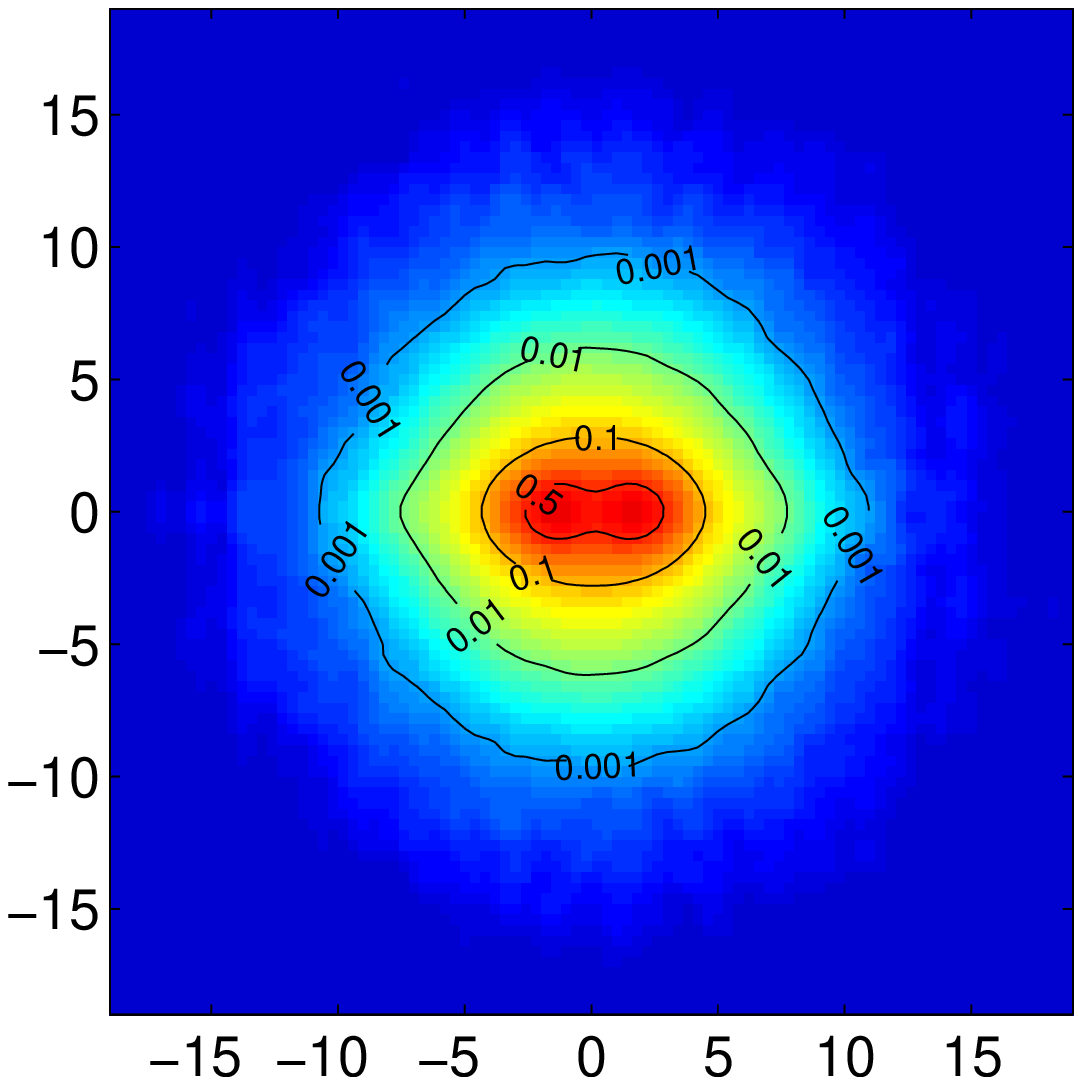}
}
\subfigure{
\includegraphics[width=\width1 \textwidth]{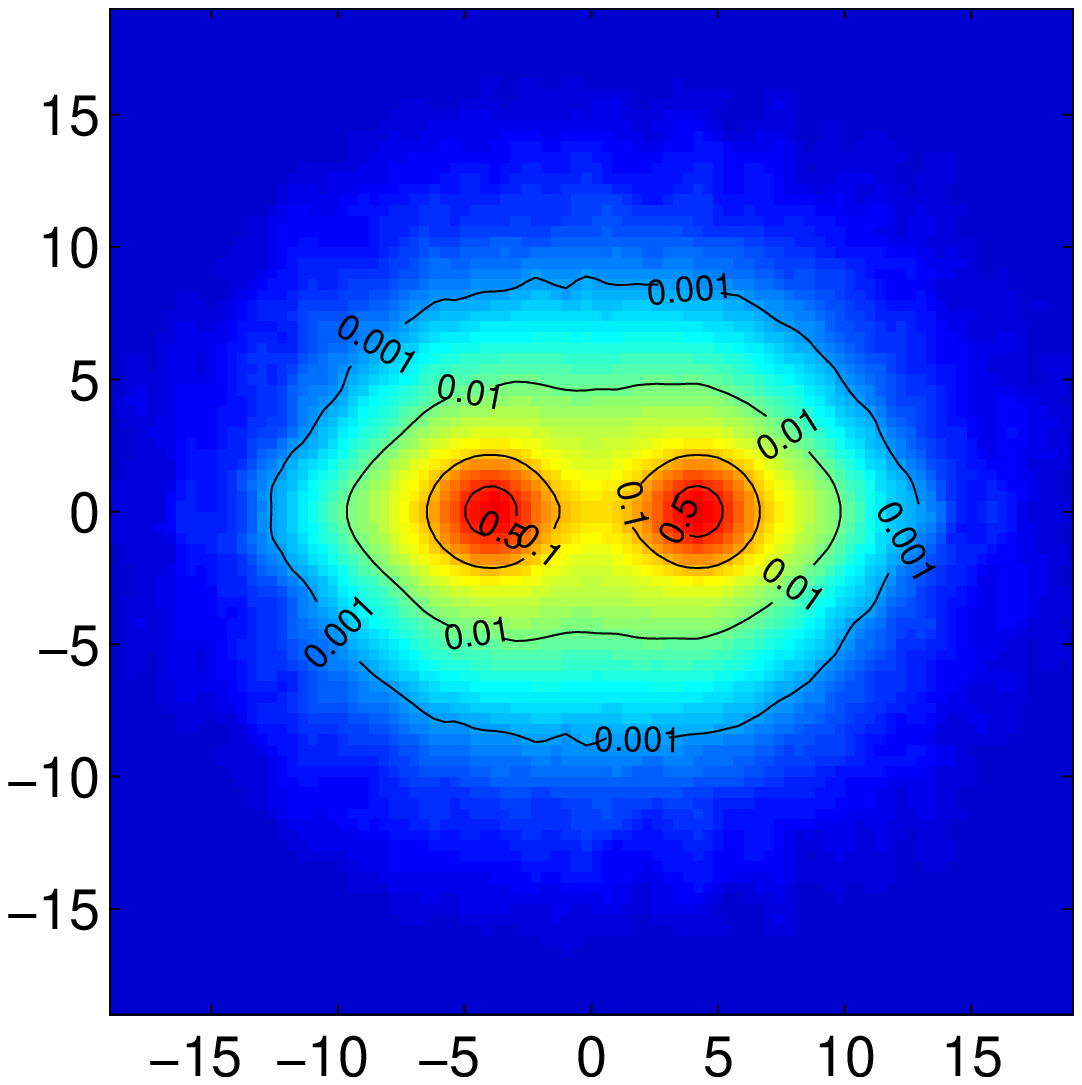}
}
\subfigure{
\includegraphics[width=\width1 \textwidth]{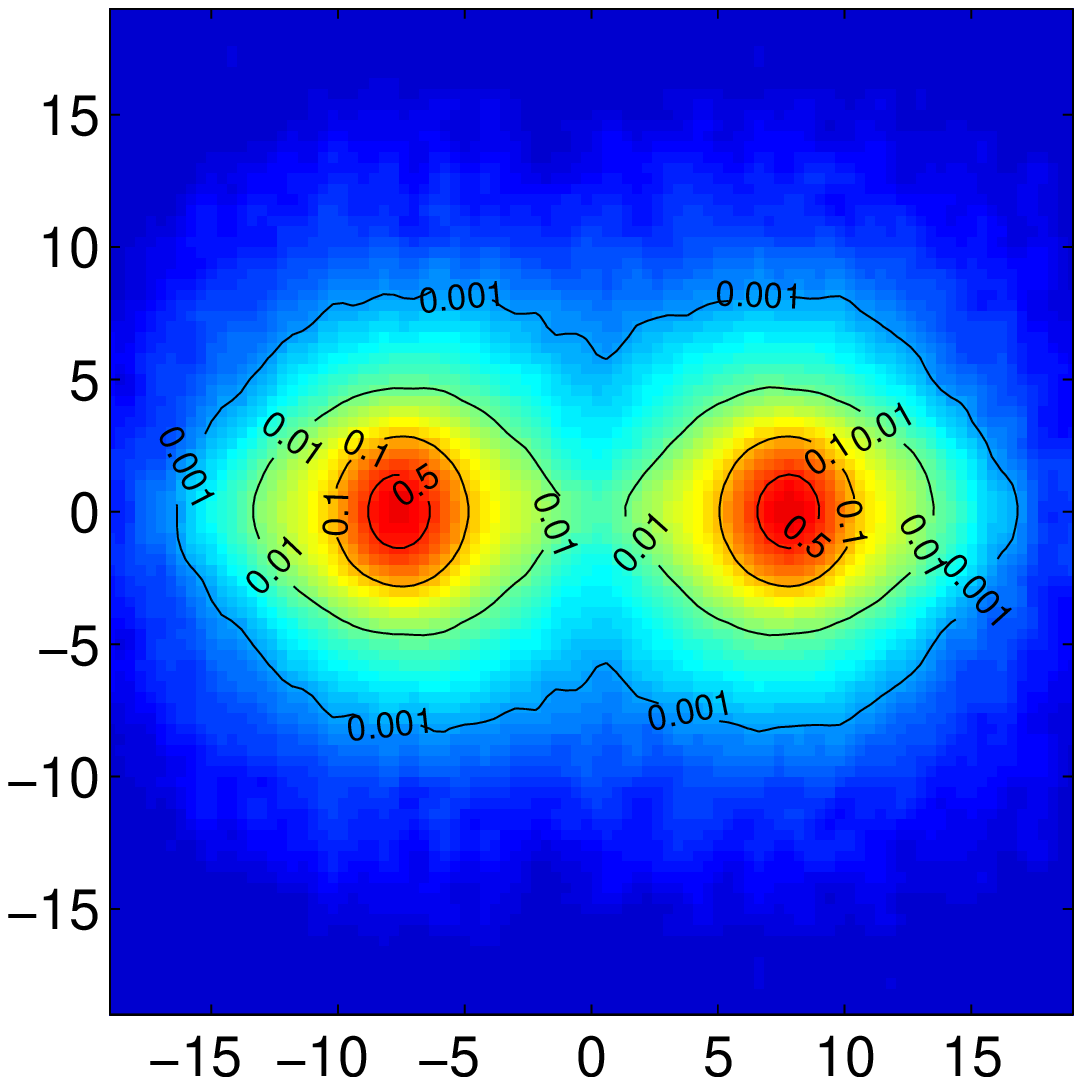}
}
\end{center}
\caption{\label{fig:obscmp40}
The difference in the post-collision mass distributions of fast symmetric galaxy cluster collisions with weakly scattering DM and $k=4.0$. Top row shows the post-collision mass distribution with weakly scattering DM and the bottom row shows the same for standard CDM. Left panels show the mass distributions at early separation stage ($\Delta r=2.5r_c$), center panels show the mass distributions at intermediate separation stage ($\Delta r=7.5r_c$), and right panels show the mass distributions at late separation stage  ($\Delta r=15 r_c$).
The distance scales are in the units of $r_c$.
The simulation parameters are as in Table \ref{table:params}.
}
\end{figure}

Figs.~\ref{fig:obscmp15}-\ref{fig:obscmp40} inspect in details the differences appearing in weakly self-scattering SIDM and CDM for kinetic parameter $k$ between 1.5 and 4.0. In each figure, the projected mass densities for the early separation ($\Delta r=2.5r_c$, where $\Delta r$ is the distance between the centers of the outgoing galaxy clusters), intermediate separation ($\Delta r=7.5r_c$) and late separation stage ($\Delta r=15r_c$) are shown in comparison with the same CDM scenario. 
The early separation stage corresponds to the colliding galaxy clusters that just barely emerged out of the collision, with the separation at just about 50\% peak-density level.
The intermediate separation stage here corresponds to the colliding galaxy clusters that nearly fully emerged out of the collision but overlap significantly within their $r_{200}$ radii. In late separation stage, the galaxy clusters are well separated by a distance exceeding $r_{200}$.

In the early separation stage, in all scenarios we see quantitative differences in the distribution of mass at central region of colliding galaxy clusters where SIDM mass distributions show substantially heavier and wider central densities, bridging outgoing galaxy clusters at high densities. This difference from CDM scenarios at 10\% to 50\% is easily noticeable. The origin of this difference is understood physically as the mass contribution coming from DM ejecta shell scattered directly above and below the collision site, along the line of sight, thus remaining near the center of the collision for a far longer times than would be observed in pure CDM.

At intermediate separations, the projection of the DM ejecta shell over the collision's center continues to contribute significantly to central region of the collision forming more prominent bridges linking the outgoing galaxy clusters, even whereas the galaxy clusters in the same CDM scenarios separate at that stage completely - middle panels of Figs.~\ref{fig:obscmp15}-\ref{fig:obscmp40}. We believe this effect will be noticeable in astrophysical reconstructions of projected mass densities observed at suitable time epochs.

At late separation stages, scattered DM ejecta shell begins to emerge as a separate component in the mass density field, introducing quantitative and qualitative changes in the mass distributions, as  discussed previously. A distinct oval-shaped shell forms at 0.1\% to 1\% peak-density levels in those scenarios, as seen in Fig.~\ref{fig:obscmp20} and Fig.~\ref{fig:obscmp40} rightmost panel. These differences are the most pronounced for collisions with high values of kinetic parameter $k=2$ and above. 
These differences are dramatically different from CDM but are also quite weak and confined to the tails of DM halos, which may make their detection challenging.

The bridge that we detect could conceivably be mimicked by the collisional intra-cluster gas as real galaxy clusters have significant gas fractions that may be slowed by ram pressure and accumulate at the center of the collision. To that extent, separation of the DM and the baryonic gas in the central mass distribution will require careful analysis including both detailed numerical simulations of the specific collision and experimental estimates of the amounts of hot central gas by using X-ray emissions. On the other hand, while the scattered DM shell manifests itself as added mass at the center of the collision at early post-collision stages, at later stages of galaxy cluster collisions such a shell expands and begins to contribute mass well beyond the central region, where the presence of intra-cluster gas will be substantially less significant. 

\subsubsection{Quantitative measures of SIDM scattering effects}

\begin{figure}[h!]
\begin{center}
\subfigure{
\includegraphics[width=0.3\textwidth]{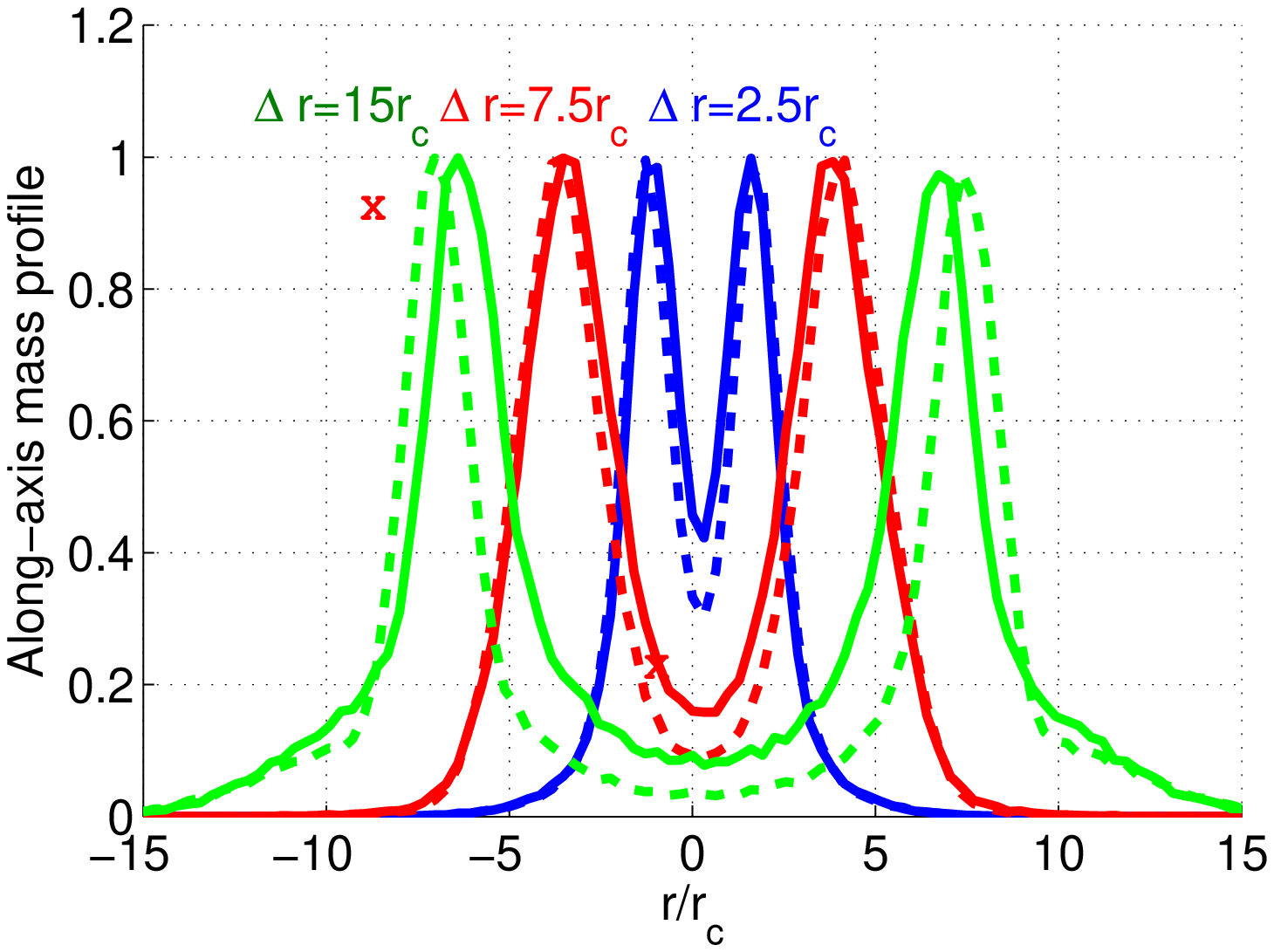}
}
\subfigure{
\includegraphics[width=0.3\textwidth]{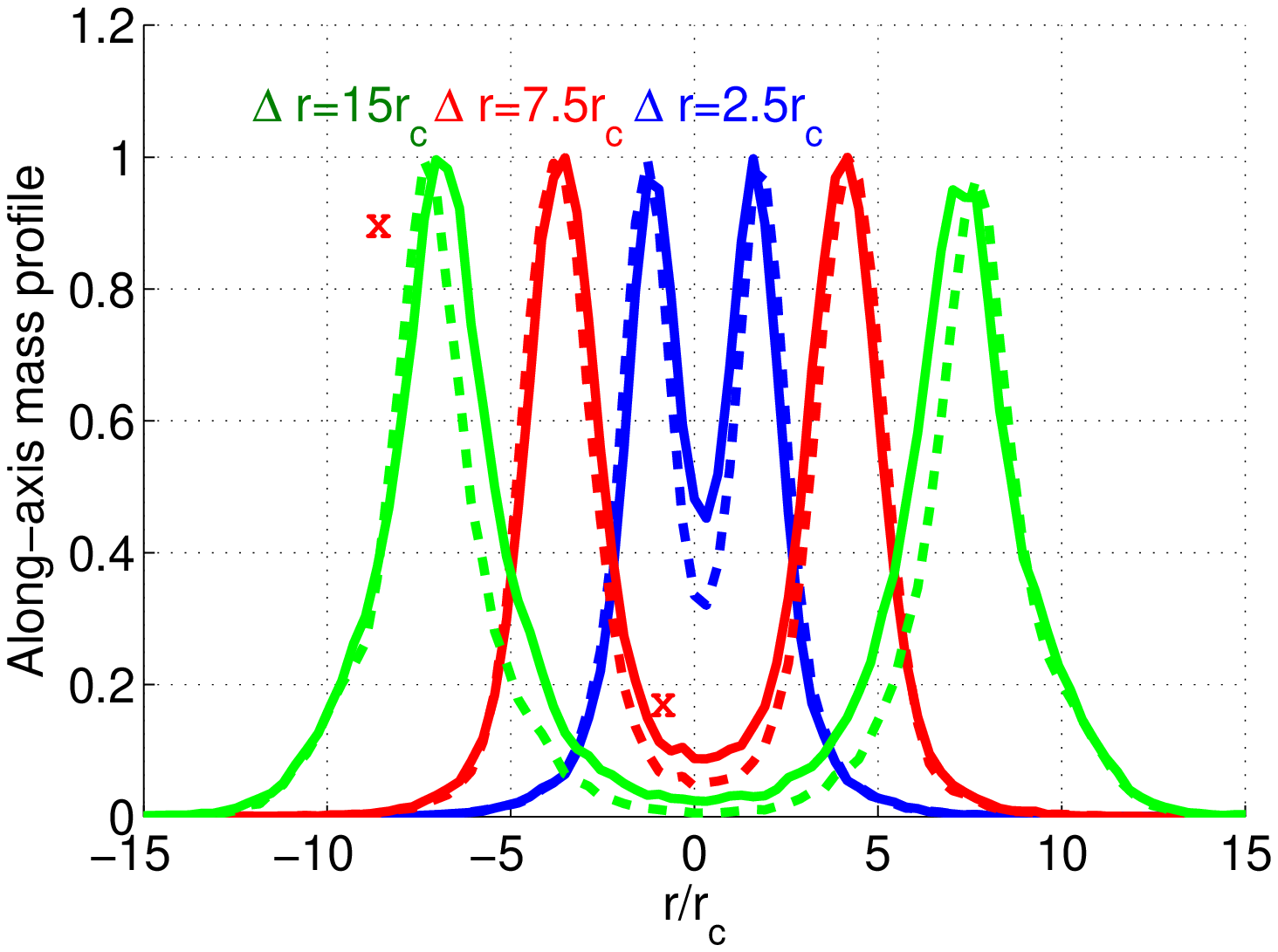}
}
\subfigure{
\includegraphics[width=0.3\textwidth]{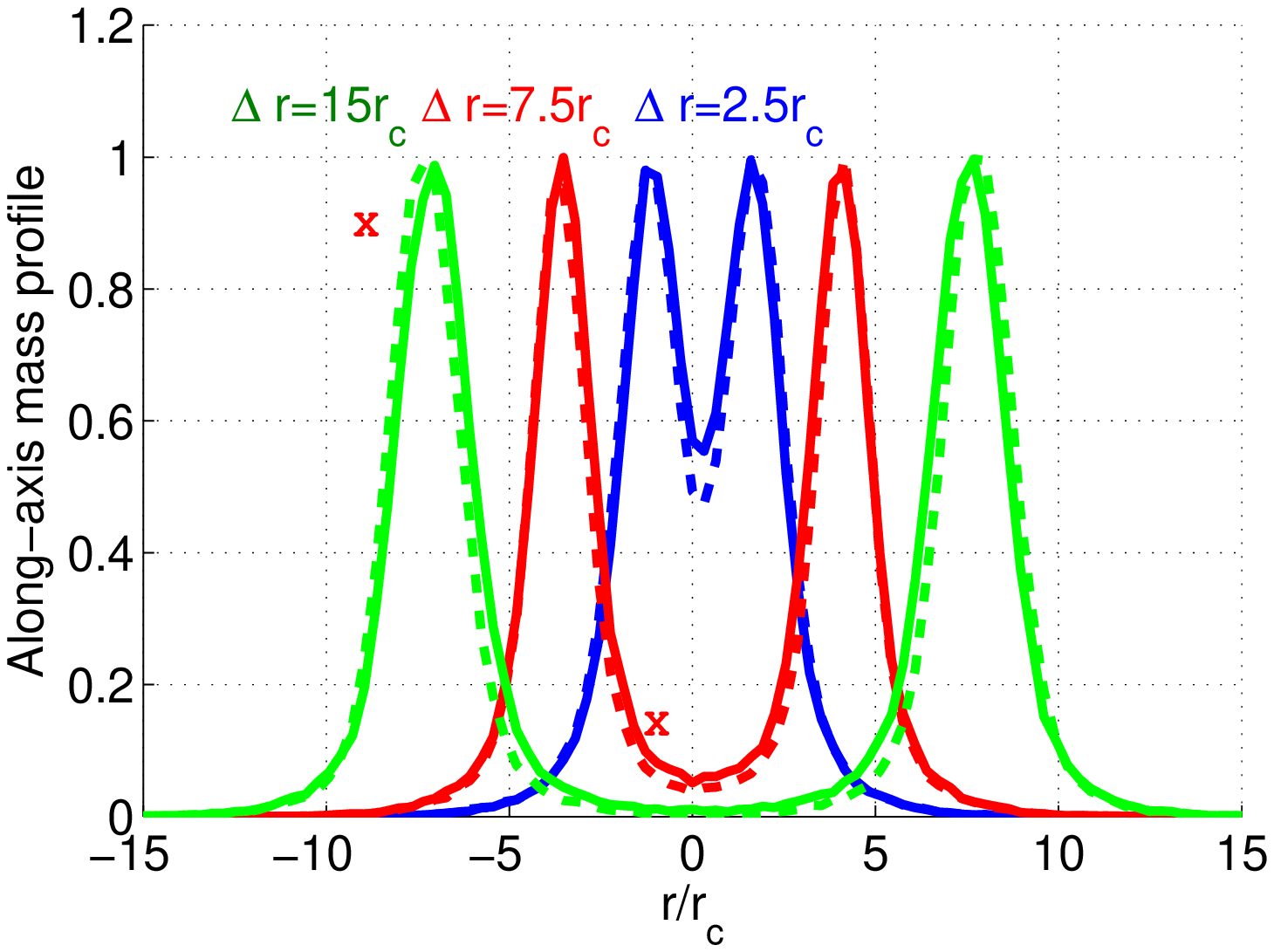}
}
\end{center}
\caption{\label{fig:axldist} The projected mass density plotted along the line connecting the centers of the colliding galaxy clusters for the collisions with different values of the kinetic parameter $k$ and different post-collision separation stages. Shown are such mass density plots for early ($\Delta r=2.5r_c$), intermediate ($\Delta r=7.5r_c$) and late ($\Delta r=15r_c$) post-collision separation stages, for weakly interacting DM (solid lines) and CDM (dashed lines), for $k=1.5$ (left), $k=2.0$ (center) and $k=4.0$ (right). 
The differences between interacting DM and CDM models focused upon in the main text are shown with the symbol ``X".
The CDM halos were first plotted at a fixed separation, and then the SIDM halos were plotted at the same post-collision time, in order to show the offsets between the centroids of the CDM and SIDM halos in galaxy cluster collisions.
}
\end{figure}

While projected mass density maps are telling about the nature of the effects introduced into galaxy cluster collisions by SIDM, specific methods of measuring such effects are advantageous for such effects' detection.
Such measures can be constructed by quantifying the projected mass densities on the line connecting the centers of the outgoing galaxy clusters and measuring angular mass distribution relative to collision center in the collisions' projected density maps. 

Fig.~\ref{fig:axldist} shows such a projected mass distribution plotted against the collision axial line. 
At early separation stages, we see that the difference in the projected mass density in central regions in SIDM and CDM can be rather significant. Such differences reach 10\% of peak-density values and can be directly measurable. 
In late stages, the central region in that scenario features as much as two times more mass than in the same CDM scenarios for collisions with lower $k$. Slower collisions also feature an appreciable lag of DM halo centroids in SIDM as opposed to CDM collision scenarios. This lag is the reflection of the lower outgoing velocity of colliding galaxy clusters with SIDM, and can be seen clearly in the left panel of Fig.~\ref{fig:axldist} reaching $0.7-0.8r_c$ at later post-collision separations.
However, the differences in ``axial'' mass density measure decreases with the increase of the kinetic parameter values $k$. In particular, whereas one continues to see a noticeable 10-15\% differences in the central mass densities in early separation stages in the middle ($k=2.0$) and right ( $k=4.0$) panels of Fig.~\ref{fig:axldist}, such differences diminish as the collision progresses. There is also no consistent lag in the outgoing DM halo positions at higher collision velocities. This is related to the change in the outgoing clusters' speed becoming negligible as the speed of the collision increases past $k=2$.

%

\begin{figure}[h!]
\begin{center}
\subfigure{
\includegraphics[width=0.3\textwidth]{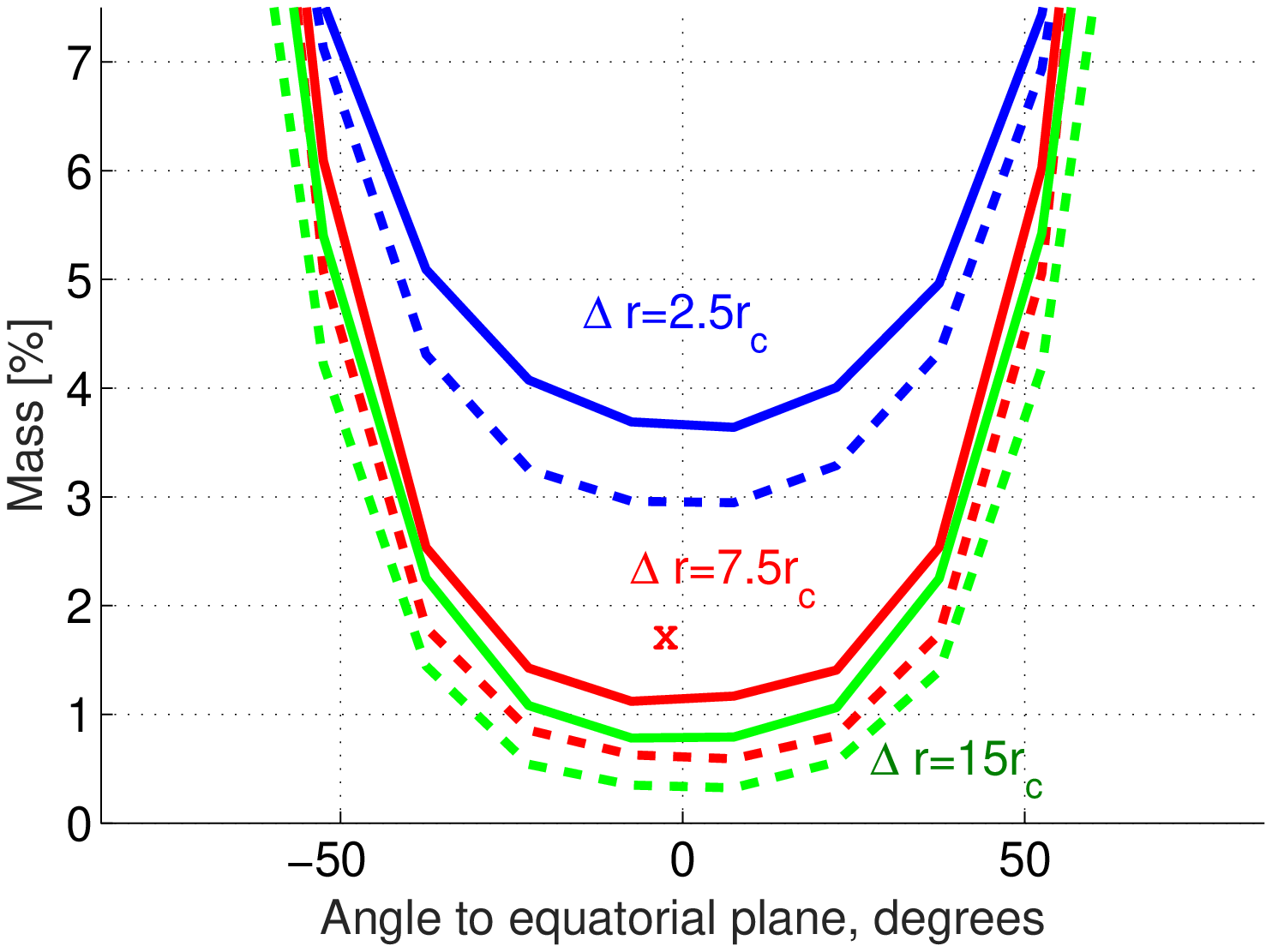}
}
\subfigure{
\includegraphics[width=0.3\textwidth]{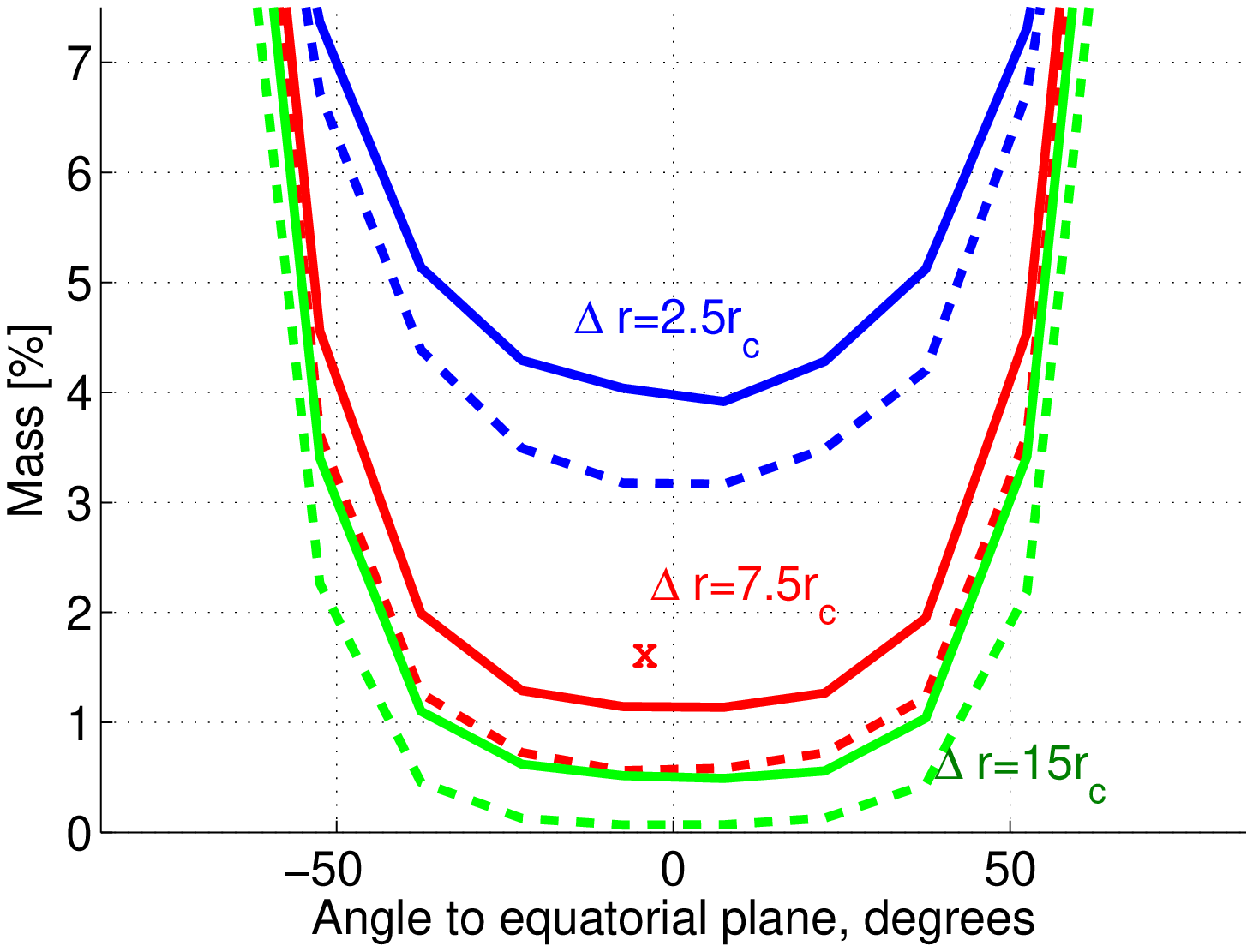}
}
\subfigure{
\includegraphics[width=0.3\textwidth]{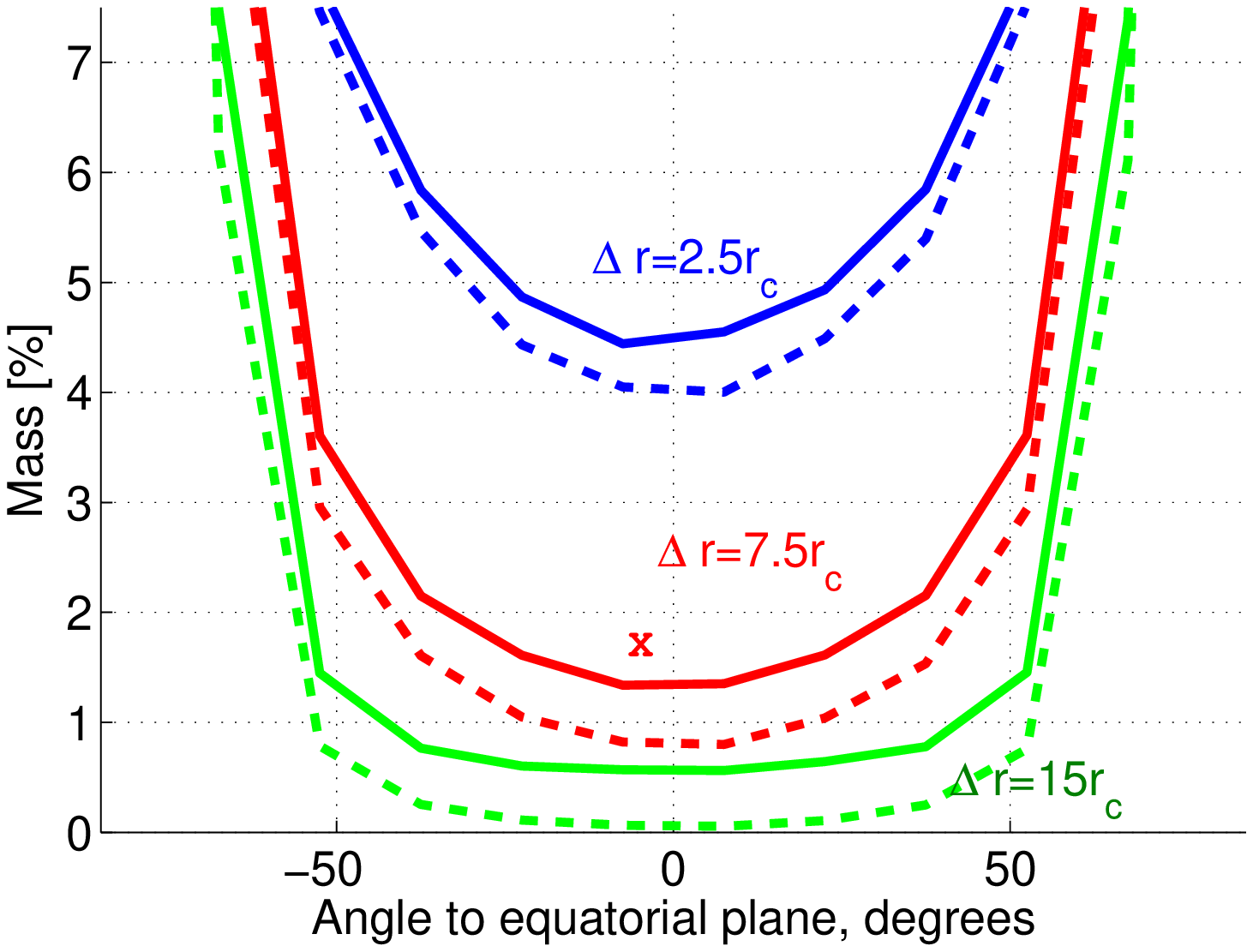}
}
\end{center}
\caption{\label{fig:angdist} The plots of the mass contained in projected mass density maps in $15^\circ$ degree sectors centered on the collision's center, as a function of the angle to the collision's equatorial plane, and shown as the percentile of the total density map's mass. Shown are the respective plots for early ($\Delta r=2.5r_c$), intermediate ($\Delta r=7.5r_c$) and late ($\Delta r=15r_c$) post-collision separation stages, for weakly interacting DM (solid lines) and CDM (dashed lines), for $k=1.5$ (left), $k=2.0$ (center) and $k=4.0$ (right). The differences between interacting DM and CDM models focused upon in the main text are shown with the symbol ``X".
}
\end{figure}

One of the differences that we observed for SIDM simulations is the shell of scattered DM material which can contribute mass at remote locations of the mass density map in directions perpendicular to the collision axis. To quantify this effect, the plot of projected mass in radial sectors vs. the scattering angle can be used. In particular, for SIDM one expects to see a close to uniform DM density in such mass-plot, 
whereas in CDM such mass will drop to zero at scattering angle close to $90^\circ$.
Fig.~\ref{fig:angdist} shows this measure for SIDM simulations and mass measured per $15^\circ$ radial sectors centered on the collision center at different angles away from the equatorial plane. In the case of CDM, we observe this measure dipping  to zero at angles close to $0^\circ$ (that is, near the equatorial plane), as expected. In SIDM, we observe the flat segment reminiscent of the scattered isotropic DM material shell. The difference in this measure is seen most prominently for simulations with high $k$ and late post-collision stage. In particular, there is no mass density contributed to sectors around the collision's equator in CDM while as much as 1\% of the total mass can be contained in the equatorial radial sector in SIDM scenario.

\begin{figure}[h!]
\begin{center}
\subfigure{
\includegraphics[width=0.35\textwidth]{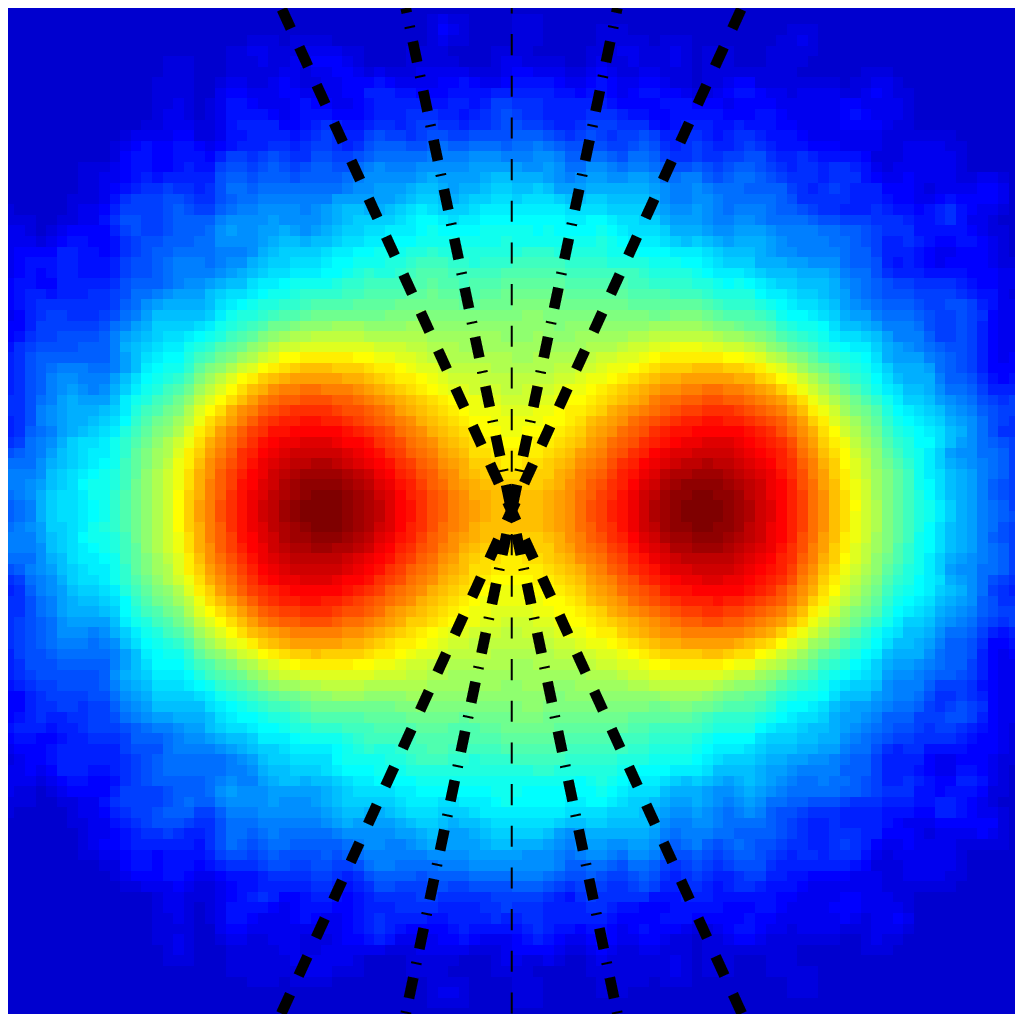}
}
\end{center}
\caption{\label{fig:sectorill} The presence of the shell of scattered DM material can be quantitatively detected by plotting the projected mass distribution in a narrow equatorial sector as the function of the distance from the collision's center. 
Shown in this illustration is such quantification, using two sectors covering $[-15^\circ,15^\circ]$ (bold dash-dotted line) and $[-30^\circ,30^\circ]$ (bold dotted line) angles around the equatorial plane (thin dashed line).
}
\end{figure}

\begin{figure}[h!]
\begin{center}
\subfigure{
\includegraphics[width=0.3\textwidth]{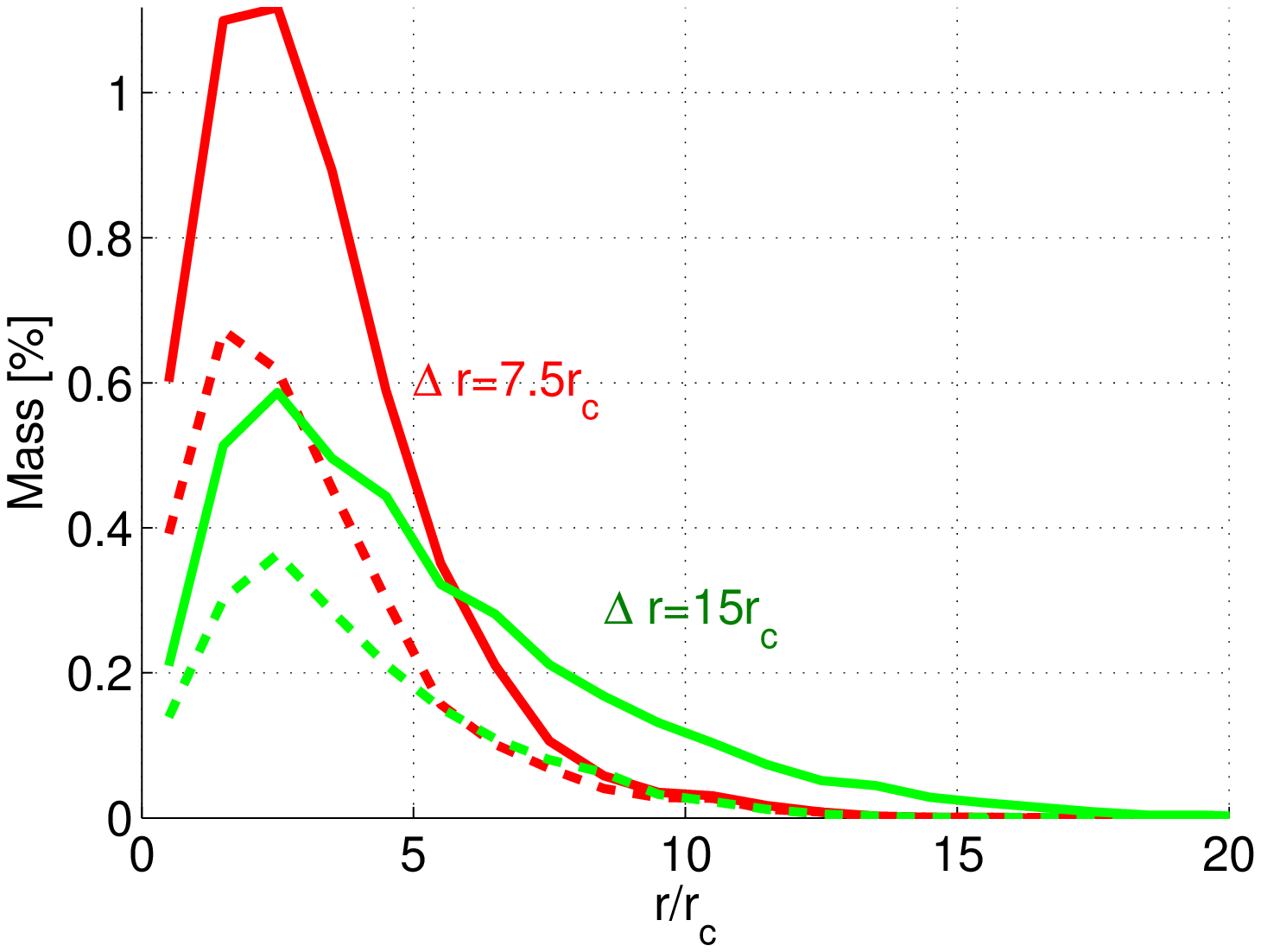}
}
\subfigure{
\includegraphics[width=0.3\textwidth]{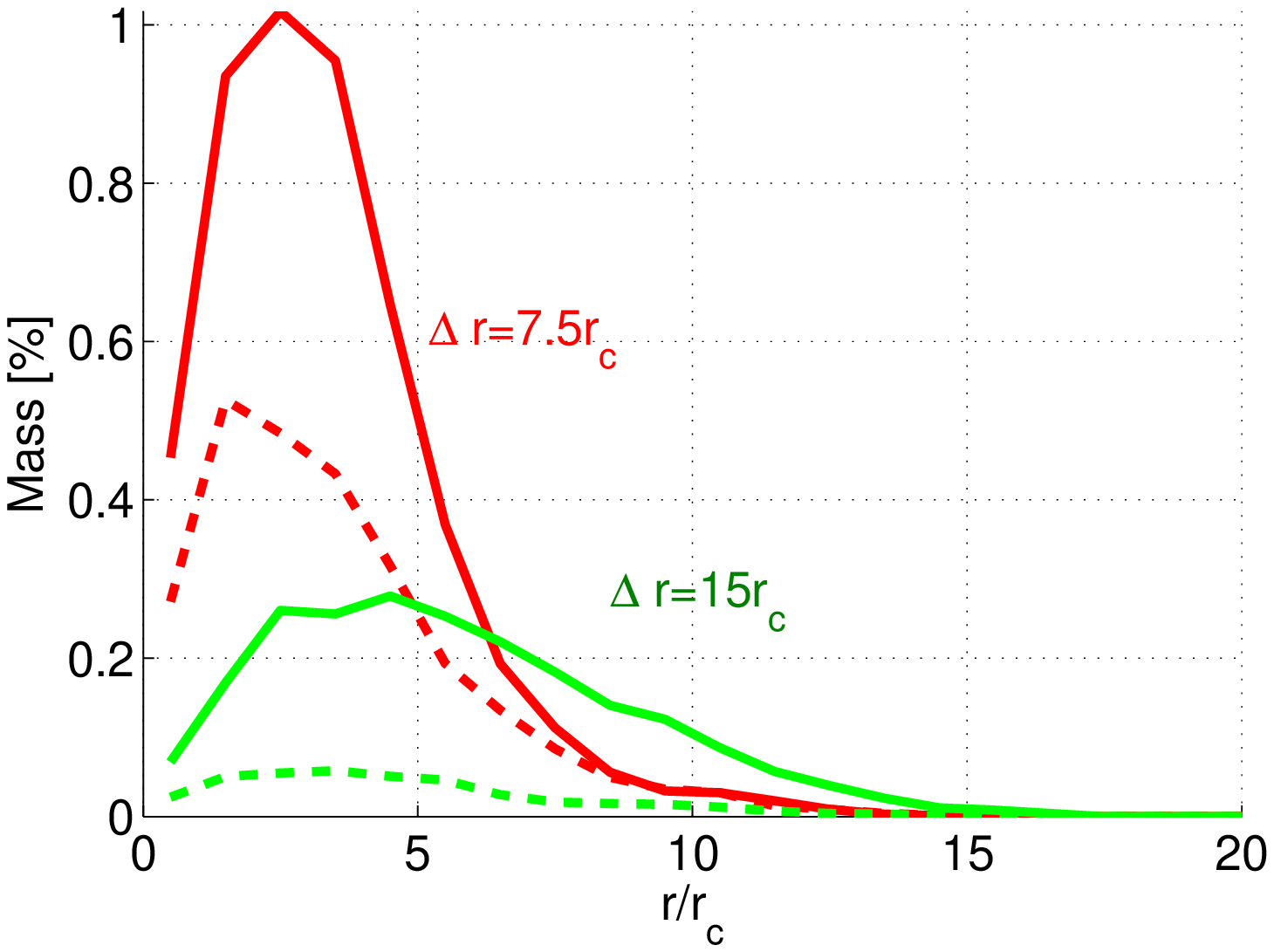}
}
\subfigure{
\includegraphics[width=0.3\textwidth]{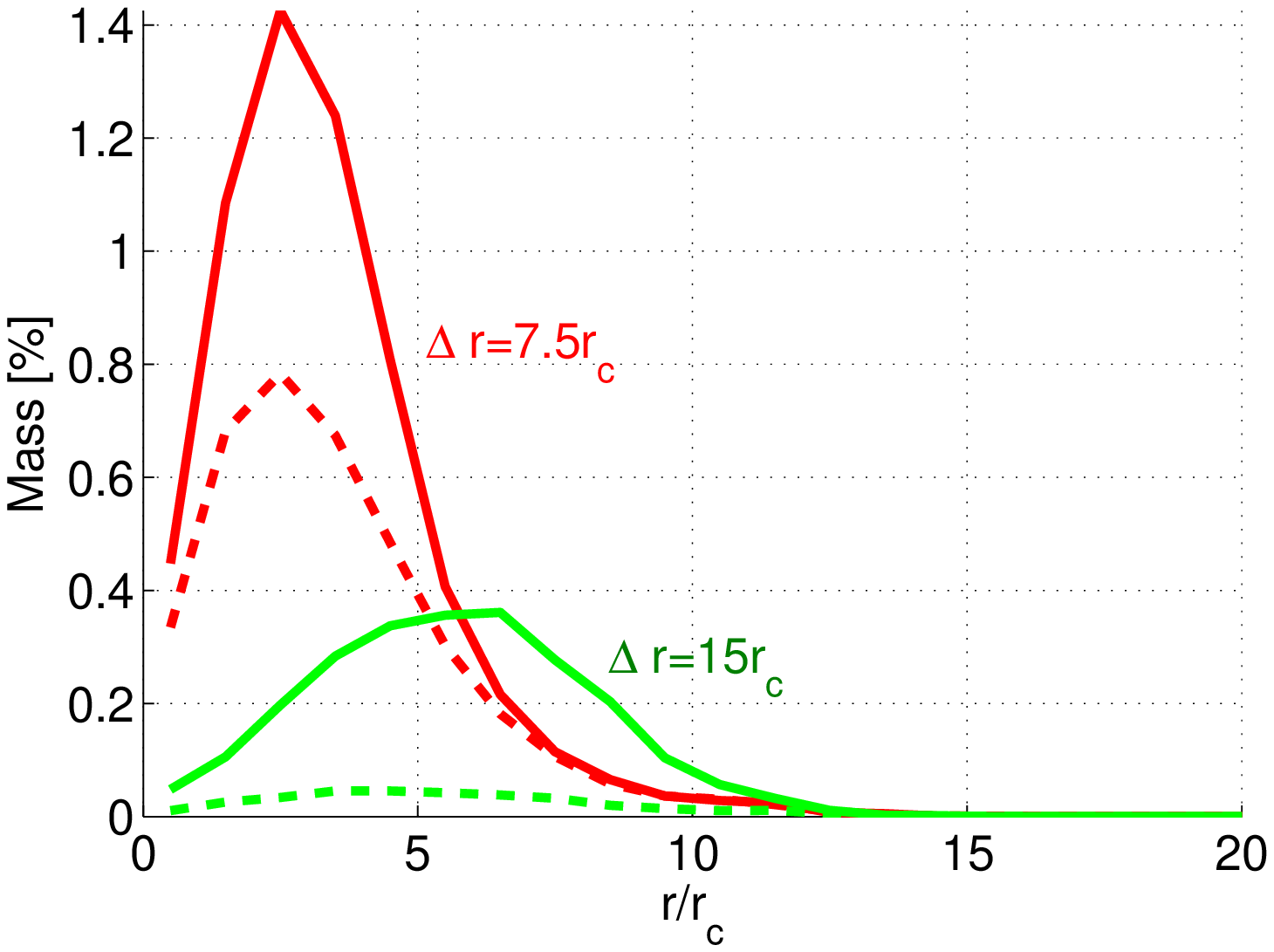}
}
\end{center}
\caption{\label{fig:sectdist30} The plot of the distribution of mass in the projected density maps inside a $[-30^\circ,30^\circ]$ sector around the collision's equatorial plane, as a function of the distance from the collision center. Shown are respective plots for intermediate ($\Delta r=7.5r_c$) and late ($\Delta r=15r_c$) post-collision separations, for weakly interacting DM (solid lines) and CDM (dashed lines), for $k=1.5$ (left), $k=2.0$ (center) and $k=4.0$ (right). 
}
\end{figure}

Yet another quantitative measure of SIDM effects is the presence of DM concentrations in galaxy cluster collisions at large scattering angles and large distances from collision center. 
As illustrated in Fig.~\ref{fig:sectorill},
This feature can be quantified using plots of the mass distribution in projected mass density maps within a narrow $15^\circ$ to $30^\circ$ radial sector built around the equatorial plane of the collision, verses the distance from the collision center. 
Fig.~\ref{fig:sectdist30} shows such measure in the units of percentile of the total collision mass contained in such a sector $[-30^\circ,30^\circ]$ per $r_c$ distance interval. Clear differences between SIDM and CDM are seen either in intermediate and late post-collision stages. The differences are the most profound in late stages of collisions of very fast galaxy clusters. In CDM in that case, there is practically no contribution in the right panel of Fig.\ref{fig:sectdist30} (dashed line for $\Delta r = 15r_c$), as expected, while for SIDM consistent concentration of DM is observed at the correct distances from the center (that is, the distance of the outgoing galaxy clusters). 
Note that the separation $\Delta r$ in Fig.~\ref{fig:sectdist30} is that between the centers of the colliding galaxy clusters, so that the distance from the clusters to the collision center is $\Delta r/2$.

\begin{figure}[h!]
\begin{center}
\subfigure{
\includegraphics[width=0.45\textwidth]{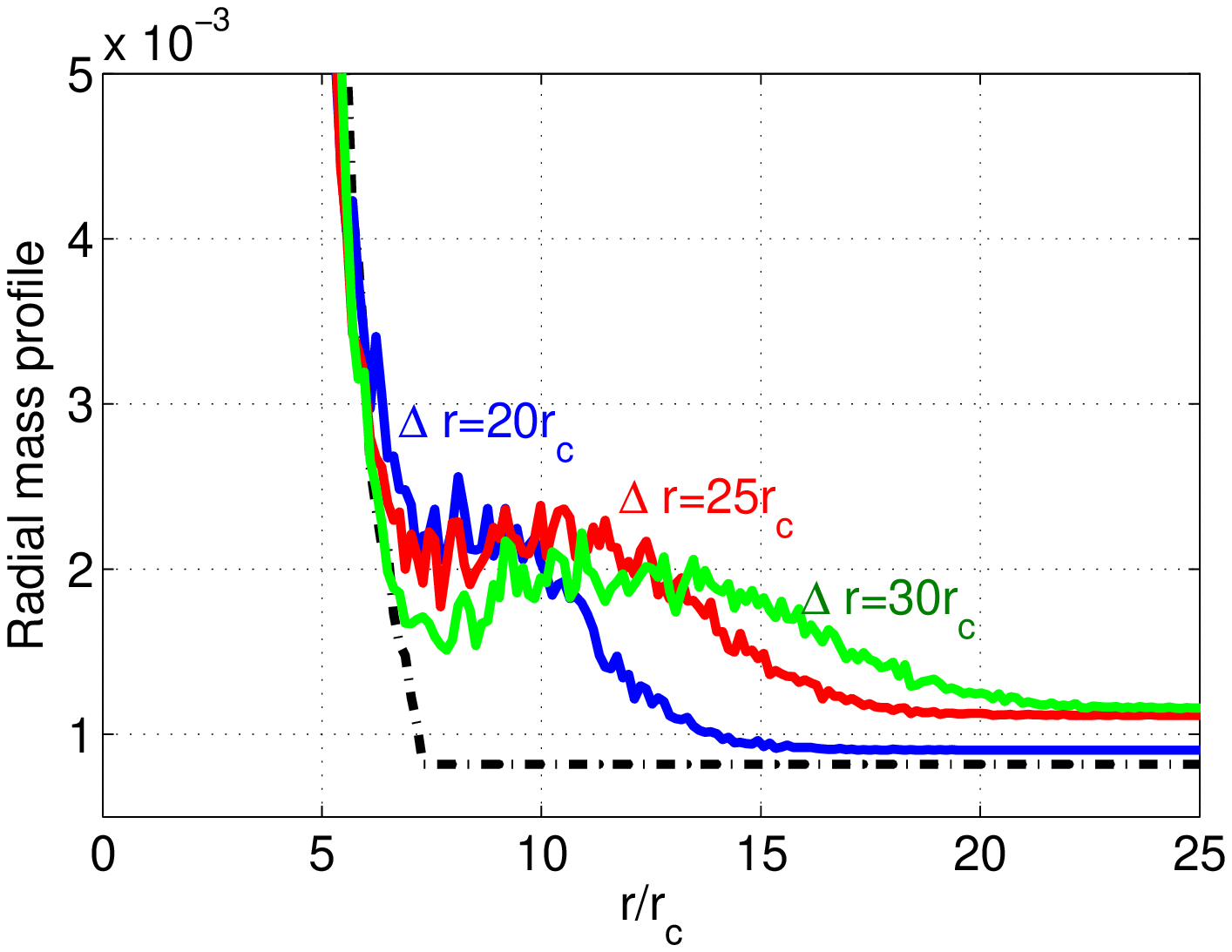}
}
\end{center}
\caption{\label{fig:raddmshell} The scattered DM shell in the tail of the radial mass profile of a very high speed galaxy cluster collision seen along its collision axis, observed at very late post-collision stages. The thick black dash-dotted line illustrates the 
behavior of the galaxy clusters' mass profile at the boundary of the galaxy clusters, near and at the background density, and, thus, the behavior of the radial mass profile expected in the absence of DM scattering.
}
\end{figure}

Finally, an interesting feature that can appear in astrophysical observations of galaxy cluster collisions due to SIDM is a ring surrounding the colliding galaxy clusters in galaxy cluster collisions observed along the collision axis. Such feature in our simulation can emerge at very late separation stages when the shell of scattered DM moves beyond the galaxy cluster's virial radius. Despite the weak magnitude of that effect, such ring-features may be the most dramatic effect of the self-interacting nature of DM. Fig.~\ref{fig:raddmshell} shows an example of such a feature in the tail of the radial mass profile of a simulation of a galaxy cluster collision observed along the axis of the collision.
The size of that feature is just about $0.2\%$ of peak-density, however, its remote location from the collision center may still render such features observable. In fact, we can approximately estimate the relative size of such a feature using the following formula,
\begin{equation}\label{eq:radshellsize}
\frac{\sigma_{shell}}{\sigma_{max}}=\beta a \frac{\log(1+c)-c/(1+c)}{r/r_c},
\end{equation}
where $a$ is the fraction of DM halos scattered into the shell, $c$ is the galaxy cluster's concentration parameter, $r$ is the current radius of the shell, and $\beta\approx 0.5$ is a numerical factor.
In formula (\ref{eq:radshellsize}), we consider that a uniform spherical shell of radius $r$, width (or shell-depth) $h$, and volume-density $\rho$, carrying a total mass $M=4\pi r^2h\rho$, appears in projected density maps as two components: a nearly uniform disk of total mass
$M_{disk}\approx 2\pi r^2h\rho=\frac12M$, and a width $h$ circular rim with the remainder of the mass, that is, having the projected surface density of  $\sigma_{shell}\approx\frac12M/(2\pi r h)$.
For a scattered DM shell carrying a fraction $a$ of the total mass of the collided DM halos, $M_{tot}\approx 2\times 4\pi\rho_0 r_c^3 (\log(1+c)-c/(1+c))$, and having a width of $h\approx 2r_c$, we thus obtain  $\sigma_{shell}\approx a\rho_0 r_c^2 (\log(1+c)-c/(1+c))/r$. For a more general, nonuniform shell of scattered DM material, we write $\sigma_{shell}= 2\beta a \rho_0 r_c^2 (\log(1+c)-c/(1+c))/r$, where $\beta\approx 0.5$ is a numerical factor defining the fraction of the mass contained in the circular rim for specific radial profile of the shell.
Considering that $\sigma_{max}\approx 2r_c\rho_0$, we then can arrive at Eq.~(\ref{eq:radshellsize}).
Eq.(\ref{eq:radshellsize}) allows us to estimate the magnitude of such a scattered DM ring-like feature more generally between 0.1\% and 2\%. 
Interestingly, Ref. \cite{Jee2007} reports observation of such a ring-like DM structure in the long-range reconstructions of the mass profile of the galaxy cluster CL 0024+17, having compatible magnitude between 1\% and 5\% .

\subsubsection{Optimal conditions for observation of SIDM effects} 

As discussed, we find that the most profound SIDM effects can be observed in galaxy cluster collisions in the space between the colliding galaxy clusters but not within them. These differences can be well described in terms of a spherical shell of singly-scattered DM particles engulfing the outgoing galaxy clusters and produced due to DM particle-particle scatterings during the central passage of the galaxy clusters. Such shells of scattered DM material may contain 10\% to 20\% of the entire collision mass without significantly distorting the outgoing galaxy clusters or their DM halos.

For galaxy cluster collisions observed during their early and intermediate post-collision separation stages, the parts of such DM shells scattered directly above and below the center of the collision, along the line of observation, contribute significantly to heavier and wider central regions of the projected mass density maps of such SIDM galaxy clusters, as shown in Figs.~\ref{fig:obscmp15}-\ref{fig:axldist}. 
These differences are very substantial and may account the changes relative to CDM scenarios of up to 10\%-20\%. Up to 2-fold increase can be observed in such central densities in late separation stages in galaxy cluster collisions with lower collision velocity.
Lag of centroids of DM halos can be also observed for SIDM and slower collisions ($k\approx 1.5$), but not fast collisions ($k\geq 2$). 
These are quantitatively measurable by means of projected mass plots on the axial line connecting the centers of the colliding galaxy clusters, as shown in Fig.~\ref{fig:axldist}.

A qualitative new feature observed for SIDM is the presence of expanding shells of scattered DM material engulfing the outgoing galaxy clusters in such galaxy cluster collisions. Such shell contributes mass densities at 0.1\%-5\% peak-density levels at very large distances and large scattering angles relative to the collision center. This feature can be quantified using the azimuthal plots of the mass distribution in projected mass density maps relative to the collision center, and the plots of mass distribution within small (for example, $30^\circ$) radial sectors surrounding the equatorial plane of the collision, as shown in Figs.~\ref{fig:angdist}-\ref{fig:sectdist30}. In such cases, significant deviations from CDM are observed in late post-collision clusters. Whenever the shell of scattered DM materials moves into the region of constant background, a ring-like DM structure can emerge in the mass density observed along the collision axis. 

Given these observations, the conditions for best chance of observing such features comprise collisions of galaxy clusters at high speed, during intermediate or late post-collision separation stages. 
With respect to the collisions' speed, the higher $k$ above 2.0 may be preferred because they lead to much lesser gravitational disruption of colliding galaxy clusters. 
The higher speed of the collision does not affect the DM particle-particle scattering redistributing DM particles azimuthally, as such speed cancels out from the DM particle-particle scattering, but reduces the efficiency of gravitational effects. 
Tables \ref{table:kinetic2} and \ref{table:kinetic} list the initial in-fall and the closest-approach speed corresponding to different values of $k$ for different total masses of the collision $M_{tot}$.
As can be seen from these tables, for very heavy galaxy clusters the initial speed required to achieve $k\approx 2-4$ is high but not prohibitively so. 
Such speeds can be achieved if the galaxy clusters collide after in-falling towards an element of the large-scale structure or within the context of galaxy clusters' relative motion within a superstructure such as galaxy supercluster. 
Examples of galaxy cluster collisions hypothesized to have occurred near a filament are known in the literature \cite{Jee2012}.
Note that while the closest-approach speeds may seem to be high on first inspection, most of such speed is the speed gained due to gravity during in-fall.


\begin{table}[ht]
\caption{The initial clusters' in-fall speeds in relation to the collision's kinetic parameter $k$ and the collision's total mass $M_{tot}$. 
}
\centering
\begin{tabular}{cccccc}
\hline\hline
k & $10^{15} M_\odot$ & $5\cdot10^{14} M_\odot$ & $10^{14} M_\odot$ & $5\cdot10^{13} M_\odot$ & $10^{13} M_\odot$ \\
\hline
1 & 0 kmps & 0 kmps & 0 kmps & 0 kmps & 0 kmps \\
2 & 2100 kmps & 1700 kmps & 1000 kmps & 800 kmps & 470 kmps \\
4 & 3600 kmps & 2900 kmps & 1700 kmps & 1400 kmps & 780 kmps \\
8 & 5600 kmps & 4400 kmps & 2500 kmps & 2100 kmps & 1200 kmps \\
\hline
\end{tabular}
\label{table:kinetic2}
\end{table}

\begin{table}[ht]
\caption{The closest approach relative collision speeds in relation to the the collision's kinetic parameter $k$ and the collision's total mass $M_{tot}$. 
}
\centering
\begin{tabular}{cccccc}
\hline\hline
k & $10^{15} M_\odot$ & $5\cdot10^{14} M_\odot$ & $10^{14} M_\odot$ & $5\cdot10^{13} M_\odot$ & $10^{13} M_\odot$ \\
\hline
1 & 4200 kmps & 3300 kmps & 1900 kmps & 1500 kmps & 900 kmps \\
2 & 5900 kmps & 4700 kmps & 2800 kmps & 2200 kmps & 1300 kmps \\
4 & 8400 kmps & 6600 kmps & 3900 kmps & 3100 kmps & 1800 kmps \\
8 & 11900 kmps & 9300 kmps & 5500 kmps & 4400 kmps & 2600 kmps \\
\hline
\end{tabular}
\label{table:kinetic}
\end{table}

\section{Summary and Discussion}
\label{sec:conclusions}

In this work, we performed study of possible configurations of post-collision mass distributions in high-speed galaxy cluster collisions with respect to different hypothetical self-interaction strengths of DM.

All such scenarios can be characterized essentially by two main parameters comprising the ratio of the kinetic and gravitational energy of the collision, $k$, and the fraction of DM halo mass scattered in the collision by DM particle-particle interactions, $a$.

With respect to the kinetic parameter $k$, we observe three main regimes of collisions. 
Collisions with very high speed, $k>2$, feature the galaxy clusters passing through each other with little gravitational disturbance. The azimuthal DM particles redistribution effect of weak DM self-scattering is clearly discernible in this setting. 
For collisions with $2>k>1$, gravitational effects produce ``fan-out'' DM ejecta mostly confined to small scattering angles in forward and backward cones of the collision. 
In that case, DM self-scattering effects can manifest themselves as discernible new mass concentrations either in central and equatorial regions of the collisions' projected mass density maps, where the efficiency of the gravitational DM scattering is the lowest.
For yet slower collisions, the kinetic energy of the colliding clusters is insufficient to provide for their post-collision separation and rapid mergers are observed with significant and disperse gravitational ejecta of complex shapes and large extents. 

With respect to the strength of DM self-interactions, we find 
that for strong DM self-scattering in which 50\% or more of DM halo mass suffers non-gravitational scattering during the passage, $\sigma_{DM}/m_{DM}>2 cm^{-2}g$,  the DM halos are destroyed in the passage. This disruption is severe and results in formation of a single common halo composed of heated DM material. As such, this outcome is far beyond the limited effects such as changes in mass-to-light ratio or a lag of DM halo centroids previously discussed in the literature. Instead, complete and rapid reorganization of the  entire DM halo of the colliding galaxy clusters is observed. 
%
For weak DM self-scattering in which 10\% to 20\% of DM halo particles suffer a non-gravitational scattering, $\sigma_{DM}/m_{DM}<0.5 cm^{-2}g$, the formation of spherical shells of DM particles is observed. This can be understood as the outcome of DM particle-particle scattering during the time of DM halos' central passage through each other, under the conditions of the mean free path of DM particles being significantly greater than the size of the halos, $a\ll 1$.
The shell of such scattered DM material can be observed in the projected mass density maps such as obtained via gravitational lensing under certain conditions. 
The said DM structure can be discerned in the projected mass density maps as DM contributions at very large scattering angles and large distances from the collision center or as extended disk and ring-like structures of the size comparable to the post-collision separation of outgoing galaxy clusters. 
Such features are forbidden in purely gravitational collisions, as discussed above.
Such DM self-scattering structures can admit up to 20\% of the collision's total mass before any significant disruptions begin to be noticeable in the main galaxy clusters' halos. 
%
The remote location of the scattered DM shell either from the outgoing galaxy clusters or central hot ICM may allow such structures to be seen experimentally despite their weak magnitude. 
The survey of the literature on weak and strong gravitational lensing reconstructions of mass profiles of colliding galaxy clusters shows that the presence of DM densities at large scattering angles and large distances indeed is a wide-spread feature of such observations. For instance, the reconstructed projected mass density of the colliding galaxy clusters A754 and A520  show both very significant off-axial concentrations of DM at scattering angles close to $90^\circ$ and at the same distance from the collision center as the collided galaxy groups \cite{Okabe2008, Jee2012}. Combined weak and strong lensing reconstructions of the mass density of the Bullet cluster shows large diffuse DM mass concentration in between the outgoing galaxy groups, having disk shape and the size once again similar to the separation of the galaxy groups \cite{Bradac2006}. 
Yet more interesting observation has been produced in recent reconstruction of the mass profile of strongly lensing galaxy cluster CL0024+017,  now figured as colliding galaxy clusters seen along the axis of the collision, performed to large distances from the center \cite{Jee2007}. Such reconstruction implicated a weak ring-like DM structure existing around the collided galaxy clusters, as shown in Fig. 7 and Fig. 10 of Ref. \cite{Jee2007}.
The structure is consistent with the features observed in simulations in this work and has the magnitude of 1-5\%.
It is possible to suggest an interpretation of that structure as the remains of a shell of scattered DM material generated in such an ancient galaxy cluster collision.

\acknowledgements
This work was supported in part by the American Physical Society International Travel Grant Award Program (APS ITGAP) and in part by the US Department of Energy under Contract NO. DE-FG02-03ER41260. 
This research also used the resources of the National Energy Research Scientific Computing Center (NERSC), which is supported by the Office of Science of the U.S. Department of Energy under Contract No. DE-AC02-05CH11231. 
YM also would like to acknowledge the support from the Bilim Akademisi---The Science Academy (Istanbul, Turkey) young investigator award under the BAGEP program.

\appendix
\section{Dark Matter Particle Scattering Algorithm and Particle Mesh Algorithm}
\label{AppA}
\algsetup{indent=0.5cm}
\begin{algorithm}[H]
\caption{DM particle-particle scattering algorithm}
\label{alg:scattering}
\begin{algorithmic}
\FORALL{$cell \in \mathcal{G}$}
  \FORALL{particles $i,j \in cell$}
    \STATE Select the pair $(i,j)$ with probability $P=\alpha|\vec{v}_i-\vec{v}_j|\Delta t$
    \IF {selected}
    \STATE $\vec{V}_{CM}\leftarrow (\vec{v}_i+\vec{v}_j)/2$
    \STATE $V_{rel}\leftarrow |\vec{v}_i-\vec{v}_j|/2$
    \STATE Choose $\vec{n}$ uniformly at random on unit sphere
    \STATE $\vec{v}_i\leftarrow\vec{V}_{CM}+\frac12V_{rel}\vec{n}$
    \STATE $\vec{v}_j\leftarrow\vec{V}_{CM}-\frac12V_{rel}\vec{n}$
    \ENDIF
  \ENDFOR
\ENDFOR
\end{algorithmic}
\end{algorithm}

\begin{algorithm}[H]
\caption{Particle Mesh algorithm for $N$-body gravitational dynamics}
\label{alg:partmesh}
\begin{algorithmic}
\REQUIRE simulation parameters $M_{tot},N1,N2,\Delta V2,\Delta R1,\Delta R2,\Delta b$
\REQUIRE $\mathcal{G}$ is a cubic 3D grid of size $D^3$
\REQUIRE $list \leftarrow \{(\vec{r}_i,\vec{v}_i)\}$ is the list of particles' position and velocity vectors
\STATE \COMMENT {Form initial particle distributions for the two colliding clusters}
\STATE $list1\leftarrow N1$ random particles from the standard equilibrium profile (Section \ref{sec:alg-initial-profile})
\STATE $list2\leftarrow N2$ random particles from the standard equilibrium profile (Section \ref{sec:alg-initial-profile})
\STATE $scaling1\leftarrow (({M_{tot}}{N1}/{(N1+N2)})/{M_{std}})^{1/3}$
\STATE $scaling2\leftarrow (({M_{tot}}{N2}/{(N1+N2)})/{M_{std}})^{1/3}$
\STATE Update all particles in $list1: \vec r_i\leftarrow \vec{r}_i \cdot scaling1,\vec{v}_i\leftarrow\vec{v}_i\cdot scaling1$
\STATE Update all particles in $list2: \vec r_i\leftarrow \vec{r}_i \cdot scaling2,\vec{v}_i\leftarrow\vec{v}_i\cdot scaling2$
\STATE Update all particles in $list1: (\vec r_i)_x\leftarrow (\vec r_i)_x-\Delta R1$, $(\vec r_i)_y\leftarrow (\vec r_i)_y-\Delta b$
\STATE Update all particles in $list2: (\vec r_i)_x\leftarrow (\vec r_i)_x+\Delta R2$
\STATE
\STATE \COMMENT {Calculate initial in-fall velocities}
\STATE $\Delta R\leftarrow \Delta R1+\Delta R2$
\STATE $E_G\leftarrow \frac{G(M_{tot})^2}{\Delta R}(\frac{N1}{N1+N2})(\frac{N2}{N1+N2})$
\STATE $V2\leftarrow \frac{2E_G}{M_{tot}}+\Delta V2$
\STATE Update all particles in $list1: (\vec v_i)_x\leftarrow (\vec v_i)_x+\sqrt{\frac{N2}{N1}V2}$
\STATE Update all particles in $list2: (\vec v_i)_x\leftarrow (\vec v_i)_x-\sqrt{\frac{N1}{N2}V2}$
\STATE
\STATE $list\leftarrow \{list1,list2\}$
\STATE
\STATE \COMMENT {Main simulation loop}
\WHILE{$t<t_{max}$}
\STATE $\Delta t\leftarrow \min(\Delta t_{max},\Delta r_{max}/\max(|\vec{v}_i|))$
\STATE Discard the particles that moved out of the bounds of $\mathcal{G}$
\STATE Distribute the particles in $list=\{\vec r_i,\vec v_i\}$  into the cells of grid $\mathcal{G}$ based on $\vec{r}_i\in cell$, $n(\vec{x})\leftarrow list\{ \vec{r}_i \}$
\STATE $\tilde{n}(\vec{k})\leftarrow FFT(n(\vec{x}))$
\STATE $\tilde{\Phi}(\vec{k})\leftarrow -4\pi\tilde{n}(\vec{k})/k^2$ (but set $\tilde{\Phi}(0)\leftarrow 0$)
\STATE $\Phi(\vec{x})\leftarrow iFFT(\tilde{\Phi}(\vec{k}))$
\STATE For all particles in $list: \vec{a}_i\leftarrow -G \nabla \Phi(\vec{r}_i)$
\STATE $\Delta t\leftarrow \min(\Delta t,\Delta v_{max}/\max(|\vec{a}_i|))$
\STATE Update all particles in $list: \vec v_i\leftarrow \vec v_i+\vec{a}_i\Delta t$
\STATE Evaluate particle-particle scatterings using Algorithm \ref{alg:scattering}
\STATE Update all particles in $list: \vec r_i\leftarrow \vec r_i+\vec{v}_i\Delta t$
\STATE $t\leftarrow t+\Delta t$
\ENDWHILE
\end{algorithmic}
\end{algorithm}

\section{Convergence and accuracy of the numerical integration method with respect to the SIDM effects}
\label{AppB}

The Particle Mesh algorithm used in this work can experience accuracy loss at small scales due to the finite size of the meshgrid used for approximating the dynamical equations. To control for this effect and its impact on the SIDM features elucidated in this work, we simulate the fast CDM and symmetric-central-weak SIDM scenarios from Table \ref{table:params} ($k=1.6$) with different sizes of the meshgrid ranging from $D=100^3$ to $D=600^3$ points, and different numbers of particles used in the simulation from $N=100\cdot 10^3$ to $N=400\cdot 10^3$. 

\begin{figure}[h!]
\begin{center}
\subfigure{
\includegraphics[width=0.3\textwidth]{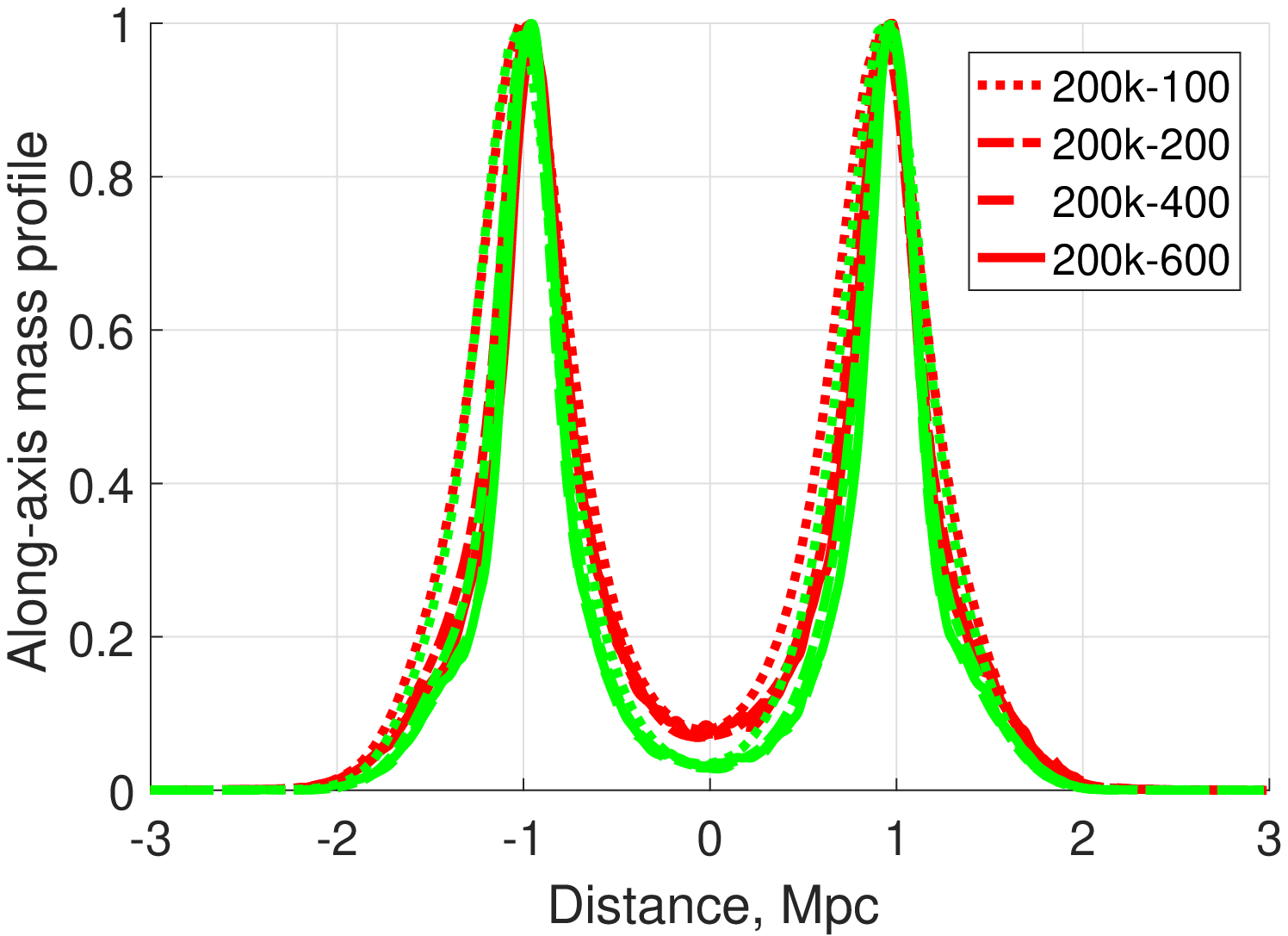}
}
\subfigure{
\includegraphics[width=0.3\textwidth]{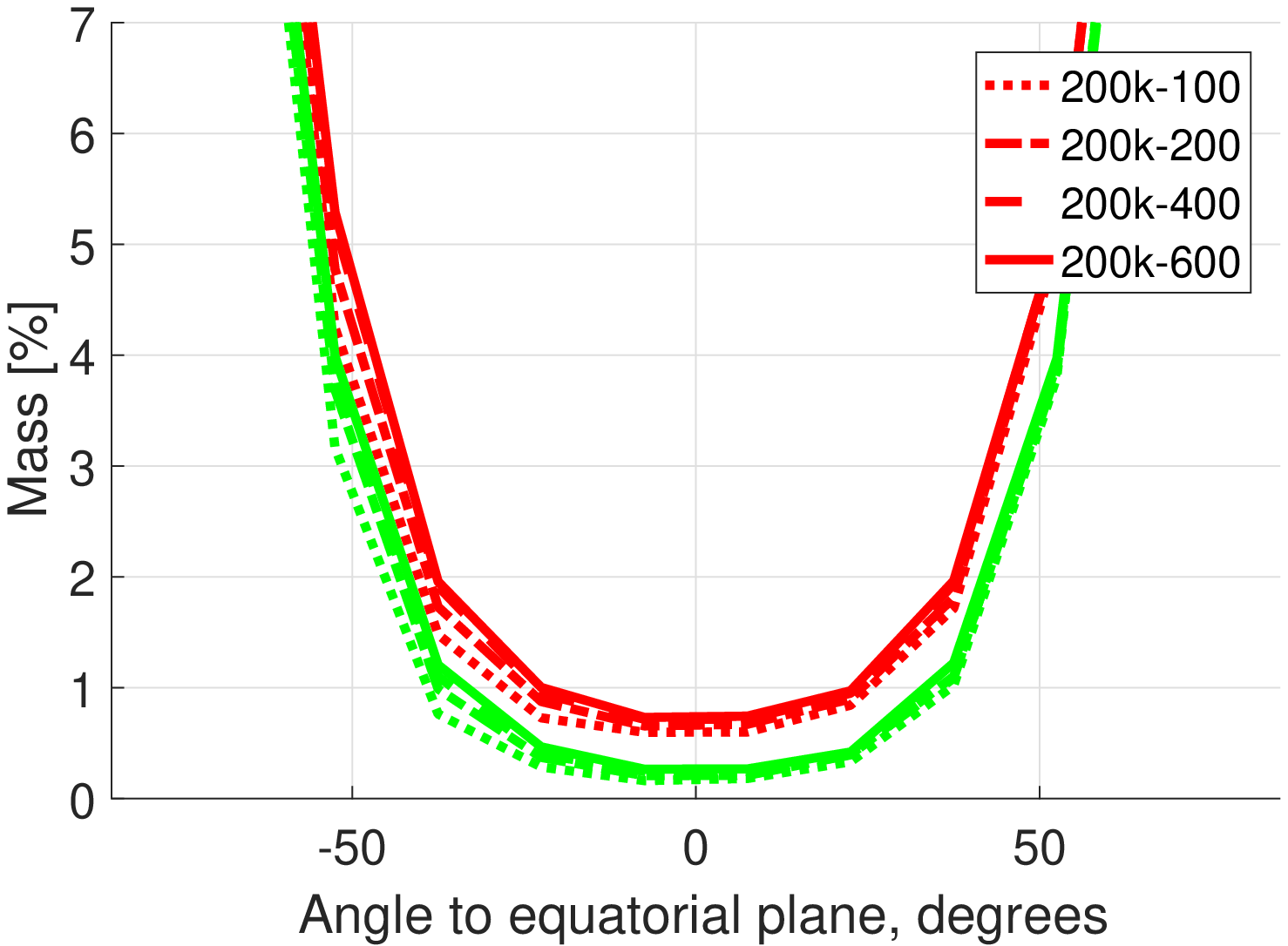}
}
\subfigure{
\includegraphics[width=0.3\textwidth]{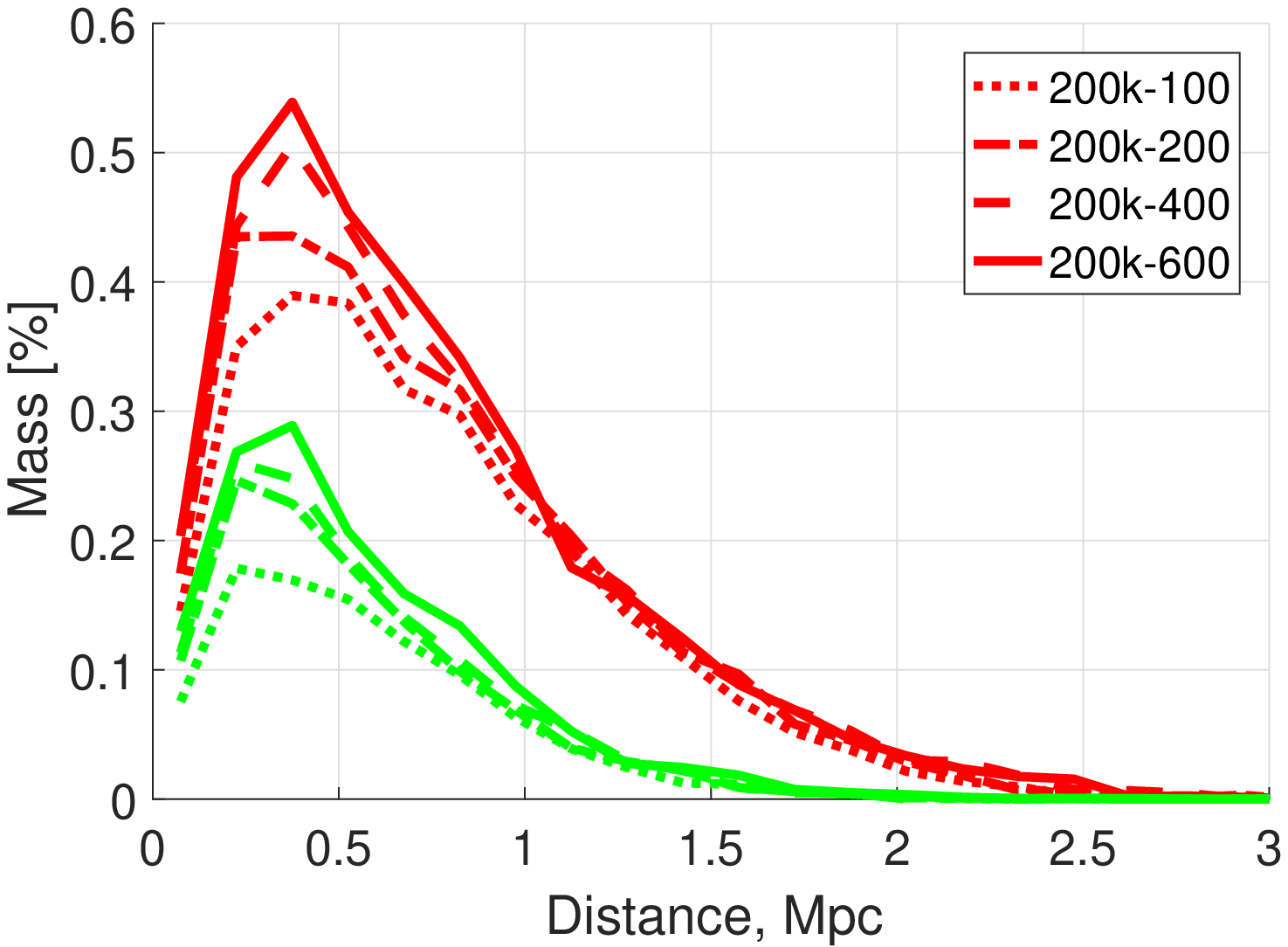}
}
\end{center}
\caption{\label{fig:convergence_res} The three SIDM-versus-CDM effects in axial (left), azimuthal (center), and radial (right) distribution of dark matter in a high speed galaxy cluster collision inspected with respect to varying Particle Mesh algorithm's grid size, using $N=200\cdot 10^3$ particles in all simmulations. 
Red lines are for SIDM and green lines are for CDM simulations. The convergence of the simulations past the grid sizes of $D=400^3$ is clearly seen.}  
\end{figure}

\begin{figure}[h!]
\begin{center}
\subfigure{
\includegraphics[width=0.3\textwidth]{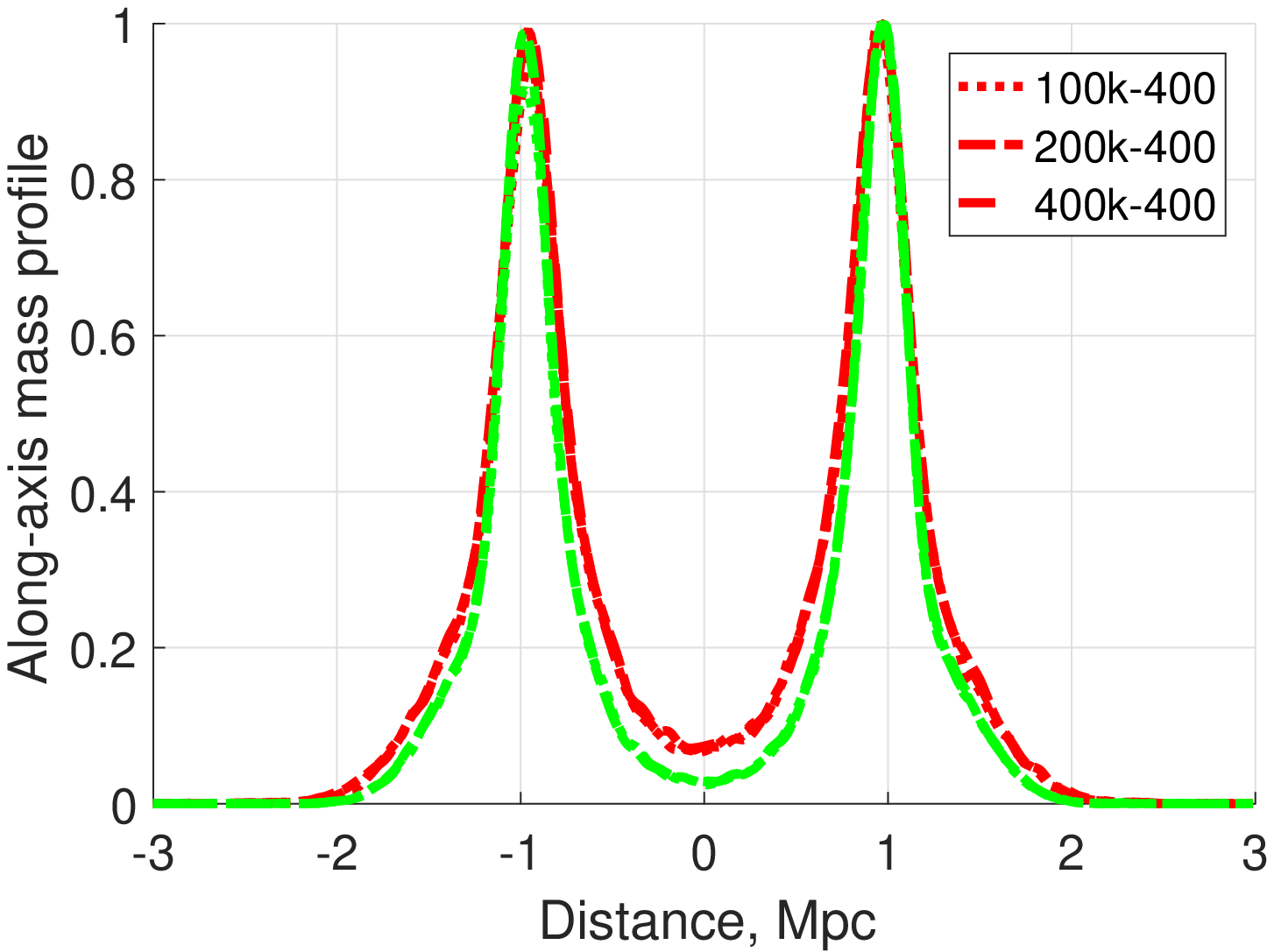}
}
\subfigure{
\includegraphics[width=0.3\textwidth]{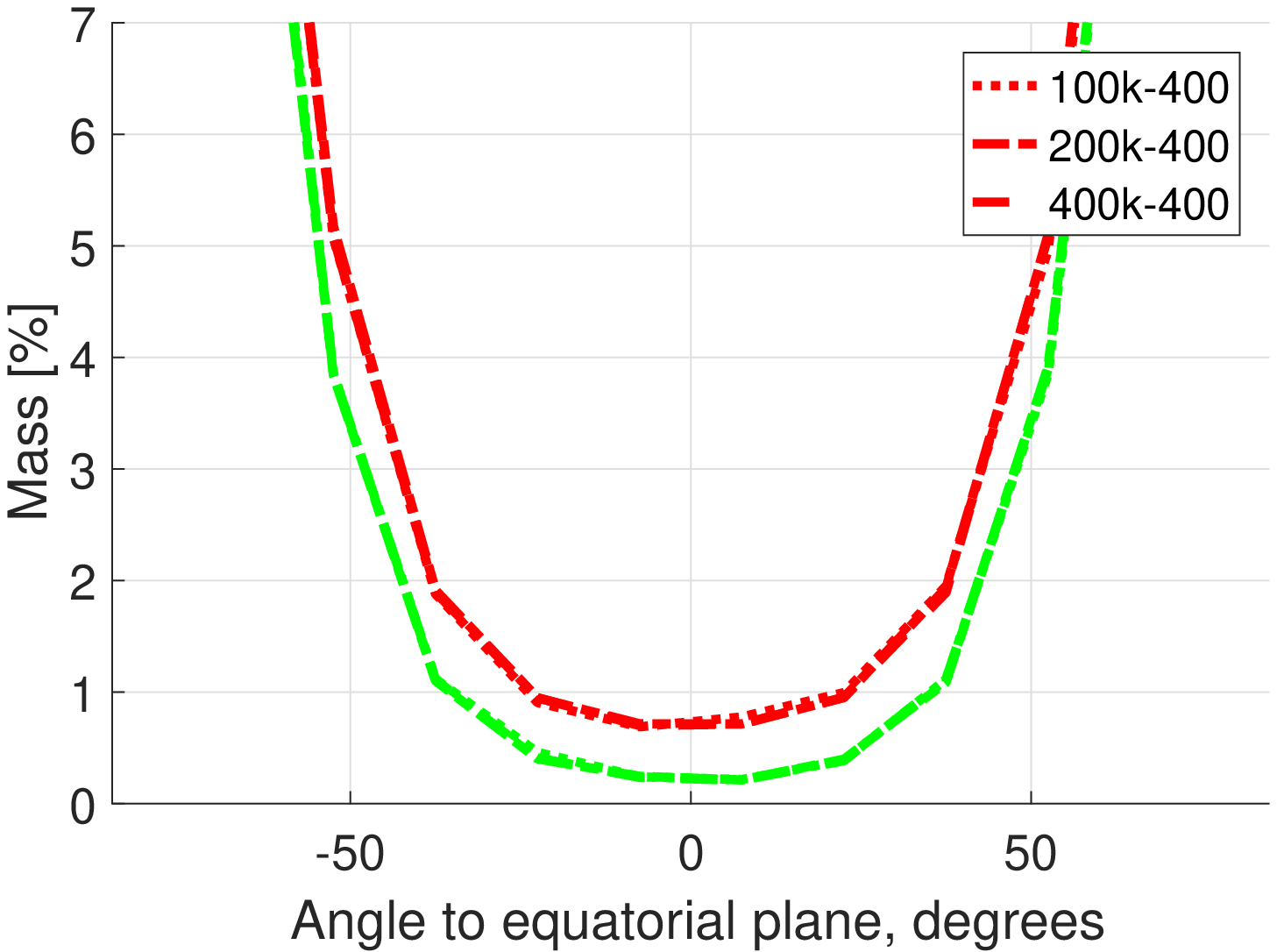}
}
\subfigure{
\includegraphics[width=0.3\textwidth]{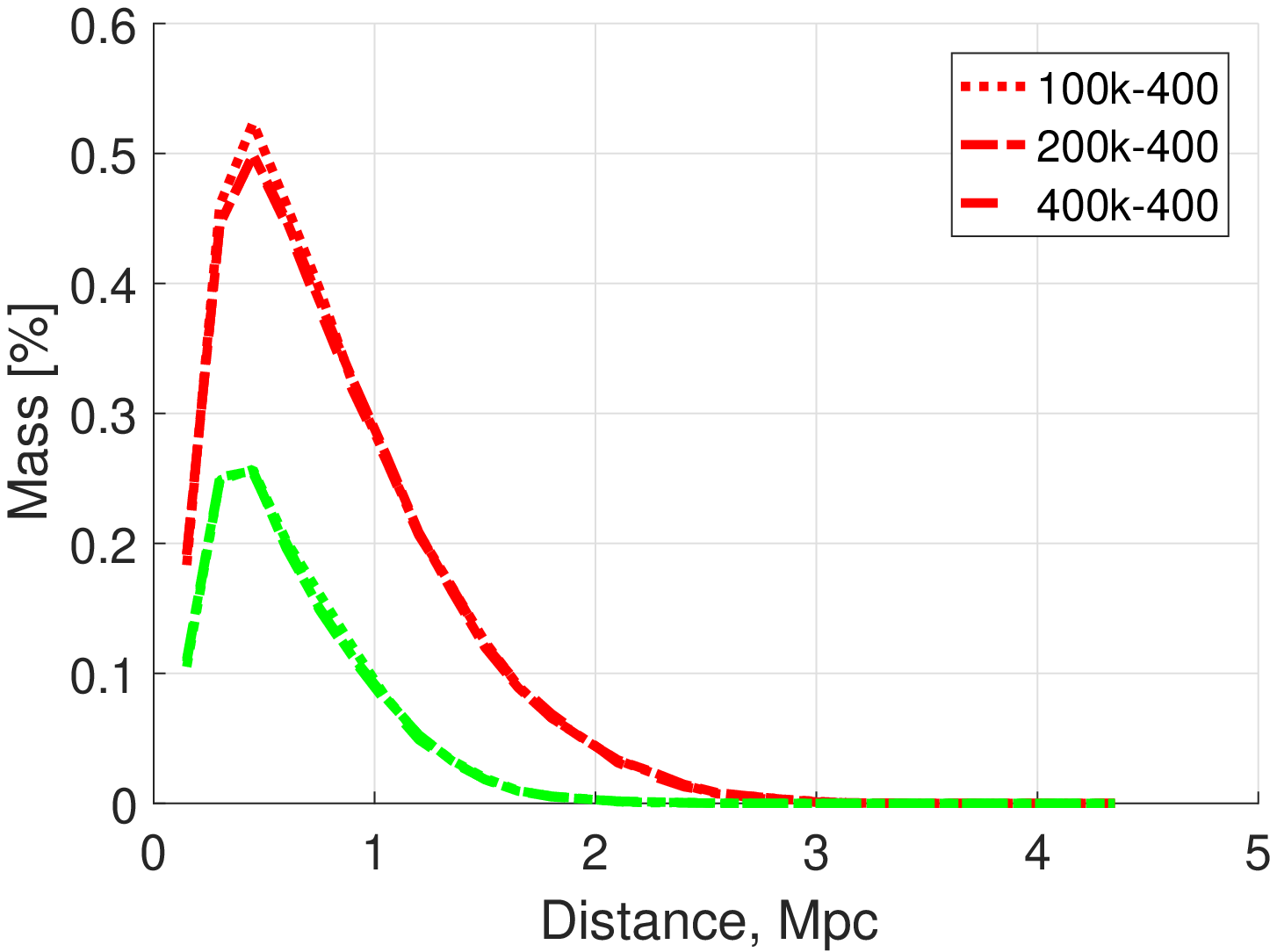}
}
\end{center}
\caption{\label{fig:convergence_mc} The three SIDM-versus-CDM effects in axial (left), azimuthal (center), and radial (right) distribution of dark matter in a high speed galaxy cluster collision inspected with respect to varying the number of particles in the simulation.
Red lines are for SIDM and green lines are for CDM simulations.
The convergence of the simulations past $N=200\cdot 10^3$ particles is clearly seen in all cases. $D=400^3$ has been used in all these simulations. 
}
\end{figure}

The results of these numerical experiments are shown in Figs. \ref{fig:convergence_res} and \ref{fig:convergence_mc}. 
The red and green curves represent the symmetric-central-weak SIDM results and the fast CDM results, respectively. Here, the distance between the centers of 
the outgoing galaxy clusters is given by $\Delta r = 15 r_c$, corresponding to the late separation stage as defined in the main text. In Fig. \ref{fig:convergence_res}, the dotted, dash-dotted, dashed and solid lines represent the different sizes of the meshgrid as $D = 100^3, 200^3, 400^3$ and $600^3$, respectively, while the number of particles is fixed at 
$N=200\cdot 10^3$. In Fig. \ref{fig:convergence_mc}, the dotted, dash-dotted and dashed lines represent the different numbers of particles used in the simulation as $N = 100\cdot 10^3,
200\cdot 10^3$ and $400\cdot 10^3$, respectively, while the size of the meshgrid is fixed at $D = 400^3$ points. 

We observe that in all cases sufficient convergence is achieved by $D=400^3$ and $N=200\cdot 10^3$, which is the choice of the parameers used in the simulations in this work.
The axial (along-the-collision-axis) mass distributions and the sector-azimuthal mass distributions, as introduced in the main text, are not very sensitive to the above simulation parameters, whereas the only significant difference in these quanities is observed for the grid size of $D=100^3$. The radial distribution of DM mass in the equatorial sector of the projected mass density maps is more sensitive, due to the original smallness of that effect. However, even in that case $D=400^3$ and $N=200\cdot 10^3$ suffice to achieve acceptable convergence. 

\bibliography{GCC}


\end{document}